\documentclass[aps,reprint,superscriptaddress,showpacs,amsmath, amssymb,prb,floatfix,longbibliography]{revtex4-2}
\usepackage{amsmath,amssymb,bm,graphicx,color,gensymb,bbold,appendix,hyperref}
\usepackage{latexsym}
\usepackage{dcolumn}
\usepackage{epstopdf}
\usepackage{ulem}
\usepackage{multirow}
\usepackage{lipsum}

\usepackage{amsthm}
\usepackage{amsfonts}

\begin{document}

\title{Fingerprinting Triangular-Lattice Antiferromagnet by Excitation Gaps}

\author{K. E. Avers}
\affiliation{Los Alamos National Laboratory, Los Alamos, NM 87545, USA}
\affiliation{Department of Physics and Astronomy, Northwestern University, Evanston, IL, USA}
\affiliation{Center for Applied Physics \& Superconducting Technologies, Northwestern University, Evanston, IL, USA}
\author{P. A. Maksimov}
\affiliation{Bogolyubov Laboratory of Theoretical Physics, Joint Institute for Nuclear Research,
Dubna, Moscow region, 141980, Russia}
\author{P. F. S. Rosa}
\affiliation{Los Alamos National Laboratory, Los Alamos, NM 87545, USA}
\author{S. M. Thomas}
\affiliation{Los Alamos National Laboratory, Los Alamos, NM 87545, USA}
\author{J.~D.~Thompson}
\affiliation{Los Alamos National Laboratory, Los Alamos, NM 87545, USA}
\author{W. P. Halperin}
\affiliation{Department of Physics and Astronomy, Northwestern University, Evanston, IL, USA}
\author{R. Movshovich}
\affiliation{Los Alamos National Laboratory, Los Alamos, NM 87545, USA}
\author{A. L. Chernyshev}
\affiliation{Department of Physics and Astronomy, University of California, Irvine, California 92697, USA}

\date{\today}

\begin{abstract}
CeCd$_3$As$_3$ is a rare-earth triangular-lattice antiferromagnet with large inter-layer separation. 
Our field-dependent heat capacity measurements at dilution fridge temperatures allow us to trace the 
field-evolution of the spin-excitation gaps throughout the 
antiferromagnetic  and paramagnetic regions. The distinct gap evolution places strong 
constraints on the microscopic pseudo-spin model, which, in return, yields
a close {\it quantitative} description of the gap behavior. 
This analysis provides crucial insights into the nature of the magnetic  state of CeCd$_3$As$_3$,
with a certainty regarding its stripe order and low-energy model parameters that sets a compelling 
paradigm for exploring and understanding the rapidly growing family of the rare-earth-based 
triangular-lattice systems.
\end{abstract}

\maketitle
Rare-earth-based quantum magnets are of great current interest, as they 
naturally combine the effects of  strong spin-orbit coupling (SOC), which entangles magnetic degrees of 
freedom and orbital orientations \cite{Witczak,Balents_Nature_2010}, with that of the geometric frustration 
of the lattices bearing a triangular motif, long anticipated as leading to exotic ground states and 
excitations \cite{Anderson_MatResBull_1973,Tsirlin_JPCM_2020}.
A remarkable recent surge in the studies of quantum anisotropic-exchange magnets 
in general and rare-earth based triangular-lattice (TL) materials in particular is propelled by an avalanche of the 
newly synthesized compounds \cite{Li_ScientificReports_2015,Sanders_MatResExp_2017, Bordelon2019, %
Baenitz_PRB_2018, Sichelschmidt_arxiv_2019, Sichelschmidt_JPCM_2019, Cevallos_MRB_2018, Li_arxiv_2018, Liu_CPL_2018, Ding_PRB_2019, Ranjith_PRB_2019, Ranjith2_PRB_2019, Xing_PRB_2019, Xing_PRM_2019, %
Xing_ACSML_2020, Scheie_arxiv_2020, Ma_arxiv_2020, Kabeya_JPSJ_2020} 
and theoretical insights into their models
\cite{Tsirlin_JPCM_2020, Li_PRB_2016, Rau_PRB_2018, Zhu_PRL_2017,Zhu_PRL_2018,Maksimov_PRX_2019,Maksimov_PRR_2020}.
Rare-earth materials provide an ideal platform for the search of  novel phases
as they  host short-range and highly anisotropic exchange interactions of 
their effective spin degrees of freedom due to strong SOC and 
highly-localized nature of the $f$-orbitals.  

In this Letter, we present low-temperature heat capacity measurements of a Ce-based 
representative member of this family of materials, CeCd$_3$As$_3$,
in a regime that has not been accessed in the earlier studies \cite{Dunsiger_PRB_2020}. 
These measurements enable us to construct 
the magnetic field--temperature ($H$--$T$) phase diagram of CeCd$_3$As$_3$ and determine the 
field-evolution of the spin-excitation gaps in its spectrum throughout the transition from a 
frustrated antiferromagnetic (AF) to a paramagnetic (PM) state. 
We augment these results by a theoretical analysis that yields a close agreement
with a distinct field-dependence of the gaps and allows us
to unequivocally identify the ground state of CeCd$_3$As$_3$ as being in a  
stripe phase. The phenomenological constraints on the key parameters of the 
microscopic model result in robust certainty regarding the ground state and 
parameter region of the general phase diagram of the anisotropic-exchange TL 
systems to which CeCd$_3$As$_3$ likely belongs. Our approach is expected to enable a better understanding 
of the rapidly growing family of rare-earth-based TL materials. 

{\it Material and Methods.}---%
CeCd$_3$As$_3$ crystallizes as thin ($\sim$1$\times$1$\times$0.2 mm), plate-like crystals in the 
hexagonal $P6_3/mmc$ space group in which Ce$^{+3}$ ions form 2D triangular lattices that are 
widely spaced along the c-axis, as shown in Supplemental Material (SM)~\cite{StripeSuppliment}.
Its low-temperature magnetization is characteristic of an easy-plane 
anisotropy with a ratio of 
Lande g-factors g$_{ab}$/g$_{c}\!\sim\!5$~\cite{StripeSuppliment, Banda_PRB_2018, Dunsiger_PRB_2020}.
Additional experimental details and crystal-field analysis of magnetic susceptibility/magnetization are provided 
in SM~\cite{StripeSuppliment}. 
Here we focus on our study of the specific heat and its theoretical modeling. 

{\it $H$--$T$ phase diagram.}---%
In zero field, CeCd$_3$As$_3$ orders at $T_N$=410(20)~mK in agreement with a 
 recent report \cite{Dunsiger_PRB_2020} of AF ordering at $T_N \!\sim\! 420$~mK.
At $T_N$, only about one quarter of  the entropy of the ground state doublet $R\ln 2$ is 
recovered \cite{StripeSuppliment}.
The substantial difference between $T_N$ and the $ab$-plane Curie-Weiss (CW) temperature 
$\theta_{CW}\!=\!-4.5$~K also suggests a large degree of frustration in CeCd$_3$As$_3$ 
\cite{Liu_Arxiv_2016}, although with the caveat that the CW temperature can only serve as a 
crude estimate of the spins' exchange strength in rare-earth materials. 

\begin{figure}[t]
\includegraphics[width=\linewidth]{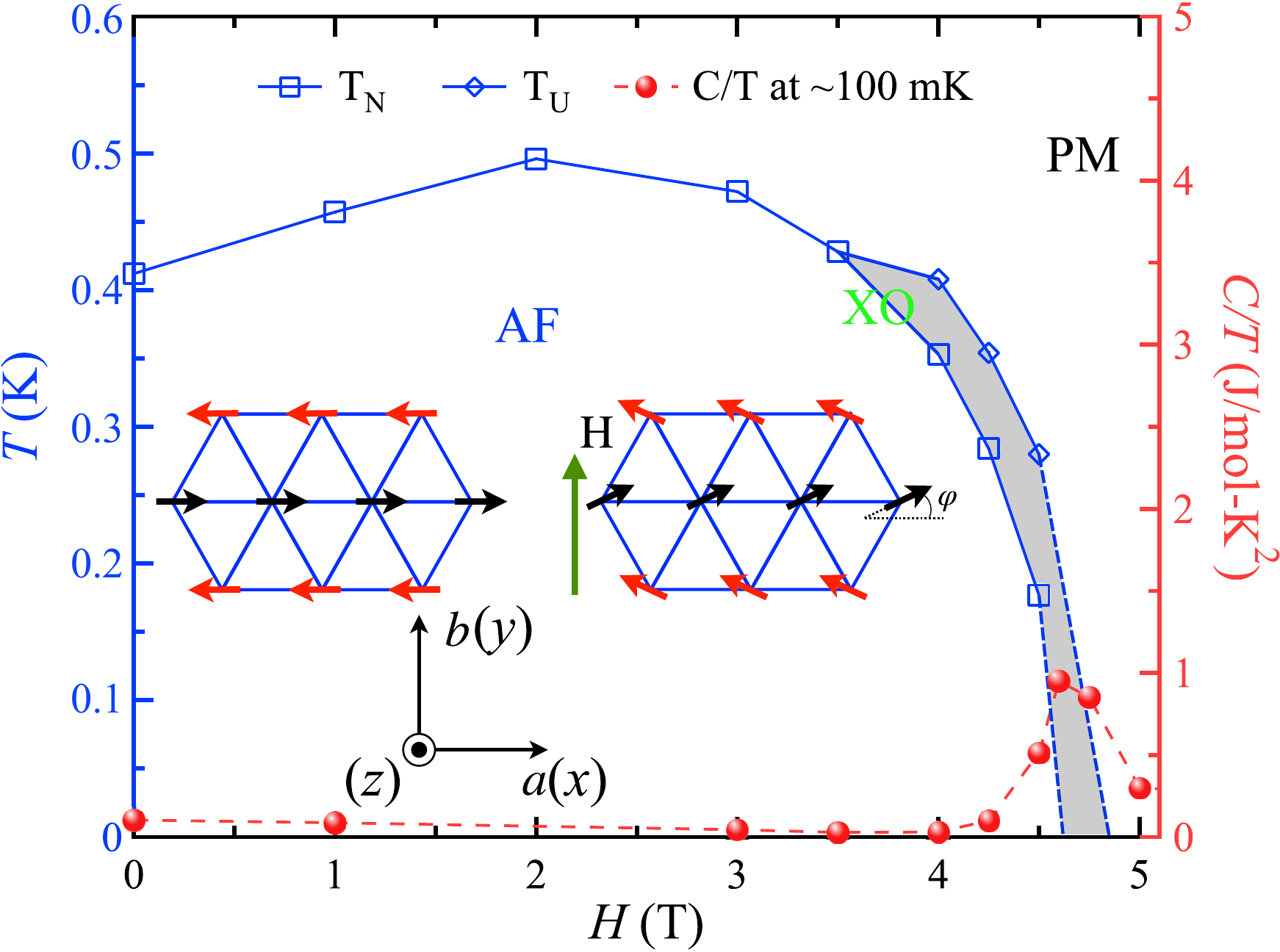}
\vskip -0.2cm
\caption{Open symbols are $T_N$ and $T_U$ vs in-plane field $H$ 
 from the heat capacity data in Figs.~\ref{fig:CpH}(a) and (b);
closed circles are $C/T$ values at 100~mK, lines are guides to the eye. 
A crossover (XO) region between $T_N$ and $T_U$ is highlighted. Insert: sketch of the stripe-$\bf x$ 
AF phase in zero and finite field, see text.}
	\label{fig:NewPhaseDiagram}
\vskip -0.4cm
\end{figure}

\begin{figure}[t]
\centering
\includegraphics[width=\linewidth]{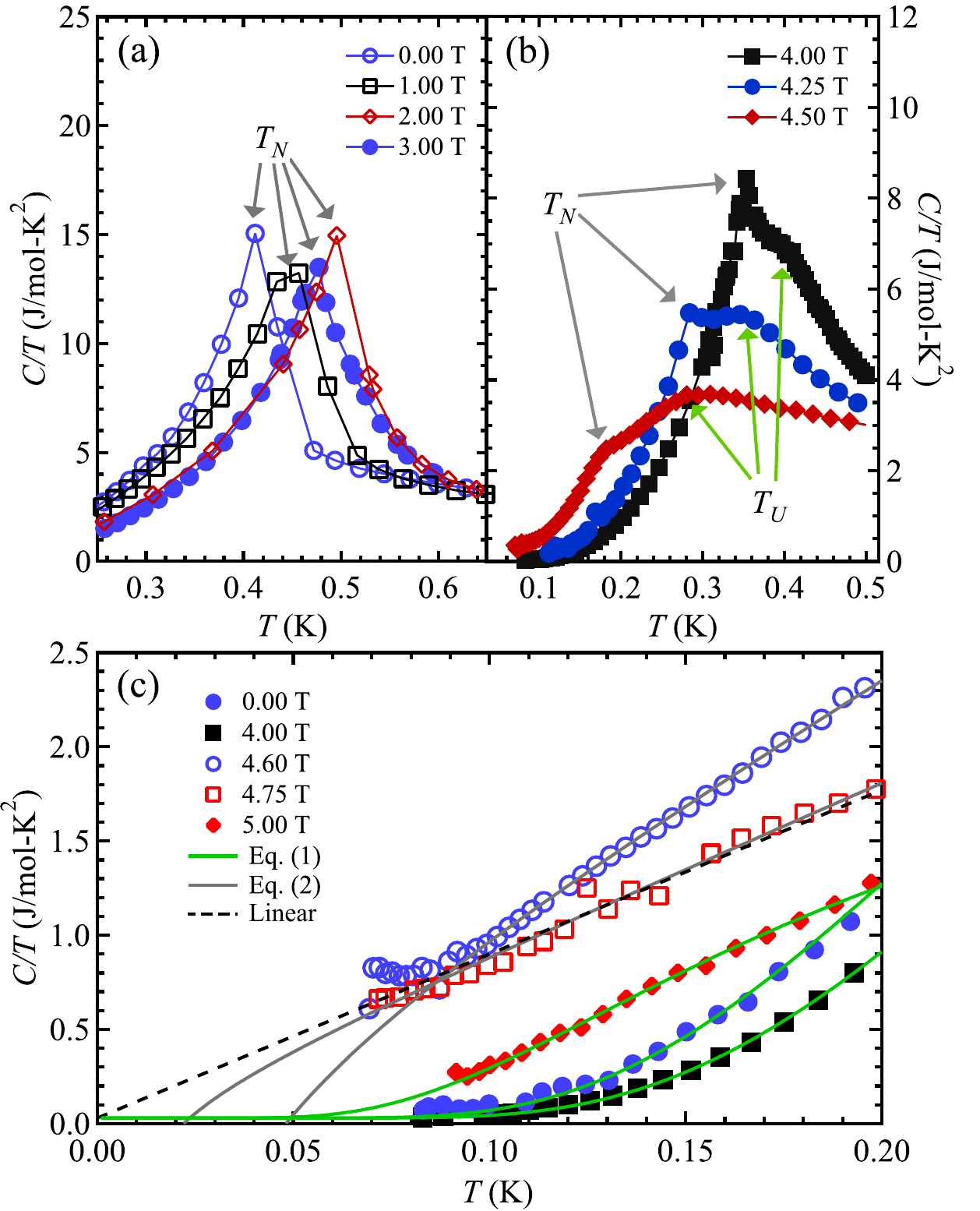}
\vskip -0.2cm
\caption{Specific heat $C/T$ vs $T$  shows:
(a)  the non-monotonic field-dependence of $T_N$ for $H\!<\!3.5$~T, 
(b)  the emergence of $T_U$ and the XO region  for $H\!>\!3.5$~T, lines are guides to the eye,
(c)  the low-$T$ dependence of $C/T$ away and at the QCP together with the fits 
from Eqs.~(\ref{eq:fullGapped}) and (\ref{eq_Cvl_as}) (lines).}
\label{fig:CpH}
\vskip -0.4cm
\end{figure}

For magnetic fields applied in the $ab$-plane, $T_N$ increases, reaches its maximum of 500(30)~mK  near  2~T, 
and is suppressed at higher fields, see Fig.~\ref{fig:NewPhaseDiagram} and Fig.~\ref{fig:CpH}(a).  
Above 3.5~T, at temperatures above the sharp feature that is 
identified with the magnetic ordering, specific heat also exhibits a shoulder-like anomaly denoted as $T_U$ in 
Fig.~\ref{fig:CpH}(b).  In Fig.~\ref{fig:NewPhaseDiagram},  we highlight the crossover (XO) region 
between these two features  
as a shaded area. Both anomalies are suppressed to zero temperature by a magnetic field near 4.7--4.8~T, 
suggesting a quantum critical point (QCP) within that field range. As is shown in Fig. \ref{fig:NewPhaseDiagram}, 
this region of the phase diagram demonstrates  a significant enhancement of the specific heat
at a reference low temperature  of 100~mK.

{\it Specific heat, $T_N$ and $T_U$.}---%
Specific heat data for CeCd$_3$As$_3$ is presented in Fig.~\ref{fig:CpH} for several field and temperature regimes. 
The data up to 3~T is shown in Fig. \ref{fig:CpH}(a). It demonstrates the non-monotonic 
$T_N$ field-dependence, characterized by an initial increase followed by a suppression. 
Such an increase indicates an enhancement of the AF order with field and is known to occur in  
several quantum magnets and their models \cite{Ishii_EPL_2011,Gvozdikova_Arxiv_1997,Sengupta_PRB_2009}, 
wherein this effect is associated with the field-induced suppression of quantum or thermal spin fluctuations or a 
reduction of frustration. Given our subsequent analysis of the nature of its magnetic ordered state, 
we ascribe the increase of $T_N$ in the case of CeCd$_3$As$_3$
to a field-induced suppression of critical fluctuations related to the phase transition, see SM \cite{StripeSuppliment}.

As is shown in Fig.~\ref{fig:CpH}(b) for the fields 4~T and above, specific heat acquires an additional 
shoulder-like feature  at $T_U\!>\!T_N$, giving rise to an XO region between the two temperatures, see also 
Fig.~\ref{fig:NewPhaseDiagram}. It appears that the line of $T_N$ transitions continues through 3.5~T
with no inflection.  Both $T_U$ and $T_N$ decrease towards zero temperature 
at higher fields, in agreement with the expected suppression of the AF order parameter to zero at a QCP. 
For the XO region of the phase diagram, we note that while some frustrated TL models 
consistently have a two-peak structure in their specific heat \cite{Schmidt_PR_2017}, the second, 
higher-temperature anomaly at $T_U$ in Fig.~\ref{fig:CpH}(b) 
could also be related to a field-induced high density of states in the magnon spectrum, 
see Ref.~\cite{StripeSuppliment}. 

{\it Specific heat, low-$T$.}---%
Fig.~\ref{fig:CpH}(c) focuses on the low-$T$ heat capacity. As is already indicated in 
Fig.~\ref{fig:NewPhaseDiagram}, the behavior of $C/T$ in this temperature range near a presumptive 
QCP is drastically different from that in other fields regions. It should also be noted that the 
QCP region is where universal scalings are expected to dictate the $T$-dependence of all thermodynamic 
quantities \cite{Zapf_RMP_2014}.

Fig.~\ref{fig:CpH}(c) also displays the specific heat as a function of $T$ for several representative fields away from the 
QCP region  (solid symbols). At $T\!=\!100$~mK, $C/T$ is small
up to 4.25~T and is suppressed again in fields greater than 5~T.
Moreover, the temperature dependence of $C/T$ in Fig.~\ref{fig:CpH}(c) in fields outside of the critical region 
clearly indicates activated behavior characteristic of gapped systems, as we elaborate below.

The enhancement in low-$T$ entropy, manifested as a build up of area under the $C/T$ vs $T$ curves, 
occurs for fields between 4.25~T and 5~T. This results in the large $C/T$ values at 100~mK shown in 
Fig. \ref{fig:NewPhaseDiagram}. The value of $C/T$ at 100~mK is substantially enhanced, 
reaching a maximum of about 1~J/mol-K$^2$ at 4.6~T. It is also accompanied by a distinct change in the 
temperature dependence, shown in Fig.~\ref{fig:CpH}(c) for 4.6~T and 4.75~T  (open symbols), 
that is indicative of a power-law in $T$. The rise in $C/T$ below 80 mK for $H$=4.6 T is likely due to our
calorimeter setup not accurately accounting for longer internal thermal relaxation times in this $H$--$T$ range.

{\it Low-$T$ asymptotes.}---%
The leading contribution to the heat capacity from a 2D gapped excitation can be obtained by approximating 
its energy as $\varepsilon_{\bf k}\!\approx\!\Delta +J  {\bf k}^2$ near the minimum gap $\Delta$ and $J$ 
parametrizing the bandwidth 
\begin{align}
\label{eq:fullGapped}
C(T)/T= A\left(x^2+2x+2\right)e^{-x} +b  +O\left(e^{-2x}\right)\, ,
\end{align}
where $x\!=\!\Delta/T$, $A \!\sim\! 1/J$, 
and a small constant offset $b$ of 0.029~J/mol-K$^2$ is introduced to account for a  
a calorimeter background heat capacity in all our fits \cite{StripeSuppliment}. 
The 2D activated behavior of Eq.~(\ref{eq:fullGapped}) fits very well all $C/T$ data 
up to 200~mK for fields away from the QCP region, 
as is demonstrated in Fig.~\ref{fig:CpH}(c), see also Ref.~\cite{StripeSuppliment}. 
Since the fit in Eq.~(\ref{eq:fullGapped}) is controlled by a single parameter $\Delta$, 
this analysis allows us to  accurately
trace the field-dependence of the lowest excitation gap in CeCd$_3$As$_3$.

For systems with continuous spin symmetries, field-induced PM to AF transition
is of the Bose-Einstein condensation type \cite{Batyev_SPJETP_1984,Zapf_RMP_2014}. 
In our case, because SOC leaves no continuous symmetries intact,  this transition 
is of a different universality class, characterized by the closing of the  excitation spectrum gap
in a relativistic manner, $\varepsilon_{\bf k}\!\approx\!\sqrt{\Delta^2+J^2  {\bf k}^2}$
\cite{Zapf_RMP_2014}. This implies an acoustic mode at the QCP,
$\varepsilon_{\bf k}\!\propto\!|{\bf k}|$ (dynamical exponent   $z\!=\!1$), leading to
a universal 2D scaling, $C(T)\!\propto\! T^2$, at the transition field.
The gap closure also explains the peak in $C/T$ vs field in Figs.~\ref{fig:NewPhaseDiagram} and \ref{fig:CpH}(c) 
near the QCP.

At 4.75~T, CeCd$_3$As$_3$ appears to be close to the QCP, as is indicated by the linear fit of $C/T$ 
(dashed line) in Fig.~\ref{fig:CpH}(c), which is indicative of gapless excitations. 
For small gaps, we obtain a modified  scaling 
\begin{align}
\label{eq_Cvl_as}
C(T)/T\approx A\,T\left(1-x^2/\alpha\right) +b \, ,
\end{align}
with $\alpha\!=\!12\zeta(3)$, valid down to $T\!\sim\!\Delta/2$ \cite{StripeSuppliment}.  
Eq. (\ref{eq_Cvl_as}) provides an excellent fit to the 4.6~T and 4.75~T data sets in Fig.~\ref{fig:CpH}(c)
(solid lines), putting an upper bound on $\Delta$ at 4.75~T of $\alt\!90$~mK and of 
$\approx\!200$~mK at 4.6~T.

\begin{figure}[t]
\centering
\includegraphics[width=\linewidth]{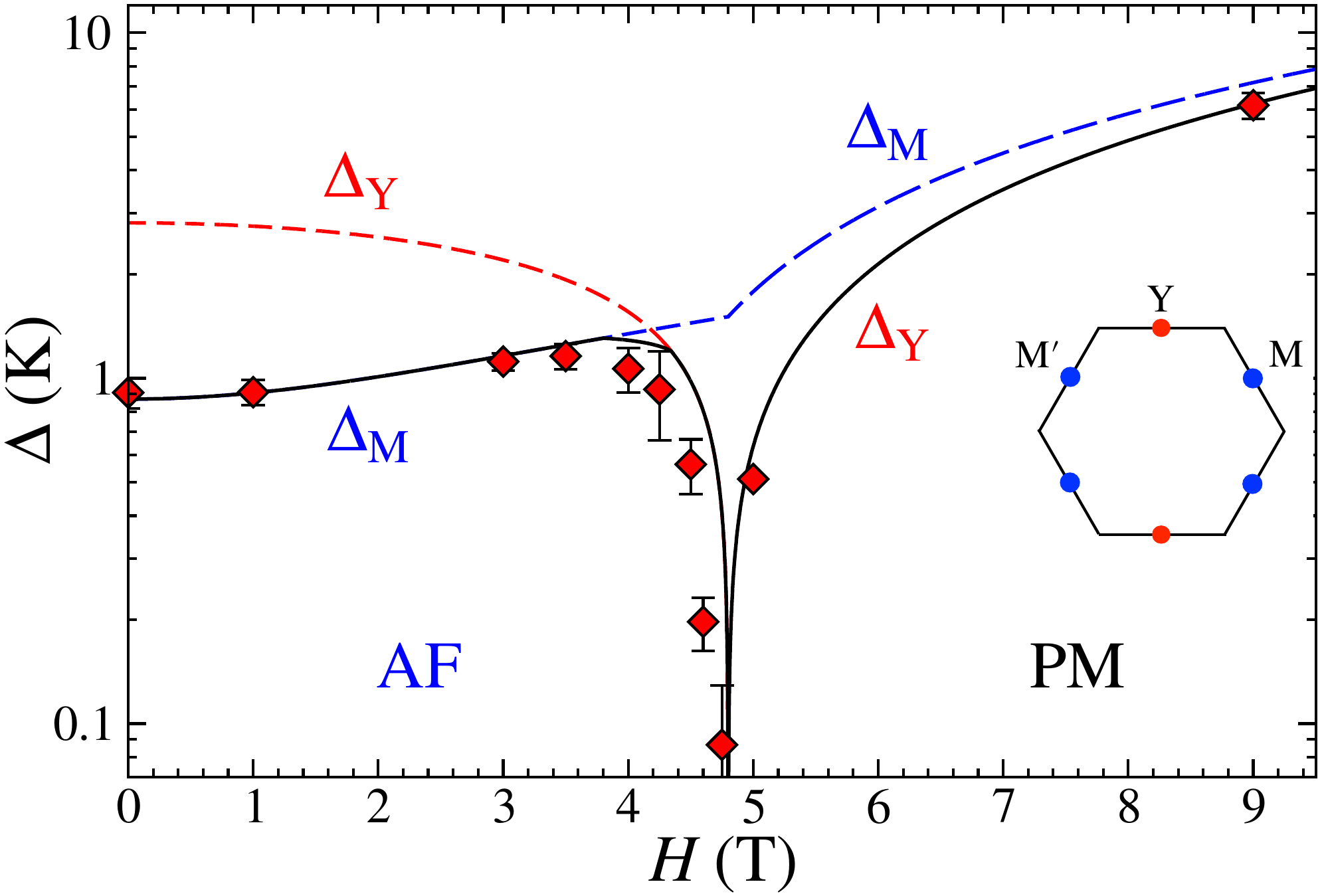}
\vskip -0.2cm
 \caption{$\Delta$ vs $H$ obtained using Eqs.~(\ref{eq:fullGapped}) and (\ref{eq_Cvl_as}) (symbols).
The excitation spectrum gaps at the ordering vector Y ($\Delta_{\rm Y}$) and complimentary 
M point ($\Delta_{\rm M}$)  for the parameters of the model (\ref{HJpm}) discussed in text. 
Inset: Brillouin zone with the Y and M points for the stripe phase in Fig.~\ref{fig:NewPhaseDiagram}.}
\label{fig:gap_vs_H}
\vskip -0.4cm
\end{figure}

{\it Gaps and other phenomenologies.}---%
The spectrum gap $\Delta$, extracted from the specific heat data 
using Eqs.~(\ref{eq:fullGapped}) and (\ref{eq_Cvl_as}), is  shown in Fig.~\ref{fig:gap_vs_H} vs $H$.  
The semi-log scale is to accommodate  $\Delta\!\approx\!6$~K at 9~T  deep  in the PM phase. 
Most of the results in Fig.~\ref{fig:gap_vs_H} use $C(T)$ data from the temperature 
window 0--200~mK and are checked for stability by varying this fitting range. Error bars combine 
internal quality of the fits with such variations, see SM~\cite{StripeSuppliment}. 

The zero-field gap of $0.91(9)$~K is compatible with the CW temperature 
\cite{Dunsiger_PRB_2020,StripeSuppliment} and  shows no sign of  closing below the QCP, 
implying that the AF phase of CeCd$_3$As$_3$ evolves continuously with $H$. 
Magnetization data  corroborate this assertion \cite{StripeSuppliment} showing no traces of 
the plateau-like phase transitions emblematic of TL magnets \cite{Starykh_RPP_2015,Seabra_PRB_2011}. 

The field-dependence of $\Delta$ demonstrates an essential feature. 
It shows a  gradual increase to about 1.1~K at 3.5~T
followed by an abrupt closing upon approaching the critical field. 
This non-monotonic behavior is an important distinguishing hallmark that allows  us
to unambiguously identify the ordered state of CeCd$_3$As$_3$.

{\it Model.}---%
For Kramers ions in layered compounds, 
crystal field effects (CEF) lead to an energy splitting of the local Hilbert space
of the rare-earth ion magnetic moment  into a series of doublets \cite{Hutchings_SSP_1964}. 
At temperatures much lower than the crystal field splitting, 
the lowest Kramers doublets,  naturally parametrized as effective pseudo-spins $S\!=\!\frac12$, 
are responsible for the dominant magnetic properties of insulating materials.
Because of the entanglement with the orbital orientations 
that are tied to the lattice due to CEF, the pairwise interactions of these pseudo-spins 
{\it a priori} retain no spin-rotational symmetries \cite{Li_PRB_2016, Tsirlin_JPCM_2020}. 
Instead, it is the discrete symmetries of the lattice that restrict possible forms of the 
bond-dependent interactions. Together with the localized nature of $f$-orbitals that limits the ranges of 
interactions, these symmetries lead to generic Hamiltonians 
that are expected to adequately describe {\it all} \, rare-earth-based Kramers compounds on a 
given lattice  \cite{Rau_PRB_2018}.

For a layered TL structure, the relevant point-group symmetry operations  
allow four terms in the Hamiltonian that can be separated into bond-independent, 
${H}^{XXZ}$, and bond-dependent, ${H}^{bd}$, parts, see Ref.~\cite{Maksimov_PRX_2019},
\begin{align}
&{\cal H}=\sum_{\langle ij\rangle}\Big({H}_{\langle ij\rangle}^{XXZ}+{H}_{\langle ij\rangle}^{bd}\Big)
+\sum_{\langle ij\rangle_2}{H}_{\langle ij\rangle}^{XXZ},\nonumber\\
\label{HJpm}
&{H}_{\langle ij\rangle_m}^{XXZ}=  J_m
 \Big(S^{x}_i S^{x}_j+S^{y}_i S^{y}_j+\bar{\Delta} S^{z}_i S^{z}_j\Big)\\
&{H}_{\langle ij\rangle}^{bd}= 2 J_{\pm \pm} \Big[ \Big( S^x_i S^x_j - S^y_i S^y_j \Big) \tilde{c}_\alpha 
-\Big( S^x_i S^y_j+S^y_i S^x_j\Big)\tilde{s}_\alpha \Big]\nonumber\\
 &\quad \quad\quad + J_{z\pm}\Big[ \Big( S^y_i S^z_j +S^z_i S^y_j \Big) \tilde{c}_\alpha 
 -\Big( S^x_i S^z_j+S^z_i S^x_j\Big)\tilde{s}_\alpha \Big],\nonumber
\end{align}
where $\tilde{c}(\tilde{s})_\alpha\!=\!\cos(\sin)\tilde{\varphi}_\alpha$,   $\tilde{\varphi}_\alpha$
are angles of the primitive vectors with the $x$ axis, $\tilde{\varphi}_\alpha\!=\!\{0,2\pi/3,-2\pi/3\}$, 
and $\{x,y,z\}$ are the crystallographic axes, see Fig.~\ref{fig:NewPhaseDiagram}.
The bond-independent exchange constants $J_m$ are $J_1$ and $J_2$ for the first-
and second-neighbor couplings, respectively.  
Following prior works \cite{Paddison_NatPhys_2019, Zhu_PRL_2018}, 
we use a minimal extension of the model  by the  $J_2$ 
term with the same $XXZ$ anisotropy  $\bar{\Delta}$. 
In an external field, Zeeman coupling 
\vspace{-0.05cm}
\begin{align}
{\cal H}_Z=-\mu_B\sum_{i}\Big[g_{ab} \Big(H_x S_i^x+H_yS_i^y\Big) +g_cH_z S_i^z\Big]
\label{HZ}
\end{align}
contains anisotropic $g$-factors  of the pseudo-spins that reflect the  build-up of the ground-state doublets 
from the states of the ${\bf J}$-multiplet of the rare-earth ions by a combined effect of SOC and CEF.  
The in-plane $g$-factor is uniform because of the TL three-fold symmetry  \cite{Li_PRL_2015}.

{\it Phase identification.}---%
As the model (\ref{HJpm}) has no spin-rotational symmetries \cite{Li_PRB_2016, Zhu_PRL_2018},
one expects gapped excitations throughout its phase diagram, but accidental degeneracies render 
most of the phases, such as well-known 120$^{\circ}$ phase and the nearby incommensurate phases, 
nearly gapless \cite{Maksimov_PRX_2019, Rau_PRL_2018}. 
Of the remaining phases,
the CeCd$_3$As$_3$ phenomenology of a single-phase field-evolution and a sizable spin-excitation gap
strongly suggests so-called stripe phases as prime contenders for its ground state. In a stripe phase, 
ferromagnetic rows of spins arrange themselves in an AF fashion, 
see inset of Fig.~\ref{fig:NewPhaseDiagram}.
In particular, as we argue in this work, the non-monotonic field-dependence of the gap is a 
hallmark of the stripe phases. 
An alternative scenario of the strongly Ising limit leads to phases and transitions 
\cite{Miyashita_JPSJ_1986,Seabra_PRB_2011} that are incompatible with the phenomenology of 
CeCd$_3$As$_3$, see SM \cite{StripeSuppliment} for more detail. 

Most importantly, the  field-evolution of the spin-excitation spectrum in the stripe phase, at the QCP, and in 
the spin-polarized PM phase are all in accord with our results for CeCd$_3$As$_3$. 
Specifically, the spectrum minima in zero field are {\it not} associated with the ordering vector 
(identified as a Y-point in Fig.~\ref{fig:gap_vs_H}), but are complementary to it (M-points in the Brillouin zone).
This feature is characteristic of the systems with significant frustrating  bond-dependent interactions
\cite{Maksimov_PRX_2019,Maksimov_PRR_2020}. In an applied field, it is the gap at the ordering vector that must
close, leading to a rather abrupt switch of the minimal gap between the M and Y-points, as is demonstrated in 
Fig.~\ref{fig:gap_vs_H} for a choice of parameters in model (\ref{HJpm}) discussed next. 

{\it Model parameters.}---%
Assuming a stripe ground state, we obtain the field-evolution of spin excitations for model (\ref{HJpm}) 
in the AF and PM phases. There are two empirical quantities that 
provide strong constraints on the model parameters: the value of the  critical field, $H_s$, which we
take as 4.8~T, allowing for  an ambiguity  in  the QCP for CeCd$_3$As$_3$, 
and the zero-field gap, $\Delta_0$, taken as 0.87~K to account for an uncertainty of the fit.  
Qualitatively, with $g_{ab}\!\approx\!2$ known from the in-plane magnetization 
data \cite{StripeSuppliment,Banda_PRB_2018}, 
the empirical value of $H_s$ strongly binds the cumulative exchange term, $J_1+J_2$, while $\Delta_0$ restricts
the $J_{\pm\pm}$ anisotropic-exchange term, see SM~\cite{StripeSuppliment} for details.
 
We find the second anisotropic-exchange term, $J_{z\pm}$,
to have a minor effect  on the spectrum \cite{Maksimov_PRX_2019} and neglect it in our consideration.
This mild simplifying assumption leaves two types of stripe states, stripe-${\bf x}$ and stripe-${\bf y}$, 
which correspond to spins along and perpendicular to the bonds of the lattice, respectively, 
indistinguishable  up to  a change of the sign of the   $J_{\pm \pm}$-term and 
a switch of the field direction from $H\!\parallel\!b$ to $H\!\parallel\!a$. 
Because  experiments in CeCd$_3$As$_3$ have not discriminated between the in-plane field directions,
we take a minimal-model approach \cite{StripeSuppliment}, by assuming $J_{\pm \pm}\!<\!0$, which corresponds to   
stripe-${\bf x}$  with $H$ along the $b$-direction shown in Fig.~\ref{fig:NewPhaseDiagram}. 
 
We have found that the excitation gaps are virtually insensitive to the value of 
the $XXZ$ anisotropy $\bar{\Delta}$, which is loosely bound in the range 0.5--1.5  by the 
out-of-plane saturation field, extrapolated from $M(H)$ data \cite{StripeSuppliment}. The situation is 
similar with the ratio $J_2/J_1$, 
which does not affect observables at the fixed total $J_1\!+\!J_2$. 
Thus, by taking $XXZ$ anisotropy $\bar{\Delta}\!=\!1$  and making an {\it ad hoc} choice of $J_2/J_1\!=\!0.1$
for the highly localized $f$-orbitals, for the empirical $H_s$ and $\Delta_0$ we obtain  
$J_1\!\approx\!1.19$~K and $J_{\pm\pm}\!\approx\!-0.31$~K \cite{StripeSuppliment}. 
With these model parameters, we derive the field-evolution of the  gaps shown in Fig.~\ref{fig:gap_vs_H}. 
 
Our results show a gradual increase of the magnon gap energy $\Delta_{\rm M}$ 
vs field at the complementary M-point and concurrent decrease of the spin-excitation energy 
$\Delta_{\rm Y}$ at the ordering vector \cite{StripeSuppliment} which inevitably takes the role 
of the global minimum of the spectrum upon approaching the QCP, all in close accord with the data 
in Fig.~\ref{fig:gap_vs_H}. 
At the QCP, the asymptotic form of the spectrum adheres to the expected relativistic form. Above the QCP, 
the gap at the Y point reopens, with the spectrum experiencing a roughly uniform, Zeeman-like shift vs $H$.
Remarkably, the high-field value of the gap  at 9~T deep in the PM phase is  
also  closely matched by the same model with no additional free parameters. 

{\it Summary.}---%
In summary, we have demonstrated that a combination of insights from the 
low-temperature specific heat data and theoretical modeling provide a comprehensive description of the 
ground state and excitations in 
a TL rare-earth  anisotropic-exchange magnet, paving the way 
to a deeper understanding of a broad class of materials. 
The phenomenological constraints on the general microscopic model have resulted in a
precise identification of CeCd$_3$As$_3$ magnetic ground state as a stripe phase and 
with a remarkable level of certainty regarding the part of the phase diagram where it belongs.

This study is of immediate relevance to KCeS$_2$ \cite{Bastien_Arxiv_2020}, 
KErSe$_2$ \cite{Xing_arxiv_2020}, and isostructural CeCd$_3$P$_3$  \cite{Lee_PRB_2019}, 
where some of the same phenomenology has been observed.
Future studies by thermodynamic and spectroscopic methods, such as low-$T$ magnetization, 
nuclear and electronic magnetic resonance, neutron  
and x-ray magnetic dichroism scattering, are expected to provide further insights into the nature of 
the crossover  region in the $H$--$T$ phase diagram and into the role of structural disorder 
in the static and dynamic properties of these materials. With more theoretical input, 
they should yield more systematic constraints on the model  
and elucidate the role of different terms in unusual magnetic 
states and excitations of anisotropic-exchange magnets.  

\begin{acknowledgments}
{\it Acknowledgments}---%
Low-temperature specific heat, magnetization, and electrical resistivity measurements were 
supported by the U.S. Department of Energy, Office of Science, National Quantum Information Science 
Research Centers, Quantum Science Center (QSC). Sample synthesis and crystal structure determination 
were supported by the Los Alamos Laboratory Directed Research and Development project 20190076ER.
Scanning electron microscope and energy dispersive X-ray measurements were performed at 
the Center for Integrated Nanotechnologies, an Office of Science User Facility operated for the U.S. 
Department of Energy (DOE) Office of Science. 
Support is acknowledged from the Northwestern-Fermilab 
Center for Applied Physics and Superconducting Technologies and  by the U.S. Department of Energy, 
Office of Science, Office of Workforce Development for Teachers and Scientists, Office of
Science Graduate Student Research (SCGSR) program (K.~E.~A).
The SCGSR program is administered by the Oak Ridge
Institute for Science and Education for the DOE under
contract number DE-SC0014664.
The work of A.~L.~C. was supported by the U.S. Department of Energy, 
Office of Science, Basic Energy Sciences under Awards No. DE-FG02-04ER46174 and DE-SC0021221.
P.~A.~M. acknowledges support from JINR Grant for young scientists 21-302-03.
A.~L.~C. would like to thank  Kavli Institute for Theoretical Physics (KITP) where this work was advanced. 
KITP is supported  by the National Science Foundation under  Grant No. NSF PHY-1748958.

\end{acknowledgments}

\bibliography{Bibby}

\end{document}



\title{Fingerprinting Triangular-Lattice Antiferromagnet by Excitation Gaps: \\ Supplemental Material}

\author{K. E. Avers}
\affiliation{Los Alamos National Laboratory, Los Alamos, NM 87545, USA}
\affiliation{Department of Physics and Astronomy, Northwestern University, Evanston, IL, USA}
\affiliation{Center for Applied Physics \& Superconducting Technologies, Northwestern University, Evanston, IL, USA}
\author{P. A. Maksimov}
\affiliation{Bogolyubov Laboratory of Theoretical Physics, Joint Institute for Nuclear Research,
Dubna, Moscow region, 141980, Russia}
\author{P. F. S. Rosa}
\affiliation{Los Alamos National Laboratory, Los Alamos, NM 87545, USA}
\author{S. M. Thomas}
\affiliation{Los Alamos National Laboratory, Los Alamos, NM 87545, USA}
\author{J.~D.~Thompson}
\affiliation{Los Alamos National Laboratory, Los Alamos, NM 87545, USA}
\author{W. P. Halperin}
\affiliation{Department of Physics and Astronomy, Northwestern University, Evanston, IL, USA}
\author{R. Movshovich}
\affiliation{Los Alamos National Laboratory, Los Alamos, NM 87545, USA}
\author{A. L. Chernyshev}
\affiliation{Department of Physics and Astronomy, University of California, Irvine, California 92697, USA}

\date{\today}

\maketitle
\tableofcontents

\section{Experimental details and  other data}

\subsection{Experimental methods}

Bulk single crystals of CeCd$_3$As$_3$ were grown by chemical vapor transport using 
iodine as a transport agent and recipe described in 
Ref.~\cite{Stoyko_InOrgChem_2011, Liu_Arxiv_2016}. 
Polycrystalline CeCd$_3$As$_3$ was first synthesized by solid state reaction. 
A stoichiometric mixture of Ce, Cd and As pieces was sealed in a quartz 
tube under partial Ar atmosphere and heated to 800 $^\circ$C for one week. 
The resulting polycrystal was sealed in a quartz tube with iodine and placed 
in the hot end of a zone furnace. The other end of the tube was held at 
700 $^\circ$C where single crystal platelets of typical size 1 mm were obtained. 
The crystallographic structure of CeCd$_3$As$_3$ was verified by single-crystal diffraction at room 
temperature, using Mo radiation in a Bruker D8 Venture diffractometer. 
Our results, shown in Table~\ref{Table-crystal1}, are consistent with previous reports
\cite{Stoyko_InOrgChem_2011, Liu_Arxiv_2016}.
Elemental analysis of our single crystals using energy-dispersive x-ray spectroscopy in a 
commercial scanning electron microscope resulted in a stoichiometry of CeCd$_{2.95(8)}$As$_{3.07(9)}$.

Heat capacity measurements were performed in a dilution refrigerator down 
to 70 mK using the heat pulse technique. A RuO thermometer was attached directly
 to one side of the sample with GE varnish. The other side of the sample was attached 
 to a sapphire substrate with a heater on the opposite side of the substrate. 
 The weak thermal link wire was silver painted directly to the sample. No evidence 
 for multiple timescale relaxation behavior was observed indicating good thermal 
 contact between the sample, RuO thermometer, heater, and sapphire substrate, as 
 well as fast internal relaxation time of the sample. No subtraction of electron nor 
 phonon heat capacity contributions from the sample, heater, sapphire, or thermometer were 
 performed because the magnetic heat capacity of CeCd$_3$As$_3$ is vastly dominant in the 
 temperature range of interest in this work. It is estimated that the magnetic degrees of 
 freedom contribute more than 99 percent of the entropy change from 0 K to 2 K. The calorimeter 
 was weakly thermally linked to a copper temperature regulation block, and a temperature 
 stabilized Lakeshore automatic bridge with active feedback PID system was employed. The sample RuO
  thermometer was previously calibrated in magnetic field up to 9~T.

\begin{table}[t]
\caption{Crystallographic data and atomic coordinates of CeCd$_{3}$As$_{3}$ determined by 
single-crystal X-ray diffraction. }
\label{Table-crystal1}
\begin{tabular*}{1\linewidth}{@{\extracolsep{\fill}}ll}
\hline
\hline
   Crystal system            &                Hexagonal    $\hspace{10.0cm}$                                                                                                    \\
   Space group               &                P6(3)/mmc (194)                                                                                                \\
   Temperature               &                296(2) K                                                                                                                            \\
   Wavelength                &                0.71073 \AA                                                                                                       \\                                                                                                          
\end{tabular*}
\begin{tabular*}{1\linewidth}{@{\extracolsep{\fill}}lc}
\hline
\hline                        
Formula weight      (g/mol)                    &    702.11                       \\
  a (\AA)\                               &    4.3970(3)          \\
     c (\AA)\                                &    21.3271(12)            \\
\hline
\hline
\end{tabular*}
\begin{tabular}{l@{\hspace{0.3cm}}c@{\hspace{0.2cm}}c@{\hspace{0.2cm}}c@{\hspace{0.6cm}}c@{\hspace{0.6cm}}c}                                
        \hline                                                    
        Atom           & Wyck.      & Occ.   &      x              &          y             &      z               \\
        \hline                                                    
        Ce                   &    2a        &    1    &    0                     &    0                         &    0           \\
         Cd(1)           &    4f        &    1    &    1/3              &    2/3                 &    0.6278             \\
         Cd(2)           &    6h        &    0.4(1)    &    0.2717              &    0.5435                 &    1/4             \\
         As(1)           &    4f        &    1    &    1/3              &    2/3                 &    0.078            \\
         As(2)           &    6h        &    0.4(1)    &    0.6036             &   0.394              & 1/4                \\
                    
      \hline                          
        \hline
\end{tabular}
\end{table}

Measurements of resistivity vs temperature were performed in zero field down to 95 mK using a 
Quantum Design physical property measurement system with an adiabatic demagnetization refrigeration 
attachment. The sample was taken from the same batch as the sample used for heat capacity 
measurements. It was glued to a copper cold finger using GE varnish with cigarette paper providing electrical 
insulation. Four 25 micron platinum wires were spot welded to the sample, and resistivity was 
measured by the 4-wire method. Multiple excitation currents were applied to ensure Joule heating 
was not significant.

Magnetization measurements were performed using a commercial Quantum Design MPMS SQUID-based magnetometer.

\vspace{-0.3cm}

\subsection{Crystal structure}
\vspace{-0.2cm}

\begin{figure}[t]
	\includegraphics[width=0.90\linewidth]{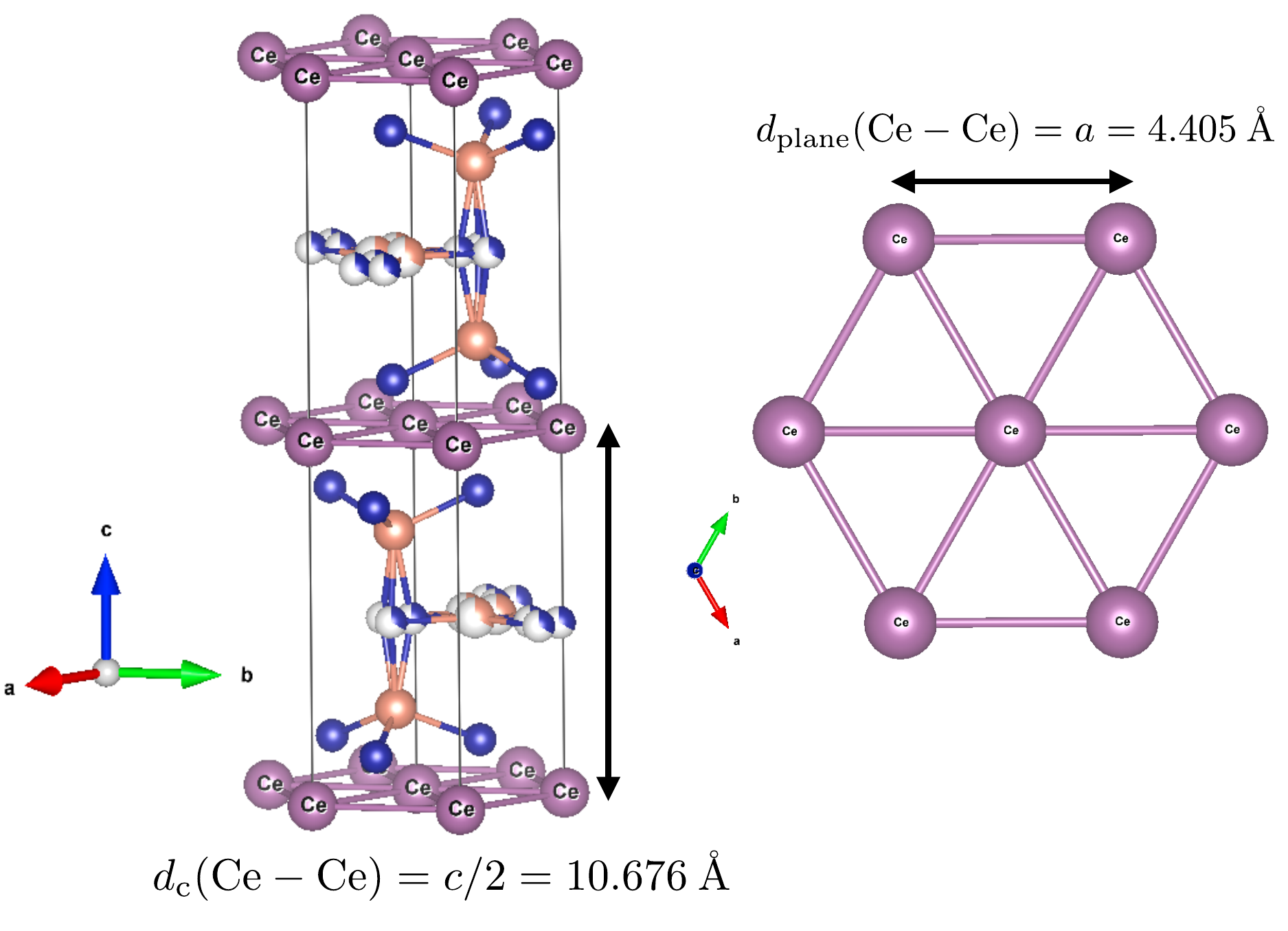}
\vskip -0.2cm
	\caption{(left) Crystal structure of CeCd$_3$As$_3$, which crystallizes in space group \#194 ($P6_{3}/mmc$). 
	Orange and blue circles represent Cd and As atoms, respectively. (right)
	The two-dimensional triangular arrangement of the magnetic Ce$^{3+}$ atoms.}
	\label{fig:Crystal}
\vskip -0.4cm
	\end{figure}

CeCd$_3$As$_3$ crystallizes in the PrZn$_3$As$_3$-type structure (space 
group $P6_3/mmc$) with lattice parameters
a=b=4.4051 \AA $ $ and c=21.3511 \AA, as shown in Fig. \ref{fig:Crystal}.  
Magnetic rare-earth Ce$^{+3}$ ions form planes of 2D triangular lattices that 
are separated from each other by layers of As and Cd atoms with an aspect 
ratio of inter-plane to intra-plane Ce spacing of approximately 2.4. 
There is only one cerium site in this structure, but both Cd and As atoms have
two sites, one of which is only 1/3 occupied.
Though this partial occupancy does not distort the Ce triangular structure directly, 
the role of disorder in this material remains poorly understood.

\begin{figure}[t]
\includegraphics[width=0.99\linewidth]{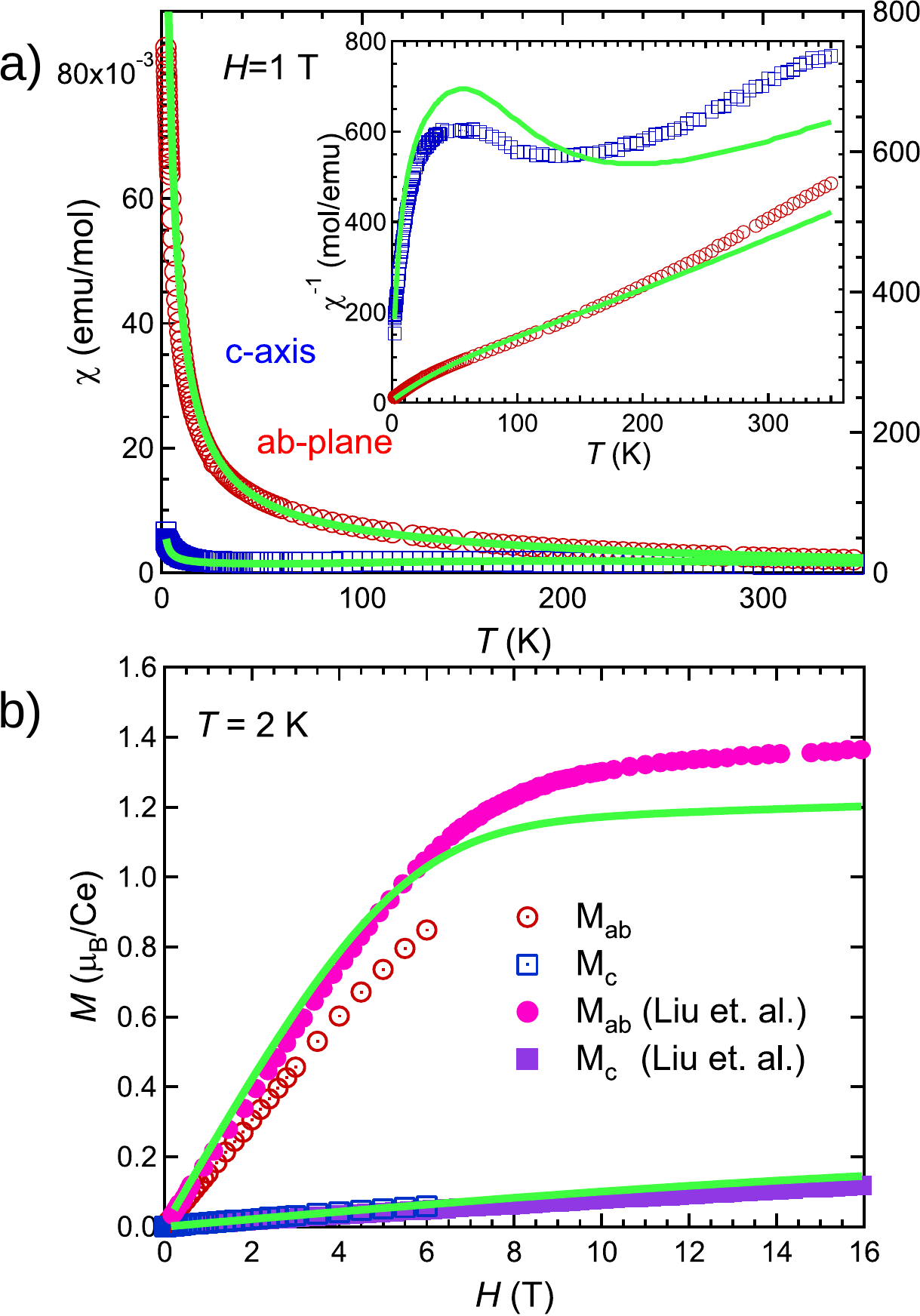}
\vskip -0.2cm
\caption{a) Susceptibility ($\chi=\frac{M}{H}$) vs temperature ($T$) taken with H=1 T along the c-axis, 
and within the ab-plane with the inverse susceptibility ($\chi^{-1}$) in the insert. The infection points in 
$\chi^{-1}$ are due to the influence of excited doublets of the Ce moment. The light green curves are fits 
to data as described in the text.  b) Magnetization ($M$) vs field ($H$) at $T$=2 K to $H$=6 T along the 
c-axis and within the ab plane alongside the data of Liu et. al. \cite{Liu_Arxiv_2016}. The light green curve 
is calculated by integrating the low temperature susceptibility. Even though the fit only reaches 
1.2 $\mu_b/Ce$ it is relatively good agreement.}
\label{fig:Magnetization}
\vskip -0.4cm
 \end{figure}

\vspace{-0.3cm}

\subsection{Magnetization and susceptibility}
\vspace{-0.2cm}

The higher temperature investigation of thermodynamic properties provide important clues 
into the nature of magnetism in CeCd$_3$As$_3$. 
The magnetic susceptibility ($\chi$) vs temperature ($T$) is presented in
 Fig.~\ref{fig:Magnetization}(a). The data were taken in an applied field of 1~T 
 both along the c-axis (blue squares), and within the ab-plane (red circles). 
 A simple Curie-Weiss law does not describe the experimental results. 
$\chi(T)^{-1}$, shown in the insert, is visibly non-monotonic for the field along the c-axis, with a local minimum 
 near 140~K. 
Careful inspection of the susceptibility for the field in the ab-plane reveals inflection 
points in its temperature behavior as well. 
Fig.~\ref{fig:Magnetization}(b) presents magnetization ($M$) vs field ($H$) results at 
 2~K for magnetic field up to 6~T applied along the c-axis 
 (open blue squares) and within the ab-plane (open red circles). 
We also plot data at 1.9~K up to 16~T
 extracted from Ref. \cite{Liu_Arxiv_2016} (filled circles and squares). The strong 
 magnetization anisotropy of roughly 10 
 between the (easy) ab-plane and the (hard) c-axis was initially taken as evidence
 for strong antiferromagnetic interactions of Ising-like spins aligned along the c-axis 
 \cite{Liu_Arxiv_2016}; however, inspection of the temperature 
 dependent susceptibility shows that this may be an inaccurate description 
 of the physics of this system.
 
As recognized in Ref.~\cite{Banda_PRB_2018}, the high-temperature non-monotonic behavior in $\chi$ vs $T$ 
can be accounted for by a non-interacting model involving the Zeeman term and the trigonal 
 crystalline electric field (CEF) Hamiltonian 
 $\mathcal{H}_{\mathrm{CEF}}= B_{2}^{0}O_{2}^{0} + B_{4}^{0}O_{4}^{0} + B_{4}^{4}O_{4}^{3}$,
 in which $B_{i}^{n}$ are the CEF parameters and $O_{i}^{n}$ are the Stevens
 equivalent operators \cite{Banda_PRB_2018}. The trigonal $\mathcal{H}_{\mathrm{CEF}}$ splits
 the $j\!=\!\frac{5}{2}$ sixfold degenerate state of Ce$^{3+}$ into three doublets, two of which are a mixture
 of $|m_{j}\!=\!\pm 5/2\rangle$ and $|\!\pm\! 1/2\rangle$ states. 
In an attempt to capture the low-temperature
magnetization of CeCd$_{3}$As$_{3}$, here we also include the spin Hamiltonian 
$\mathcal{H}_{\mathrm{spin}}= \tilde{J}_1 \sum {\bf j}_{i} \cdot {\bf j}_{j}$, in which 
$\tilde{J}_1>0$ represents AF interactions between nearest neighbors, 
and ${\bf j}_i$ is the total angular momentum operator on site $i$.
Following Ref.~\cite{Pagliuso_JAP_2006}, we employ a mean-field 
approximation,
which allows the spin Hamiltonian to be written simply as 
$z\tilde{J}_1\sum_i{\bf j}_{i} \cdot \langle{\bf j}\rangle$, 
with $z\!=\!6$ being the number of nearest neighbors.
Our results give a CEF doublet hierarchy with the lowest energy doublet dominated by the $|\!\pm\!1/2\rangle$ 
states. This ground state doublet gives rise to an $g$-factor anisotropy that causes the moments to lie in 
the ab-plane at low temperatures, which explains the aforementioned strong magnetization anisotropy. 
The two excited doublets that would allow the magnetic moment to point out of the ab-plane are separated 
from the ground state by 372~K, and 545~K, respectively, and hence are not populated at the low 
temperatures where the AF phase emerges. They may be responsible for the observed inflection points in $\chi(T)$.
  
Considering a CEF plus  nearest-neighbor  interaction Hamiltonian does not describe the data well,   
but including additional mean field antiferromagnetic interactions 
($\tilde{J}_{1}=0.8$~K and $\tilde{J}_{2}=0.6$~K) produces a reasonable, although not unique, fit shown as 
light green curves in Fig \ref{fig:Magnetization}(a) with CEF parameters ($B_{2}^{0} = 11.5$~K, 
$B_{4}^{0} = -1.4$~K, and $B_{4}^{3} = 12$~K) in good agreement with Ref.~\cite{Banda_PRB_2018}. 
The  need for two mean field antiferromagnetic interactions is indicative of  additional 
frustration, not present in the nearest-neighbor-only model.
The fits reproduce the inflection points in $\chi(T)^{-1}$, although suffer in accuracy for the c-axis direction. 
This mismatch is likely because $|\chi(T)|$ is quite small along that direction leading to increased 
measurement error.
This demonstrates that the inflections are due to the CEF doublet hierarchy, and not impurities as suggested 
in \cite{Liu_Arxiv_2016}.  
The inclusion of these interactions does indeed allow an accurate calculation of the $M$ vs $H$ behavior 
matching Liu et. al. quite well for low fields. 
The saturation of magnetization and its magnitude at 6 T for the field in the ab-plane is slightly mismatched, 
but still qualitatively reproduces the effect. 
The saturation magnetization of $\sim$ 1.2 $\mu_B/$Ce in the ab-plane is consistent with the lowest energy 
Ce$^{+3}$ doublet and the results in Ref.~\cite{Banda_PRB_2018}.
One possible source of the discrepancy of our data with that of Ref.~\cite{Liu_Arxiv_2016} is in the 
mass normalization due to small crystal sizes. This is also consistent with the observation
that $C(T)$ reported in that work is larger than ours. We have carefully verified the accuracy
of such mass determination on our end.

Despite the measured susceptibility not being described by a Curie-Weiss law for the entire temperature range, 
it is still possible to locally fit the data to a Curie-Weiss law at low temperature in order to extract the 
effective Weiss temperature. 
This will provide an average measure and sign of the interaction strength.
We performed Curie-Weiss fits to the $\chi$ for $T<10$ K for c-axis, and for $T<25$ K for the ab-plane, where 
the system is in the ground state doublet and far away from any inflection points. 
We obtain effective Weiss temperatures ($\Theta$) of -5.1 K and -4.5 K, 
respectively, which are consistent with previous results \cite{Liu_Arxiv_2016}. 

\begin{table}[t]
\caption{The crystal electric field eigenfunctions, and energy level relative to the ground state of the
Ce$^{+3}$ obtained from the model as described in the main text. Take note of the $\pm$, and $\mp$ 
that indicate each level is doubly degenerate in zero field.}
\label{tab:CEFlevels}
  \begin{center}
    \begin{tabular}{c | c} 
		 \hline  
	   \ \ \  $|0\rangle=0.32|\!\pm5/2\rangle +0.95|\!\mp1/2\rangle$ \ \ \ \ \ \ & \ \ \ E=0 K\ \ \ \\
		\ \ \ 	$|1\rangle=0.95|\!\pm5/2\rangle +0.32|\!\mp1/2\rangle$ \ \ \ \ \ \ & \ \ \ \ \ \ E=372 K\ \ \ \\
		\ \ \ 	$|2\rangle=|\!\pm3/2\rangle$ \ \ \ \ \ \ \ \ \ \ \ \ \ \ \ \ \ \ \ \ \ \ \ \ \ \ \ \ \ \ &\ \ \  \ \ \ \ E=545 K \ \ \  \\ 
			\hline
\end{tabular}
\end{center}
\vskip -0.4cm
\end{table} 


\subsubsection{Crystal electric field levels}

The crystal electric field levels of the $j\!=\!\frac{5}{2}$ Ce$^{+3}$ are doubly degenerate due to 
Kramer's theorem for half-integer spin systems. The levels, $|n=0,1,2\rangle$ in increasing order of 
energy as expressed as the eigen functions of $\hat{j}_z$ are listed in Table~\ref{tab:CEFlevels}. 
Our results are also similar to KErSe$_2$ and CsErSe$_2$ in which the CEF levels were measured by 
powder neutron diffraction \cite{Scheie_arxiv_2020}.


\subsection{Heat capacity}
\vspace{-0.3cm}

\subsubsection{Zero field entropy}
\vspace{-0.2cm}

\begin{figure}[t]
\includegraphics[width=0.99\linewidth]{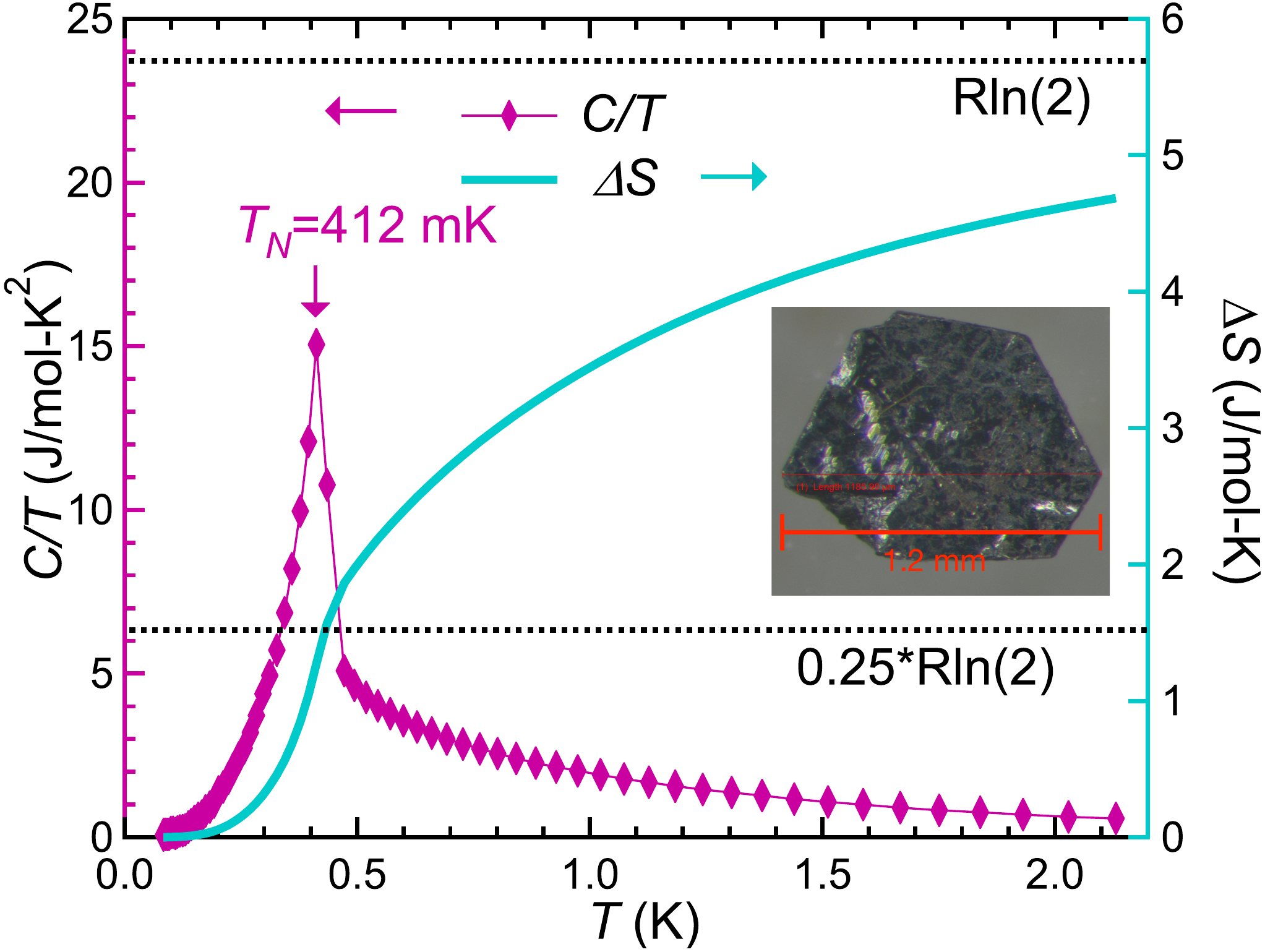}
\vskip -0.2cm
\caption{Zero-field specific heat and integrated entropy data as a function of temperature. 
At the antiferromagnetic phase transition with $T_N$=412 mK, the entropy is only 25\% of Rln2. 
A photograph of the sample is in the inset.}
\label{fig:ZeroFieldandEtrop}
\vskip -0.4cm
\end{figure}

The effects of magnetic frustration become evident in the heat capacity of CeCd$_{3}$As$_{3}$. 
The zero-field heat capacity ($C$/$T$) and the associated change in entropy ($\Delta S$) as a function 
of temperature are plotted in Fig.~\ref{fig:ZeroFieldandEtrop} with a photograph of the sample in the inset. 
$C$/$T$ data display a sharp lambda peak at $T_N$ = 412 mK indicating a phase transition from the 
high-temperature paramagnetic state to a low-temperature AF state.
We also observe the effect of frustration in the entropy change, $\Delta S$, plotted as the cyan curve in 
Fig.~\ref{fig:ZeroFieldandEtrop}. 
By taking the integral of $C$/$T$, one obtains $\Delta S$ from 0 K to 2.2 K to be nearly 80\% of Rln2, 
whereas the change in entropy from 0 K to $T_N$ for CeCd$_3$As$_3$ is only approximately 25\% of Rln2, 
which is a signature of magnetic frustration,  see also Ref.~\cite{Dunsiger_PRB_2020}. 
The Rln(2) limit is likely reached at about 8 K, in agreement 
with Ref.~\cite{Dunsiger_PRB_2020}, which, however, did not accumulate all low-$T$ entropy from 
below $\sim$0.4K.

We note that for an unfrustrated system, one should recover at least 50\% Rln2 of entropy from $T$ =~0 to $T_N$, 
depending on the symmetry of the system, e.g. Heisenberg, Ising, or XY \cite{Gopal_1966}.
This is the case in Ce$_2$PdGe$_6$ \cite{Fan_PRB_2004}, or the nuclear antiferromagnet 
$^{3}$He \cite{Halperin_LTP_1978}.

\begin{figure}[t]
\includegraphics[width=0.99\linewidth]{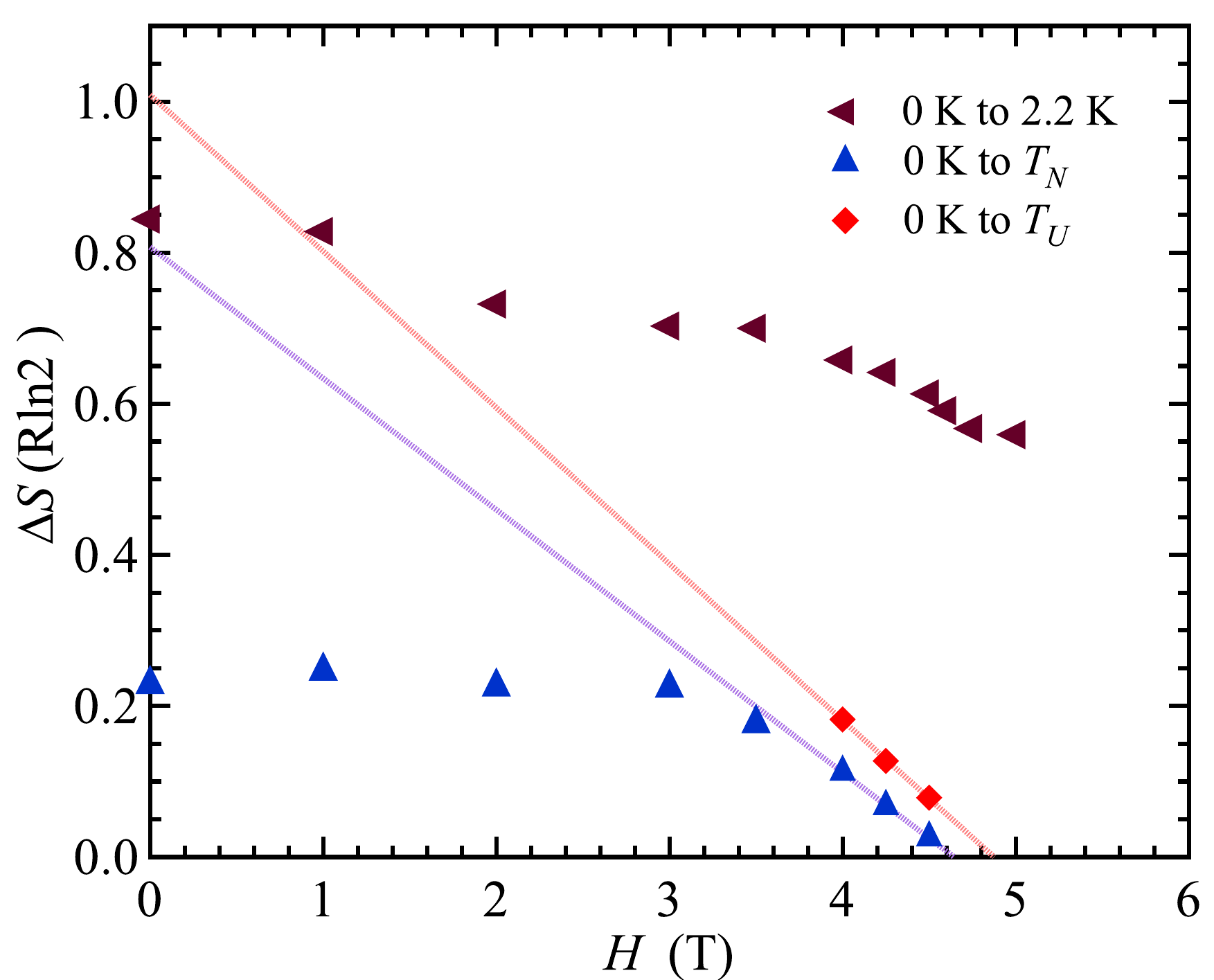}
\vskip -0.2cm
\caption{Entropy change as a function of field for various intervals of temperature integration.}
\label{fig:EntropVsField}
\vskip -0.4cm
\end{figure}

\vspace{-0.3cm}

\subsubsection{Entropy vs field}
\vspace{-0.2cm}

The entropy change vs field for relevant temperature integration ranges is shown in Fig.~\ref{fig:EntropVsField}. 
The total entropy change from 0 K to 2.2 K monotonically decreases as a function of field. 
In contrast, the entropy change from 0 K to $T_N$ is nearly constant. Once the $T_U$ 
feature described in the main text emerges
in the $C/T$, the entropy change rapidly decreases as field increases. 
The red and purple lines are linear fits to the 0 K to $T_U$ and 0 K to $T_N$ points, respectively. 
The entropy change trends towards zero at $\sim$ 4.8 T and $\sim$ 4.6 T for $T_U$ and $T_N$, 
respectively. The linear fits interpolate to $\sim$ Rln(2) and $\sim$80 \% Rln(2) at zero field for 
$T_U$ and $T_N$, respectively. 

\vspace{-0.3cm}

\subsubsection{High-temperature specific heat}
\vspace{-0.2cm}

\begin{figure}[b]
\vskip -0.4cm
\includegraphics[width=0.99\linewidth]{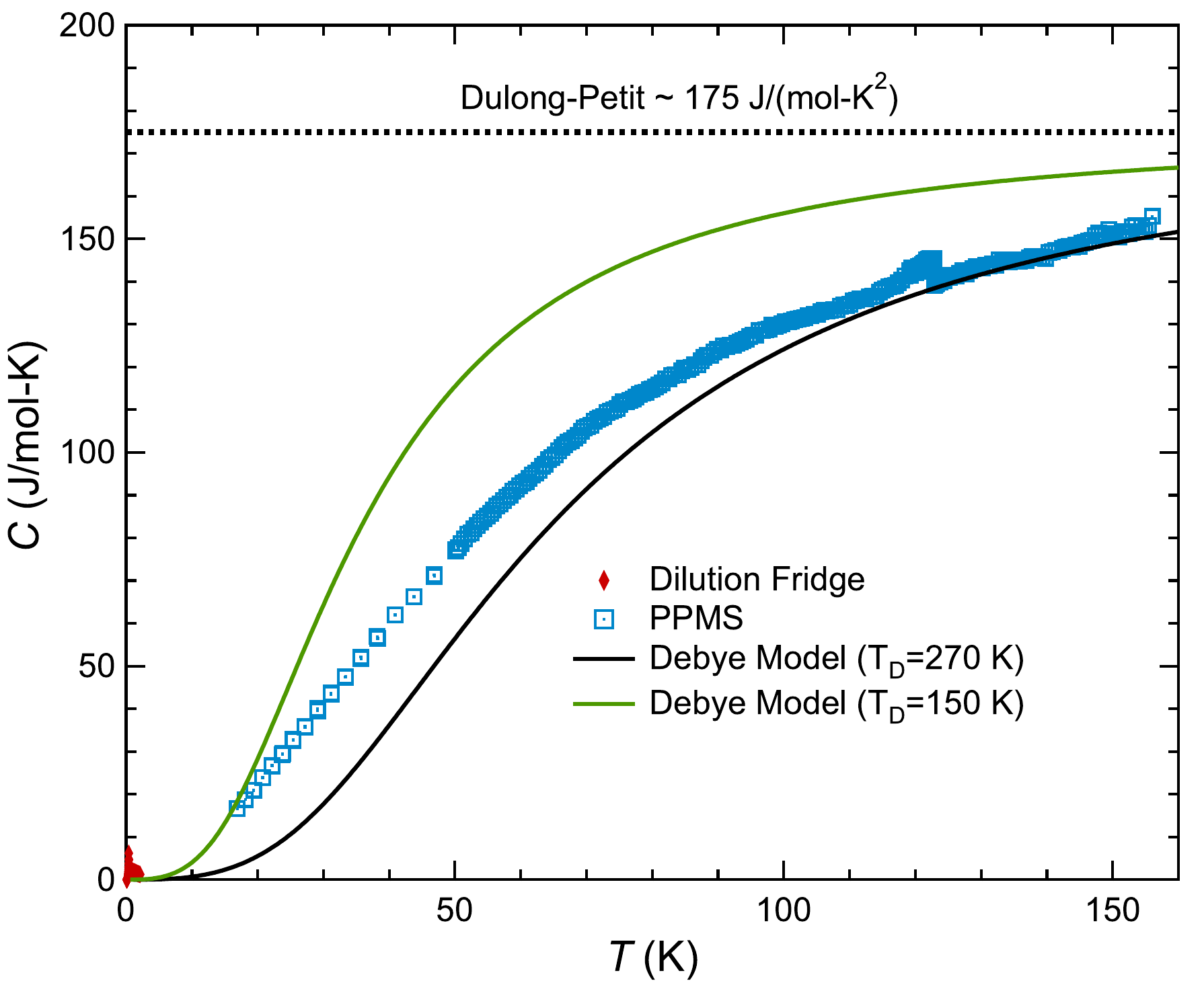}
\vskip -0.2cm
\caption{Specific heat from the low- and high-$T$ measurements.
Debye-model results for two representative Debye temperatures, 
Dulong-Petit limit is also indicated.}
\label{fig:Debye}
\end{figure}

In Fig.~\ref{fig:Debye}, we present our specific heat data for CeCd$_{3}$As$_{3}$ for the temperatures 
up to 160 K. In agreement with the prior work on this and related compounds, 
see Refs.~\cite{Lee_PRB_2019,Dunsiger_PRB_2020}, 
the one-parameter Debye model fails to provide adequate description
of $C(T)$ at intermediate temperatures, which indicates  possible 
nearly dispersionless phonon modes \cite{Gopal_1966}. 
However, as this model is expected 
to provide an adequate estimate of the low-$T$ and high-$T$ limit of the lattice contribution to $C(T)$, 
such an exposition is still instructive.

\begin{figure*}[t]
\centering
\includegraphics[width=\linewidth]{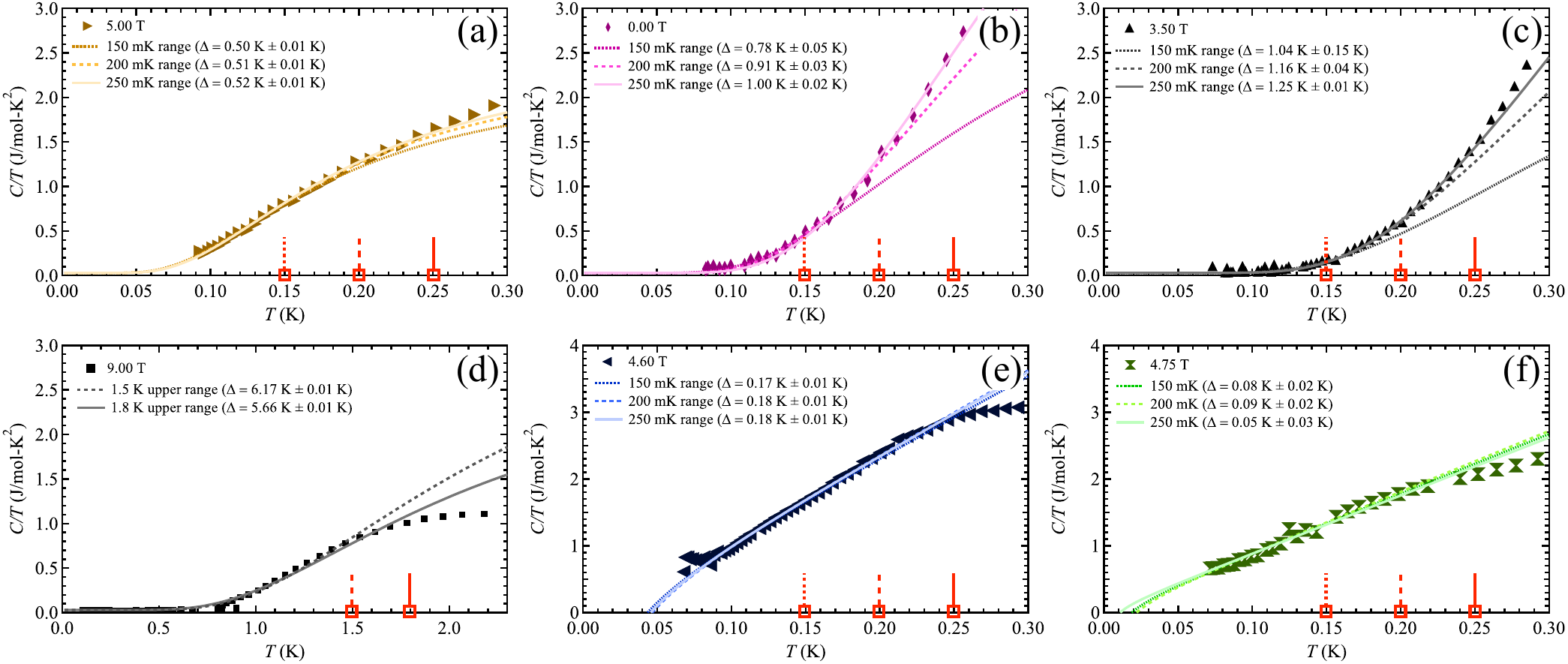}
\vskip -0.2cm
\caption{$C/T$ vs $T$ for selected field values and their fits using asymptotic expressions 
for the gapped spectra in Sec.~\ref{Sec_asympt} [Eqs.~(1) and (2) of the main text] 
for several fitting ranges. Fields 5~T and 9~T, (a) and (d),  are for the PM phase, 
4.6~T and 4.75~T, (e) and (f),  are from the QCP region, 0~T and  3.5~T, (b) and (c), are from the AF phase. 
Extracted values of $\Delta$ and their ``intrinsic'' errors 
for each  fitting range are listed; upper limits of the ranges are
highlighted by  riser bars on the $T$-axis. 
The lower $T$-limit for fitting is 0~K for all  fits except the 4.6~T data in which we exclude the tail below 
90~mK due to thermal equilibrium concerns, the upper limits are 1.5~K and 1.8~K for (d), and 
150~mK, 200~mK, and 250~mK for the rest,   with the dotted, 
dashed, and solid lines showing  corresponding fits.}
\label{fig_all_fits}
\vskip -0.4cm
\end{figure*}

In Fig.~\ref{fig:Debye}, we plot the calculated Debye $C(T)$ for two Debye temperatures, 
$T_D$, of 150 K and 270 K, which can be argued to provide reasonable bounds on the overall phonon energy scale. 
The expected upper limit of the lattice contribution at $T\!\gg\!T_D$ is also indicated. 
Then a simple estimate using the low-temperature Debye approximation ($\sim T^3$) 
with a reasonable $T_D$ of
220 K gives the lattice contribution at 2 K, the upper limit of most of our $C(T)$ dilution fridge 
measurements, 
as $C_{ph}(T=2\,{\rm K})\approx 0.01$ J/mole-K. This is, indeed, several orders of magnitude lower than 
the magnetic contribution reported in this work, as is mentioned above.

While our data in Fig.~\ref{fig:Debye} is generally in a good agreement with the recent work of 
Ref.~\cite{Dunsiger_PRB_2020}, there is a small anomaly in it at $T_s\!\sim\! 125$ K, 
somewhat lower than a similar feature reported in \cite{Dunsiger_PRB_2020}. 
This feature is indicative of a structural
transition. There is also an affiliated anomaly in resistivity, discussed below. 
However, there are no discernible indications of this transition in the 
magnetic susceptibility. While this is certainly not conclusive and a careful study of  
whether this transition can cause any direct changes to the magnetic Ce-planes is called for, 
this observation contributes to the expectation that the magnetism of 
the rare-earth $f$-orbitals is not significantly affected by this weak anomaly.

\vspace{-0.3cm}

\subsubsection{Spin-excitation gaps from the low-$T$ specific heat}
\vspace{-0.2cm}

Here we use theoretical asymptotic expressions 
for the spectral gaps  obtained in Sec.~\ref{Sec_asympt} (Eqs.~(1) and (2) of the main text) 
and demonstrate the robustness of their extracted values with respect to the temperature range. 

\begin{figure}[t]
\includegraphics[width=0.99\linewidth]{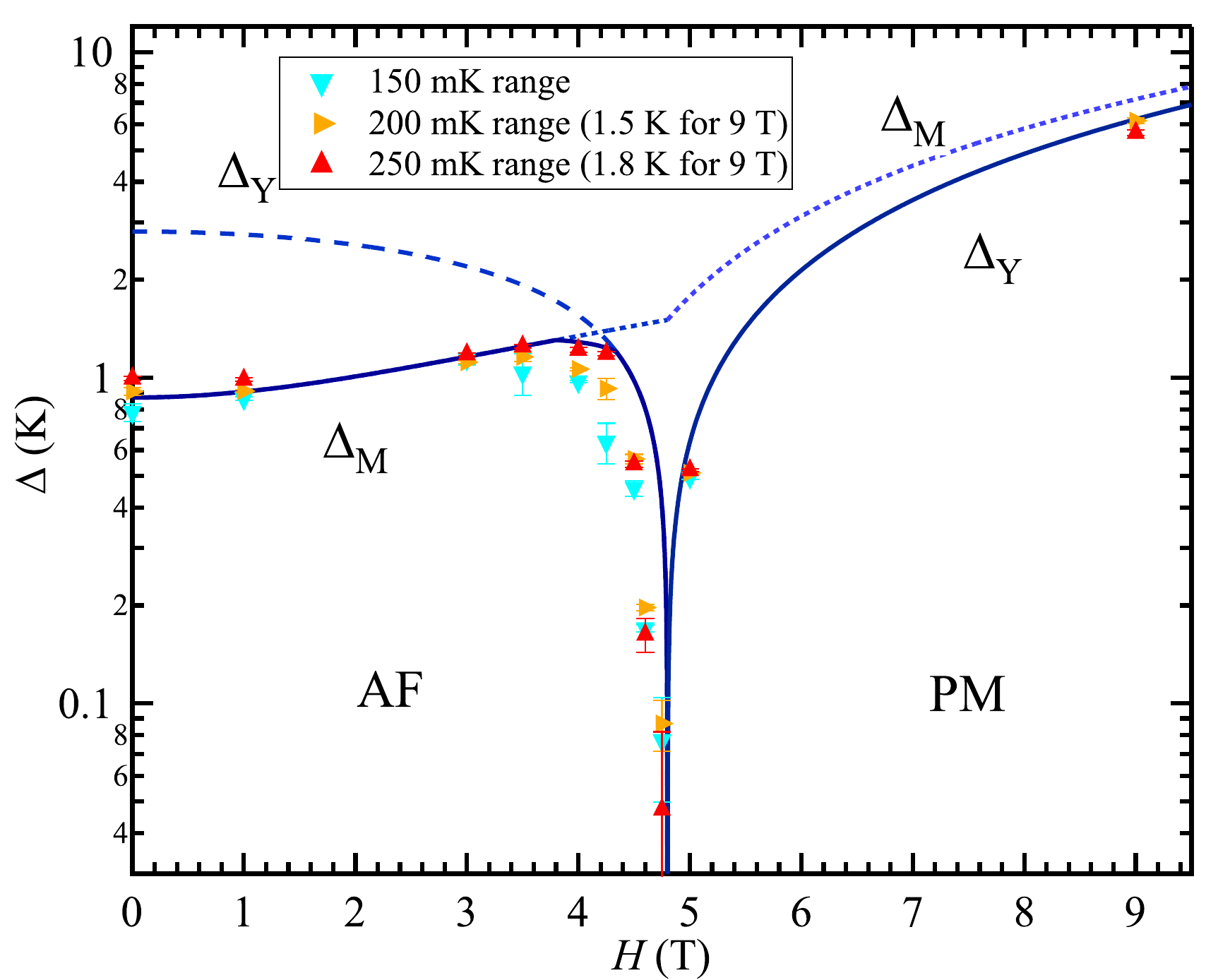}
\vskip -0.3cm
\caption{Same as Fig.~3 of the main text with the data for gap values 
$\Delta$  and their intrinsic error bars extracted from each fit as in Fig.~\ref{fig_all_fits}.}
\vskip -0.4cm
\label{fig:gaps_summary}
\end{figure}

Our Fig.~\ref{fig_all_fits} explicates the details of extracting spectral gap values $\Delta$ from 
the $C(T)$ data. It shows the results for $C/T$ vs $T$ for the field values from the paramagnetic phase, 
5~T and 9~T, Fig.~\ref{fig_all_fits}(a) and (d), from the QCP region, 4.6~T and 4.75~T, 
Fig.~\ref{fig_all_fits}(e) and (f), and for the two representative fields, $H\!=\!0$~T and  3.5~T,
from the AF ordered phase, Fig.~\ref{fig_all_fits}(b) and (c), respectively. Each panel lists the extracted 
value of $\Delta$ for each of the chosen temperature fitting range, with the upper limit of the latter 
also highlighted by the vertical riser bars on the $T$-axis. It also shows the variance of $\Delta$ 
for that temperature fitting range that 
characterizes the quality of each fit, which will be referred to as ``intrinsic'' error. 
The lower $T$-limit for fitting is 0~K for all  fits except the 4.6~T data in which we exclude the tail below 
90~mK due to thermal equilibrium concerns. The upper $T$-limits, with an exception of the 
9~T data in Fig.~\ref{fig_all_fits}(d),   are 150~mK, 200~mK, and 250~mK, with the dotted, 
dashed, and solid lines showing  corresponding fits.  For the 
9~T data in Fig.~\ref{fig_all_fits}(d), the upper limits for the fitting ranges are 1.5~K and 1.8~K, dashed 
and solid lines, respectively. 
The gap values $\Delta$ with their ``intrinsic'' error bars extracted from each fit are summarized 
in Fig.~\ref{fig:gaps_summary}. 

Several comments are in order. The 5~T data, Fig.~\ref{fig_all_fits}(a), yield $\Delta\!\approx\!0.51(1)$~K 
essentially independently of the fitting range. This is because in the PM phase and for $\Delta$ much 
smaller than the  spin-excitation bandwidth, $C(T)$ is expected to be dominated by the population of the 
lowest-gap excitation and, thus, is well-described by Eq.~(\ref{eq_Cvk2_final})  [Eq.~(1) of the main text]. 
The second set of data from the PM phase, 9~T in Fig.~\ref{fig_all_fits}(d), exhibit more variation because 
the gap in this case is at least the same value as the bandwidth and also possibly because the $T$-range is 
larger than the exchange energies.

The resultant gap values are also very stable for the fields from the QCP region, 4.6~T and 4.75~T, 
Figs.~\ref{fig_all_fits}(e) and (f), described by the small-gap asymptotic expression (\ref{eq_Cvl_as}) 
[Eq.~(2) of the main text], largely for the same reasons as the 5~T results. There is a bit more bias for the 
 4.75~T set, but this is because the gap is smaller than the lowest temperature for which the data is available. 

In the AF phase, 0~T and 3.5~T, Figs.~\ref{fig_all_fits}(b) and (c), the gap is large ($\sim \!1$~K).  
It is clear from Figs.~\ref{fig_all_fits}(b) and (c) that this makes the lowest shown $T$-range 
not representative, because its 150~mK cut-off limits the exponential tail to a small subset 
of the data that is also a subject of more significant intrinsic variance, especially in case of 3.5~T 
data where the gap is about 25\% higher than in zero field. Thus, while we still show their 
respective $\Delta$ values in the summary plot in Fig.~\ref{fig:gaps_summary} for completeness, 
we disregard the  150~mK-range results from our analysis for the strongly gapped data sets.

One can see that the 200~mK and 250~mK-range results in Figs.~\ref{fig_all_fits}(b) and (c)
are significantly more consistent. 
We refrain, however, from the further increase of the fitting ranges in the case of the AF-phase data. 
This is because their resulting fit may be biased by the critical fluctuations of the order parameter 
due to ordering at $T_N\!\sim\!500$~mK, which are unaccounted for by the theory in 
Eq.~(\ref{eq_Cvk2_final})  [Eq.~(1) of the main text]. 
Instead, we introduce generous estimates on the error bars as the geometric average of the intrinsic 
variance and the difference of the gap values of the two fitting ranges of 200~mK and 250~mK (1.5~K
and 1.8~K for 9~T set). As one can see from Fig.~3 of the main text and from Fig.~\ref{fig_gap_exp}, the
resultant errors are still rather small for most of the data.

\vspace{-0.3cm}

\subsection{Electrical resistivity}
\vspace{-0.3cm}

\begin{figure}[t]
\includegraphics[width=0.99\linewidth]{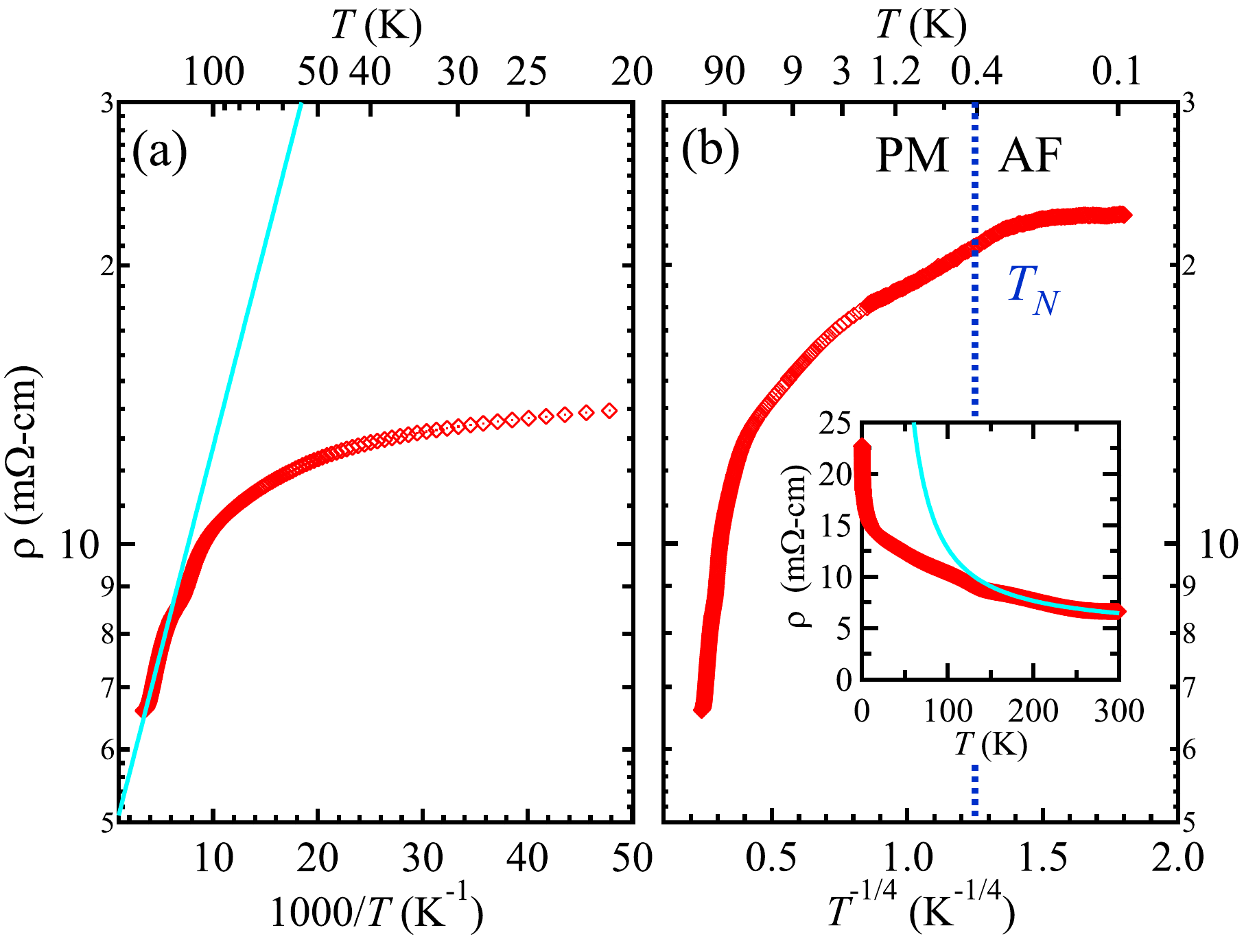}
\vskip -0.2cm
\caption{Electrical resistivity of CeCd$_3$As$_3$ as a function of temperature.
(a) log($\rho$) vs $1000/T$, (b) log($\rho$) vs $T^{-1/4}$, with inset showing   $\rho$ vs $T$. 
Activated behavior in (a), $\rho=\rho_0 e^{E_g/2T}$ with the fit of $E_g\approx 200$~K, 
and variable range hoping $\rho\sim e^{(T_0 /T)^{1/4}}$ in (b) are suggested. Dashed line is
the   zero field $T_N$ of the PM-AF transition.}
\vskip -0.4cm
\label{fig:Resistivity}
\end{figure}

The temperature dependence of electrical resistivity $\rho(T)$ of CeCd$_{3}$As$_{3}$ is presented in 
Fig.~\ref{fig:Resistivity}. It shows a semiconducting behavior at most temperatures 
with room-temperature $\rho$  approximately 6.6~$m\Omega$-cm,
suggesting low carrier concentration. 
Although no satisfactory fit range for a single gap semiconductor  could be found,
the activated  Arrhenius fit $\sim e^{E_g/2T}$ of the high-temperature tail yields a gap of 
$E_g\approx 200$ K (20 meV). However, the situation seems similar to the case of CeCd$_{3}$P$_{3}$,
where electrical transport suggested a similarly small activation energy $\sim 40$ meV, while 
the direct spectroscopic energy gap was found  to be significantly larger, $\sim 750$ meV, see
Ref.~\cite{Higuchi_Arxiv_2016}. This suggests that the entire family of these materials are indirect 
band-gap semiconductors and also raises an issue of impurity dopants shifting chemical potential 
away from the gap. In that context, we note that the resistivity of CeCd$_{3}$As$_{3}$
reported in Ref.~\cite{Dunsiger_PRB_2020} is significantly more metallic.

A small dip  centered at about 130~K can be observed in Fig.~\ref{fig:Resistivity}(a)
in agreement with Ref.~\cite{Dunsiger_PRB_2020} and our $C(T)$ results in Fig.~\ref{fig:Debye}.
A similar anomaly in $\rho$ is observed in 
CeCd$_3$P$_3$ and non-magnetic LaCd$_3$P$_3$ \cite{Lee_PRB_2019} ruling 
out a magnetic origin. It is likely related to either a subtle change in phonon modes 
unique to the crystal structure these compounds share or a structural transition. 

A log($\rho$) vs $T^{-1/4}$ is shown in Fig. \ref{fig:Resistivity}(b) with the dashed line indicating the 
location of the zero field $T_N$=412 mK PM-AF transition.
The lack of any feature  in $\rho(T)$ at $T_N$ suggests an extremely 
weak coupling between charge carries and the magnetic Ce$^{+3}$ ion.
The residual resistivity at $T\rightarrow 0$ is approximately 23~$m\Omega$-cm.
The somewhat linear log($\rho$) vs $T^{-1/4}$ behavior between 2.0 K and 0.3 K suggests that variable 
range hoping, $\sim e^{(T_0 /T)^{1/4}}$, may be playing a role at the lowest temperatures, 
but again the lack of any sizable fit range makes conclusions on the trap energy $T_0$ difficult.
Multiple excitation currents were attempted, and no Joule heating effect could be observed.

\vspace{-0.3cm}

\subsection{Comparison with CeCd$_3$P$_3$}
\vspace{-0.2cm}

Finally, we compare our results to those of isostructural CeCd$_3$P$_3$, 
which has been synthesized recently in
single crystal form \cite{Lee_PRB_2019} by the same group that reported the AF transition in 
CeCd$_3$As$_3$ \cite{Dunsiger_PRB_2020}. 
First, the magnetic susceptibility of CeCd$_3$P$_3$ is qualitatively and near quantitatively indistinguishable 
from both our and their results on CeCd$_3$As$_3$, which suggests nearly identical local CEF environments. 
Second, CeCd$_3$P$_3$ orders antiferromagnetically at $T_N=410$~mK at zero field, and the transition 
also increases to 430 mK under a 1.5 T easy-plane 
field, although it is important to emphasize that this field induced increase in $T_N$ for Ce triangular lattices 
can depend sensitively on precise field direction within the ab-plane as shown in the case of 
KCeS$_2$ \cite{Bastien_Arxiv_2020}. 
The authors were unable to measure heat capacity below 370 mK so no comparison of the residual heat 
capacity, the emergence of any crossover region, nor the extraction of the spin-wave gap could be done, 
although such a measurement would be worthwhile. 
In both compounds they obverse a momentary field induced decrease of $T_N$ for fields below 0.05 T. 
This is very likely to be domain selection since planer stripe magnetic structures in triangular lattices have 
three degenerate domains in zero field.
There are some discrepancies that could play a role in elucidating the behavior of these systems. 
The CeCd$_3$As$_3$ single crystals they made exhibit the same field induced increase of $T_N$ we observed,
however their CeCd$_3$As$_3$ made by flux growth demonstrate electrically conductive behavior whereas 
ours made by chemical vapor transport are semiconducting (see Fig. \ref{fig:Resistivity}).
However, the very low carrier density of their samples suggests that both compounds likely sit close to a 
metal-insulator transition. 
It is self-evident at this point that exact nature of electrical conductivity in these systems is very sensitive to 
minor details of how they are grown and will require further investigation. 
The lack of any noticeable difference between our results in terms of heat capacity, magnetization, and 
susceptibility demonstrates that the charge carries or lack thereof have minimal impact on the magnetic degrees 
of freedom in these compounds. There are also other closely related lanthanide compounds that deserve 
scrutiny \cite{Kabeya_JPSJ_2020}.

\vspace{-0.3cm}

\section{Theoretical model}
\label{Sec_model}
\vspace{-0.2cm}

\begin{figure}[b]
\vskip -0.2cm
\centering
\includegraphics[width=0.99\linewidth]{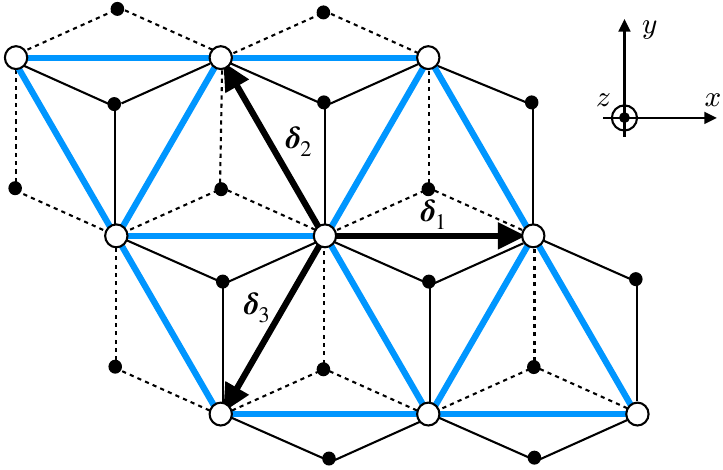}
\vskip -0.2cm
\caption{A sketch of the triangular-lattice layer of magnetic ions (empty circles) embedded in the 
octahedra of ligands (black dots) with the primitive vectors. Thick (blue) bonds are between magnetic ions 
and ion-ligand bonds are the thin solid (dashed) lines for above (below) the plane.}
\label{fig_struct}
\vskip -0.4cm
\end{figure}

For a typical layered triangular-lattice structure, the relevant point-group symmetry operations  
are the $C_3$ (120{\degree}) rotation around the $z$ axis, $C_2$ (180{\degree}) rotation around each bond, 
site inversion symmetry $\mathcal{I}$, and two translations, 
$\mathcal{T}_1$ and $\mathcal{T}_2$ along  ${\bm \delta}_1$ and ${\bm \delta}_2$, 
respectively \cite{Li_PRB_2016},
see Fig.~\ref{fig_struct}. These symmetries allow four terms in the nearest-neighbor 
Hamiltonian that can be separated into bond-independent ($XXZ$)  
and bond-dependent parts, 
\begin{align}
{\cal H}=&\sum_{\langle ij\rangle}\Big({H}_{\langle ij\rangle}^{\bar{\Delta}}+{H}_{\langle ij\rangle}^{bd}\Big)
+\sum_{\langle ij\rangle_2}{H}_{\langle ij\rangle}^{\bar{\Delta}},\nonumber\\
\label{HJpm}
{H}_{\langle ij\rangle_m}^{\bar{\Delta}}=& \ J_m
 \Big(S^{x}_i S^{x}_j+S^{y}_i S^{y}_j+\bar{\Delta} S^{z}_i S^{z}_j\Big)\\
{H}_{\langle ij\rangle}^{bd}=&\ 2 J_{\pm \pm} \Big[ \Big( S^x_i S^x_j - S^y_i S^y_j \Big) \tilde{c}_\alpha 
-\Big( S^x_i S^y_j+S^y_i S^x_j\Big)\tilde{s}_\alpha \Big]\nonumber\\
 &+ J_{z\pm}\Big[ \Big( S^y_i S^z_j +S^z_i S^y_j \Big) \tilde{c}_\alpha 
 -\Big( S^x_i S^z_j+S^z_i S^x_j\Big)\tilde{s}_\alpha \Big],\nonumber
\end{align}
where $\tilde{c}(\tilde{s})_\alpha\!=\!\cos(\sin)\tilde{\varphi}_\alpha$, the bond angles $\tilde{\varphi}_\alpha$
are that of the primitive vectors ${\bm \delta}_\alpha$ with the $x$ axis, 
$\tilde{\varphi}_\alpha\!=\!\{0,2\pi/3,-2\pi/3\}$, 
and spin projections are in crystallographic axes that are tied to the lattice, see Fig.~\ref{fig_struct}.  The
bond-independent exchange constants $J_m$ are $J_1$ and $J_2$ for the nearest- and second-nearest-neighbor
couplings, respectively.  Following previous considerations \cite{Paddison_NatPhys_2019, Zhu_PRL_2018}, 
we use a minimal generalization of the 
nearest-neighbor model by augmenting it with the second-nearest-neighbor $XXZ$ 
term with the same anisotropy parameter $\bar{\Delta}$. 

In an external field, the standard Zeeman coupling 
\begin{align}
{\cal H}_Z=-\mu_B\sum_{i}\Big[g_{ab} \Big(H_x S_x+H_yS_y\Big) +g_zH_z S_z\Big],
\label{HZ}
\end{align}
contain anisotropic $g$-factors  of the pseudo-spins that reflect the  build-up of the ground-state doublets 
from the states of the original ${\bf J}$-multiplet of the rare-earth ions due to a combined effect of 
spin-orbit coupling
and CEF. The in-plane $g$-factor is uniform because of the three-fold symmetry of the lattice \cite{Li_PRL_2015}. 

\begin{figure*}[t]
\centering
\includegraphics[width=0.99\linewidth]{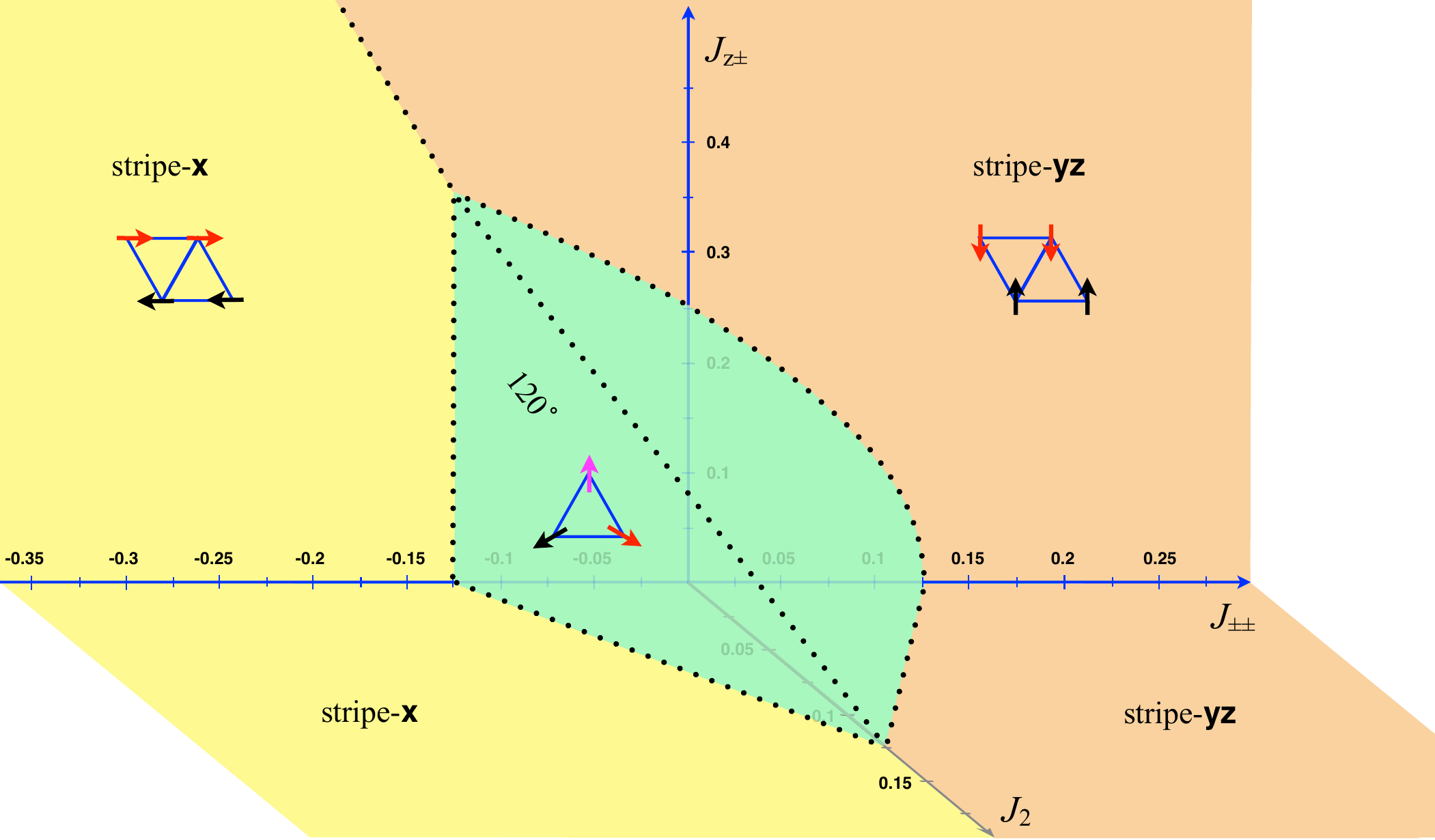}
\caption{Classical $J_{\pm \pm}$--$J_{z\pm}$--$J_2$ phase diagram of the Hamiltonian from the main text 
with $J_2$ term for representative $\bar{\Delta}\!=\!1$ obtained by the energy minimization 
for the commensurate single-$\mathbf{Q}$ states.
All couplings  in units of $J_1\!>\!0$.}
\label{fig_PhDbig}
\vskip -0.4cm
\end{figure*}

\vspace{-0.3cm}

\section{Phases}
\label{Sec_phases}
\vspace{-0.2cm}

Fig.~\ref{fig_PhDbig} shows a section of the 
classical $J_{\pm \pm}$--$J_{z\pm}$--$J_2$ 3D phase diagram of the model 
(\ref{HJpm}) for a representative choice of $\bar{\Delta}\!=\!1$ and antiferromagnetic $J_1$.
All couplings are in units of $J_1$ and  the Hamiltonian is invariant under 
$J_{z\pm}\!\rightarrow \!-J_{z\pm}$ \cite{Maksimov_PRX_2019}.
It is obtained by the energy minimization for the commensurate single-$\mathbf{Q}$ states
and, thus, ignores more complicated multiple-$\mathbf{Q}$ states that occur near some of the phase boundaries
as discussed in Ref.~\cite{Maksimov_PRX_2019}. Since they are unimportant for our present consideration 
we ignore them as well.
This phase diagram is essentially the same for the other values of the easy-plane, or ``XY-like'' 
$0\!\leq\!\bar{\Delta}\!\leq\!1$, aside from the  $120{\degree}$ phase extending 
to somewhat larger values of $J_{z\pm}$ \cite{Zhu_PRL_2018, Maksimov_PRX_2019}.

As is discussed in Refs.~\cite{Zhu_PRL_2018, Maksimov_PRX_2019}, the $120{\degree}$ phase 
is favored by the $XXZ$ part 
of the Hamiltonian while the stripe phases are favored by the bond-dependent 
$J_{\pm\pm}$ and $J_{z\pm}$  as well as by the $J_2$ term. 
The stripe phases differ by the mutual orientation of spins and bonds. In the stripe-${\bf x}$ phase, 
favored  by the negative $J_{\pm\pm}$, spins are fully in plane and along one of the bonds.
In the stripe-${\bf yz}$ phase, spins are perpendicular to one of the bonds and are  tilted out of 
plane for the non-zero $J_{z\pm}$. There is  an obvious three-fold degeneracy between the  
stripe states of different orientation of the ``ferromagnetic'' bonds along ${\bm \delta}_{1,2,3}$. 

The key message of Fig.~\ref{fig_PhDbig} is that there are only three ordered antiferromagnetic 
phases in the phase diagram of the Hamiltonian for  $J_1\!>\!0$, with or without the $J_2$-term. 
For the ``Ising-like'' $\bar{\Delta}\!>\!1$ and finite $J_{\pm\pm}$ and $J_{z\pm}$, stripe phases survive and 
continue to occupy much of the parameter space. In this limit, the $XXZ$ and bond-dependent 
anisotropic terms  also compete, resulting in a transition to a different stripe phase with the spin pointing 
out of the plane along the $z$ axis.
The full phase diagram for $\bar{\Delta}\!>\!1$ also contains ferrimagnetic ``Y'' phase in place of the 
$120{\degree}$ phase \cite{Miyashita_JPSJ_1986} and, generally, has a complicated cascade of the
field-induced phases \cite{Seabra_PRB_2011}.

\section{Case of ${\rm CeCd_3As_3}$}
\label{Sec_CeCdAs}

The essential empirical facts about CeCd$_3$As$_3$ are the following. 
It orders antiferromagnetically at $T_N\!\approx\!0.4$~K. The zero-field specific heat 
shows activated behavior with a sizable excitation gap $\Delta\!\approx\!1$~K, which 
is compatible to an estimate of a characteristic superexchange constant that one can infer from 
the extrapolated Curie-Weiss temperatures ($\Theta_{CW}\!\sim\!4$~K) \cite{Liu_Arxiv_2016}. 

The magnetization field-dependence, $M(H)$, shows a monotonic 
increase for both in-plane and out-of-plane field direction before reaching a saturation with 
some residual Van Vleck slope.
Due to a strong easy-plane magnetic anisotropy \cite{Banda_PRB_2018},
the actual saturation in the $c$-direction is beyond the measured field range,
which prevents an accurate determination of the saturation  field in that direction. 
The estimates of the corresponding in-plane and out-of-plane $g$-factors in (\ref{HZ}) give 
a factor of $\sim\!5$ between them, according to Ref.~\cite{Banda_PRB_2018}. 

Most importantly, although taken above the ordering temperature, magnetization curves 
in both principal directions show no traces of the plateau-like features or of any other phase transitions 
that are emblematic of the triangular-lattice magnets \cite{Starykh_RPP_2015, Seabra_PRB_2011}. 
We note that the presence of such plateau-like phases in the field-evolution would have definitely 
made the 120$\degree$ part of the phase diagram a prime suspect for the zero-field ground state.
This lack of the other field-induced phases is also strongly corroborated by the specific heat 
field-dependence in the in-plane field, which shows no sign of a closing of the excitation gap before 
the critical point to the paramagnetic state is reached at  $H_s^{ab}\!\approx\!4.6$--4.8~T.  
Together, these facts strongly suggest that the $H$--$T$ phase diagram of CeCd$_3$As$_3$
contains a single magnetically-ordered phase that evolves continuously from the $H\!=\!0$ state. 

\begin{figure}[t]
\centering
\includegraphics[width=0.99\linewidth]{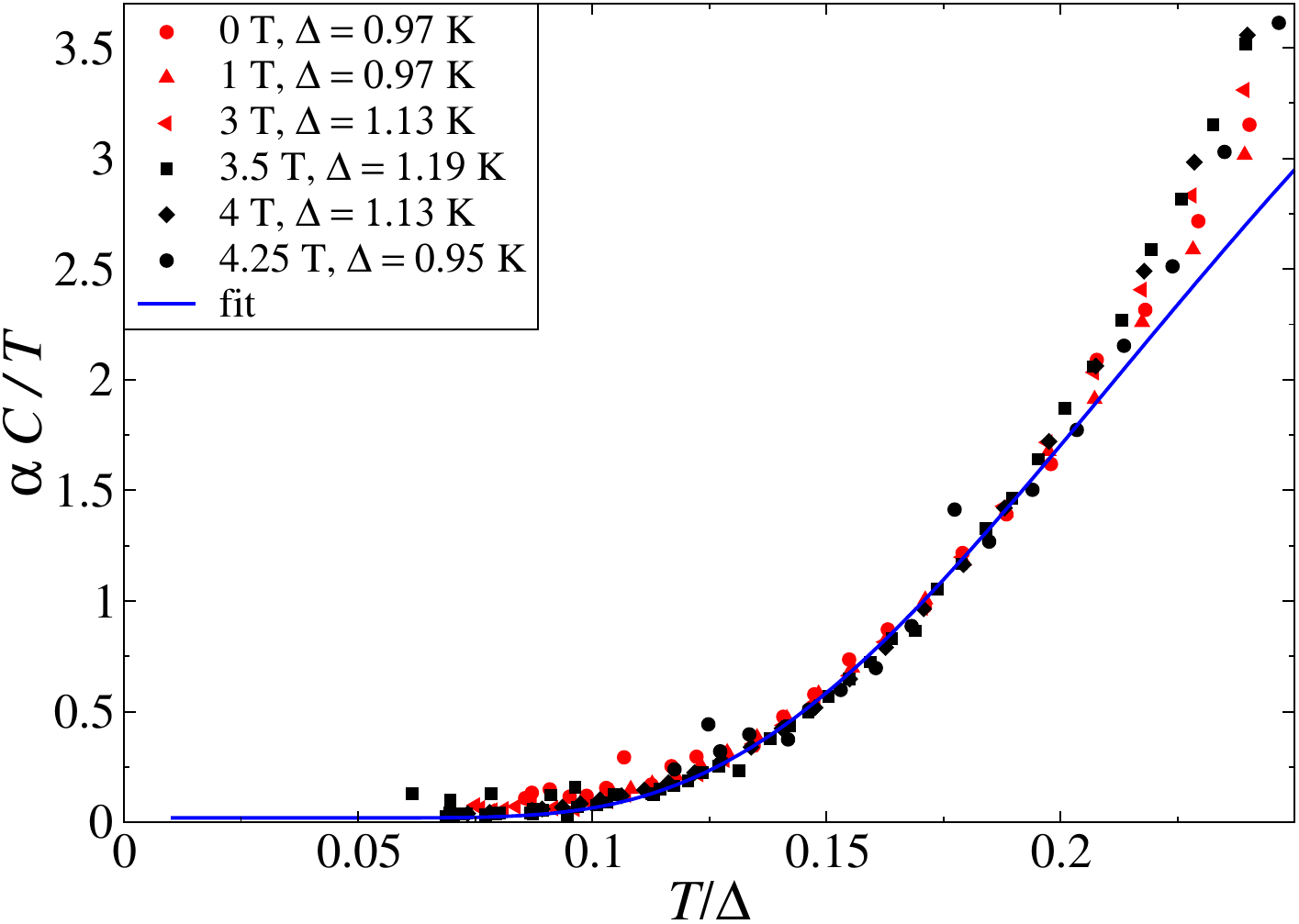}
\caption{Data collapse of the $C(T)/T$ vs $T$ data, where $C(T)$ is the specific heat at various fields 
below $H_s^{ab}$, with $T$ scaled with the gap $\Delta$ and $\alpha$  an {\it ad hoc} multiplicative constant. 
Solid line is the fit by the leading term in the 2D activated behavior, $Ax^2e^{-x}$, where $x\!=\!\Delta/T$.}
\label{fig_collapse}
\vskip -0.2cm
\end{figure}

To substantiate this statement, we show the specific heat  vs temperature data collapse for the fields 
$0\!\leq\!H\!\alt\!H_s^{ab}$ using the rescaling of the temperature by the gap $\Delta$ and an 
{\it ad hoc} multiplicative constant $\alpha$ for  $C(T)/T$ at different fields, see Fig.~\ref{fig_collapse}. 
The single fit is by the leading term in the 2D activated behavior, $Ax^2e^{-x}$, where $x\!=\!\Delta/T$, 
see Sec.~\ref{Sec_asympt} for details. 

In addition, the field-dependence of the excitation gap extracted from the specific heat data in the main text
demonstrates a seemingly more subtle, but an essential feature. It has a noticeable gradual increase by 20\%-60\%  
(depending on the fit)  from its zero-field value, followed by a rather abrupt closing upon approaching 
the saturation field. This characteristic behavior turned out to be an important 
distinguishing hallmark of the field-induced transformations in the magnetic excitation spectrum.

The other significant observations include a characteristic $T^2$ behavior of the specific heat at the 
field-induced transition to the paramagnetic state and an initial moderate increase 
of the N\'{e}el temperature vs field that is indicative of a suppression of critical fluctuations.

The key observations that are most important for the subsequent discussion are 
the gapped ground state and the single-phase character of the  ordered phase.

\section{Phase identification}
\label{Sec_ID}

Because of the combined SOC and CEF effects, 
the Hamiltonian has no continuous spin-rotational symmetries.  Therefore, one should generally expect
that {\it all} its ordered phases host gapped spin excitations. 
However, the 120${\degree}$ as well as the ferromagnetic states of the classical model 
exhibit  accidental continuous degeneracies \cite{Maksimov_PRX_2019}. In simple terms, their 
ground state energies have 
no contribution from the bond-dependent terms,  the orientation of their spin configurations is not fixed
beyond the one dictated by the $XXZ$ term, and  their spectra are gapless. The gaps open and 
spin directions get chosen as a result of a quantum order-by-disorder effect, but the gap 
magnitude is typically a small fraction of the exchange constant \cite{Rau_PRL_2018}.

This consideration suggests the stripe phases as strong contenders for the ground state of 
CeCd$_3$As$_3$ as they do allow for a sizable spin-excitation gap. The second argument that 
will be explored later is the simplicity  of the field-evolution of the 
zero-field stripe phases for the most part of the phase diagram.
That is, for a stripe state with a considerable gap, spins continuously 
tilt in a field until reaching a paramagnetic state without encountering     intermediate states. 
Lastly, the spectrum structure in the stripe phases is both peculiar and characteristic of the 
systems with significant frustrating or bond-dependent interactions. Because of the presence of 
an accidental degeneracy elsewhere in the phase diagram, the spectral minima
are {\it not} associated with the ordering vector, but are complementary to it 
\cite{Maksimov_PRX_2019, Maksimov_PRR_2020}.
Such a structure has a significant bearing on the field-induced transformations in the magnetic excitation 
spectrum, which, as we will demonstrate below, are in accord with the experimental observations
in CeCd$_3$As$_3$.

\begin{figure}
\centering
\includegraphics[width=0.99\linewidth]{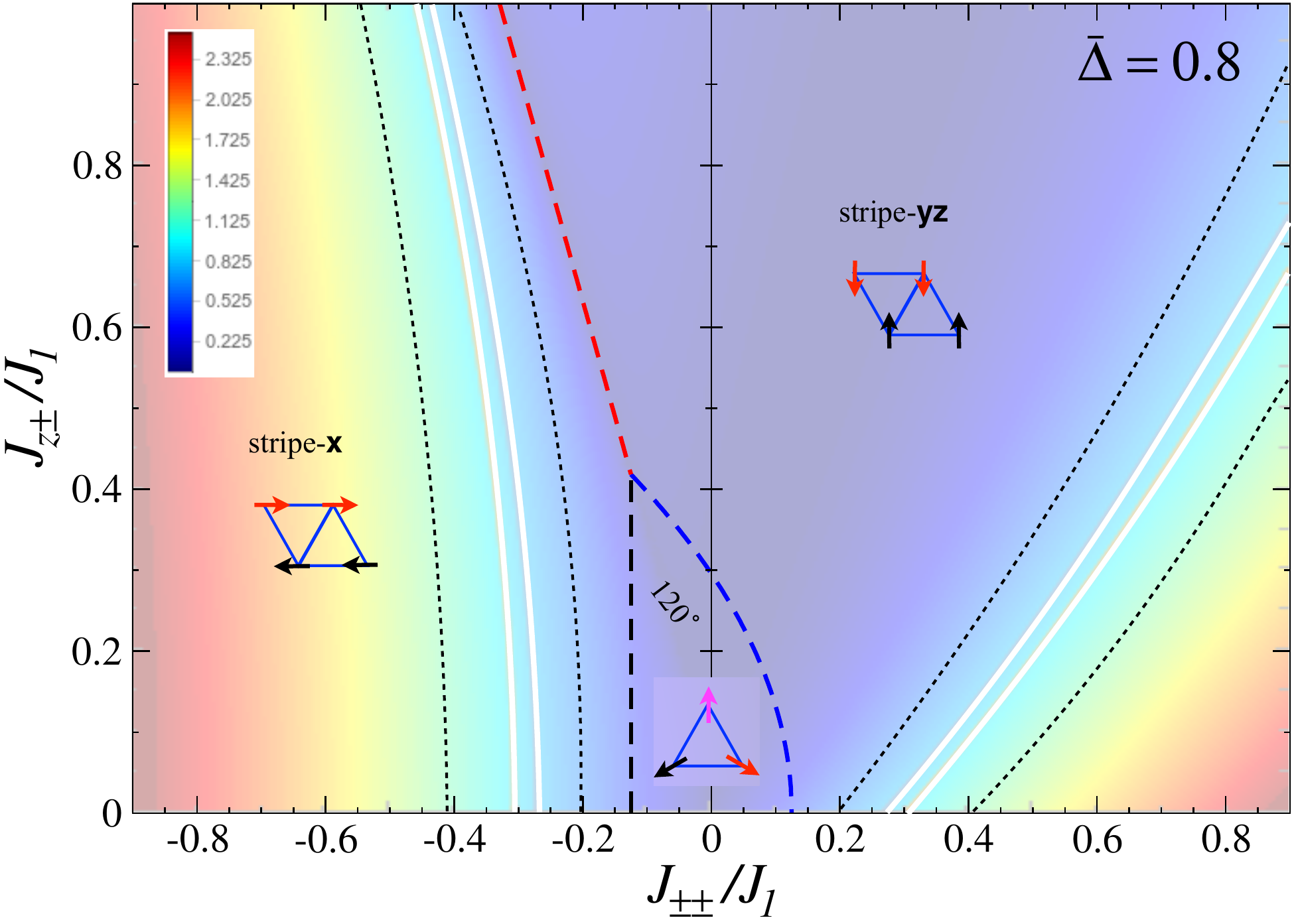}
\caption{The 2D $J_{\pm \pm}$--$J_{z\pm}$ phase diagram of the Hamiltonain 
for  $\bar{\Delta}\!=\!0.8$,  $J_2\!=\!0$, all in units of $J_1\!>\!0$. Phases are indicated. 
The  intensity plot shows  the gap in the spin excitation spectrum. 
The white solid lines are for $E_{gap}\!=\!1$~K  for  $J_1\!=\!1$~K and 1.2~K. 
The thin black dashed lines are the projections of the outer boundaries of such regions for 
$\bar{\Delta}\!=\!0.5$ and 1.5.}
\label{fig_gap_region}
\vskip -0.2cm
\end{figure}

To be quantitative,  we would like to demonstrate which regions of the phase diagram of the Hamiltonian 
CeCd$_3$As$_3$ can possibly belong. For that,
in Fig.~\ref{fig_gap_region} we present a 2D version of Fig.~\ref{fig_PhDbig},  the 
$J_{\pm \pm}$--$J_{z\pm}$ 2D phase diagram of the Hamiltonian 
for a representative choice of $\bar{\Delta}\!=\!0.8$, antiferromagnetic $J_1$, and $J_2\!=\!0$,
which shows the same three phases, the $120{\degree}$ and two stripe phases. 
The underlying intensity plot indicates the size of the gap in the spin excitation spectrum, all in units of $J_1$.
The two solid white lines on each side show the boundaries on the 
range of $J_{z\pm}$ and  $J_{\pm \pm}$  for the choice of the zero-field gap 
$E_{gap}\!=\!1$~K for $J_1$ varying between 1~K and 1.2~K. 
The two sets of the thin black dashed lines for positive and negative $J_{\pm \pm}$ show the projections 
of the outer boundaries of such ranges from the same phase diagrams but for different values of the $XXZ$ 
anisotropy, 
$\bar{\Delta}\!=\!0.5$ and $\bar{\Delta}\!=\!1.5$, altogether giving a sense of how much the model parameters 
are constrained by the gap value.

Naively, an alternative scenario for the gapped ground state that may seem to be  a much more 
straightforward option is 
the Ising-like state. One can expect it to occur for $\bar{\Delta}\!>\!1$ in the absence of the bond-dependent terms.
First, as was mentioned above, the $XXZ$ and bond-dependent anisotropies may compete in this case,
so the  stripe phases survive for larger values of $|J_{\pm\pm}|(J_{z\pm})$, leaving our 
consideration for them intact. 
Second, for $\bar{\Delta}\!>\!1$ the gapless $120{\degree}$ phase  converts into a still gapless 
ferrimagnetic ``Y'' phase, 
in which coplanar spins form a deformed ``Y'' with one spin pointing perpendicular to the plane and two tilted away 
from it with the mutual angles that differ from $120{\degree}$, hence the 
ferrimagnetism \cite{Miyashita_JPSJ_1986}.
Since the latter is not observed, and the state is also gapless, this phase can be ruled out. 
The last option from this ``Ising domain'' 
is the ``stripe-${\bf z}$'' phase, which occurs at finite $J_2$ and $\bar{\Delta}\!>\!1$ \cite{Seabra_PRB_2011} 
and should be stable for a range of $J_{\pm\pm}$ and $J_{z\pm}$.
However, there are two arguments against CeCd$_3$As$_3$ being in this phase. 
This phase should exhibit a complicated cascade of the first-order 
spin-flop-like transitions for the field in the $c$-direction, see Ref.~\cite{Seabra_PRB_2011}, 
occurring already in the low enough fields to be be visible in the available finite-temperature $M(H)$ data. 
No traces of such transitions are seen experimentally.  Second, we have also calculated the 
field-dependence of the gap for a representative set of parameters from this phase, see Sec.~\ref{Sec_gaps}. 
As opposed to the stripe-${\bf x}({\bf yz})$ case, the gap behavior vs field is monotonic and 
is not compatible  with the phenomenology of the CeCd$_3$As$_3$.

\section{Stripe phase}
\label{Sec_stripe_theory}

Although the qualitative arguments presented above strongly point toward CeCd$_3$As$_3$ being 
in a stripe part of the phase diagram, the model description requires defining, or restricting, of  
five parameters: two superexchange constants $J_1$ and $J_2$, $XXZ$ 
anisotropy constant $\bar{\Delta}$, and the bond-dependent terms $J_{\pm\pm}$ and $J_{z\pm}$.

\subsection{A minimal model}

The ground states and excitation spectra of the stripe phases of the Hamiltonian have been 
thoroughly considered in zero field \cite{Maksimov_PRX_2019, Zhu_PRL_2017} and above the saturation 
field for both the out-of-plane \cite{Paddison_NatPhys_2019, Zhu_PRL_2017} 
and in-plane \cite{Li_Arxiv_2016} field directions. The high-field description 
is reasonably straightforward and gives clear insights into the role of different terms of the 
Hamiltonian. In particular, $J_{z\pm}$ does not contribute to the spin-wave
spectrum in the polarized phase in the out-of-plane field and the saturation fields $H_s$ in any 
of the three principal direction, $a(x_0)$, $b(y_0)$, or $c(z_0)$, are also independent of it.
For the in-plane field,  the $J_{z\pm}$ term does contribute to the spectrum in the polarized state, 
but its kinematic form is such that it vanishes at all relevant high symmetry points. 
For the zero-field spectrum, one can also verify that adding a $J_{z\pm}$ term that is comparable to 
$J_{\pm\pm}$ does not provide any qualitative changes to it \cite{Maksimov_PRX_2019}. These 
insights give a significant reason to neglect the $J_{z\pm}$ term from the consideration altogether.

A different qualitative incentive that makes such a move highly desirable is the significantly simpler spin-wave 
algebra that allows to reduce the rank of the matrices to diagonalize and to derive many expressions in a simple 
analytical form. We, therefore, will not resist this temptation and, in the following, will put $J_{z\pm}\!=\!0$.

\begin{figure*}
\centering
\includegraphics[width=0.75\linewidth]{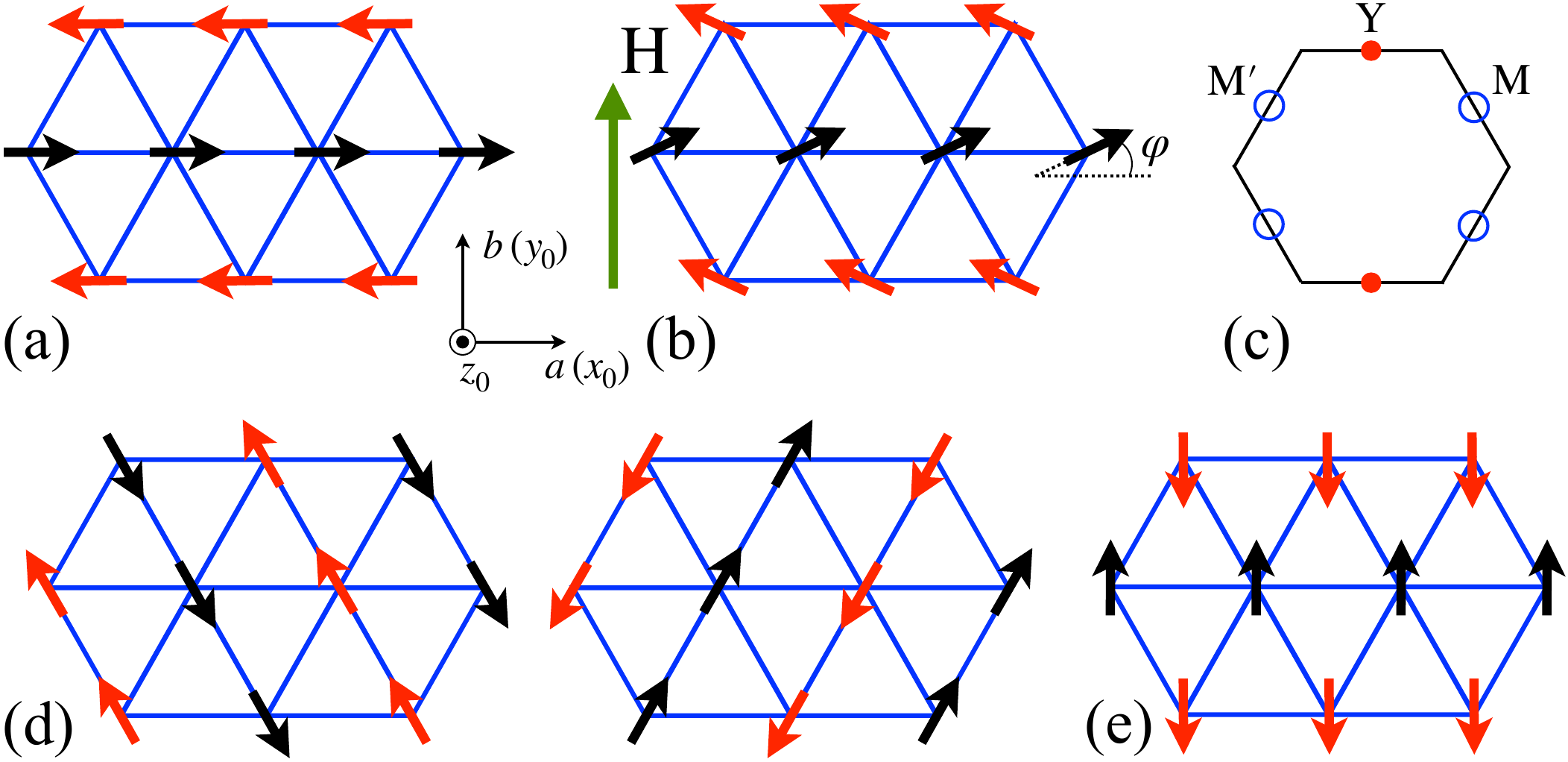}
\caption{(a) One of the three domains of the stripe-${\bf x}$ phase. (b)
Spin configuration of that domain in the $H\!\parallel\! b$ field. 
(c) Brillouin zone with the ordering vectors of the stripe phases Y, M, and M$^{\prime}$.
(d) Two  domains of the stripe-${\bf x}$ phase with the ordering vectors at M and M$^{\prime}$.
(e)  Spin configuration of  the stripe-${\bf yz}$ phase.}
\label{fig_domains}
\vskip -0.2cm
\end{figure*}

The second useful observation \cite{Zhu_PRL_2017} is that the linear combination of  the superexchange constants 
$J_1$ and $J_2$, in which they enter in the expressions for observables such as saturation fields and energy gaps at 
high symmetry points as well as into the classical energy of the state, is the same: $J_1+J_2$.
This means that we cannot constrain them independently and they simply define an overall scale of the exchange 
matrix. Thus, for the following consideration we will simply fix their ratio to a reasonable value that is 
consistent with the expectations for localized $f$-orbitals: $J_2/J_1\!=\!0.1$. We note that completely neglecting
$J_2$ leads to some additional subtle degeneracies in the spectrum both at zero and at the saturation
field that are not generic and are useful to avoid by keeping $J_2/J_1$ finite.
 
These choices lead to significant simplifications and leave us with three terms to constrain,
$J_1$, $\bar{\Delta}$, and $J_{\pm\pm}$. 

\subsection{Stripe domains}

For $J_{z\pm}\!=\!0$,   spin configurations in the stripe-${\bf x}$ and stripe-${\bf yz}$ phases 
are within the triangular lattice plane ($x_0$--$y_0$ or $ab$ plane), aligned with the bond 
($J_{\pm\pm}\!<\!0$) and  perpendicular to it ($J_{\pm\pm}\!>\!0$), respectively, 
see Fig.~\ref{fig_domains}(a) and (e). Their corresponding
energies and excitations in zero field are fully symmetric with respect to 
$J_{\pm\pm} \!\rightarrow\! -J_{\pm\pm}$ \cite{Zhu_PRL_2017}. As was mentioned in Sec.~\ref{Sec_phases},
there is a three-fold degeneracy of the stripe states in zero field, illustrated in  Fig.~\ref{fig_domains}(d), which
shows two configurations that are degenerate with the one in Fig.~\ref{fig_domains}(a). 
The domains of all three states should be present in a material.

There are two principal in-plane field directions, $H\!\parallel\! b$ and $H\!\parallel\! a$. 
Theoretically,  one of the three domains with the spin configuration that is ``most transverse'' to the external field  
will be selected  by an infinitesimal field. For instance,  for the stripe-${\bf x}$ phase in the $H\!\parallel\! b$ 
field, the domain in Fig.~\ref{fig_domains}(a) is energetically favored  over the domains in Fig.~\ref{fig_domains}(d).
In the $H\!\parallel\! a$ field, {\it two} domains in Fig.~\ref{fig_domains}(d) are preferred over the one in 
Fig.~\ref{fig_domains}(a) and remain degenerate till the saturation field.
Experimentally, the domain selection often occurs at some small but finite field because of the lower symmetry
of the spin system due to lattice distortions, domain surface energy, disorder pinning, and other real-life
complications.  The former reason is known to take place in case of $\alpha$-RuCl$_3$ 
\cite{Maksimov_PRR_2020} and a small-field transition has been observed in CeCd$_3$P$_3$ \cite{Lee_PRB_2019}, 
a material that is related to the present case.

We also note that  for $J_{z\pm}\!=\!0$, the finite-field consideration of the stripe-${\bf x}$ and stripe-${\bf yz}$ 
phases is fully symmetric under the simultaneous change of sign of $J_{\pm\pm}$ 
and switching the roles of $H\!\parallel\! b$ with $H\!\parallel\! a$.
 
\subsection{Saturation fields and magnetization}

To be specific, we will focus on the case of  $J_{\pm\pm}\!<\!0$ that corresponds to the stripe-${\bf x}$ phase
and select the field direction $H\!\parallel\! b$. In this case, the spins cant gradually toward the field as
is shown in Fig.~\ref{fig_domains}(b). The corresponding ordering vector for this domain is associated 
with the Y-point in the Brillouin zone in Fig.~\ref{fig_domains}(c), while the M- and the M$^{\prime}$-points 
are referred to as complementary to it. We also remark that this choice is also the simplest from the 
analytical point of view as the canting angle   in Fig.~\ref{fig_domains}(b) is the same for all the spins.
For comparison, the choice of $H\!\parallel\! a$ for the same $J_{\pm\pm}\!<\!0$ case would not only
lead to a coexistence of two domains from Fig.~\ref{fig_domains}(d), but there are four distinct spin 
tilt angles in this case that need to be found numerically.

The saturation fields for the principal field directions can be straightforwardly found from vanishing 
of the high-field spectrum gap at the ordering vector at the transition \cite{Zhu_PRL_2017, Li_Arxiv_2016}. 
For the in-plane field directions for the Hamiltonian with the Zeeman term, taking $S\!=\!\frac12$, this yields 
\begin{align}
\label{Hs}
h_s^{(b)}=g_{ab}\mu_B H_s^{(b)} &=4\Big( J_1+J_2 -J_{\pm\pm}\Big), \\ 
h_s^{(a)}=g_{ab}\mu_B H_s^{(a)}&=4\left( J_1+J_2 -\frac12 J_{\pm\pm}\right), \nonumber
\end{align}
note that $H_s^{(b)}\!>\!H_s^{(a)}$ since $J_{\pm\pm}\!<\!0$.
For the stripe-${\bf yz}$ case, one needs to switch the sign of $J_{\pm\pm}$ in (\ref{Hs}) and 
$H_s^{(b)}$ with $H_s^{(a)}$. 

Importantly, despite the uniform in-plane $g$-factor, the 
saturation fields are different for the two principal field directions, with the difference 
$\Delta h_s\!=\!2|J_{\pm\pm}|$. This is a general consequence of the bond-dependent interactions in 
the anisotropic-exchange materials \cite{Maksimov_PRR_2020}.
Note that the recent work on a different Ce-based triangular-lattice material, KCeS$_2$, has found a  
clear indication of the splitting of the transition lines in the $H$--$T$ phase diagram for 
the two in-plane field directions \cite{Bastien_Arxiv_2020}, providing an evidence of the same trend.  

Another important feature of Eq.~(\ref{Hs}) is that for the fixed $J_2/J_1$ only two parameters of the 
Hamiltonian define $H_s^{(a/b)}$, $J_1$ and  $J_{\pm\pm}$, 
thus offering a strong constraint on them from the empirical value of the in-plane $H_s$.
For the out-of-plane field, the saturation field is given by \cite{Zhu_PRL_2017}
\begin{align}
\label{Hsc}
h_s^{(c)}=g_{c}\mu_B H_s^{(c)}=\Big(J_1+J_2\Big)\Big(3\bar{\Delta}+1\Big) +4|J_{\pm\pm}|, 
\end{align}
that strongly depends on the $XXZ$ anisotropy $\bar{\Delta}$. 

\begin{figure}[t]
\centering
\includegraphics[width=0.99\linewidth]{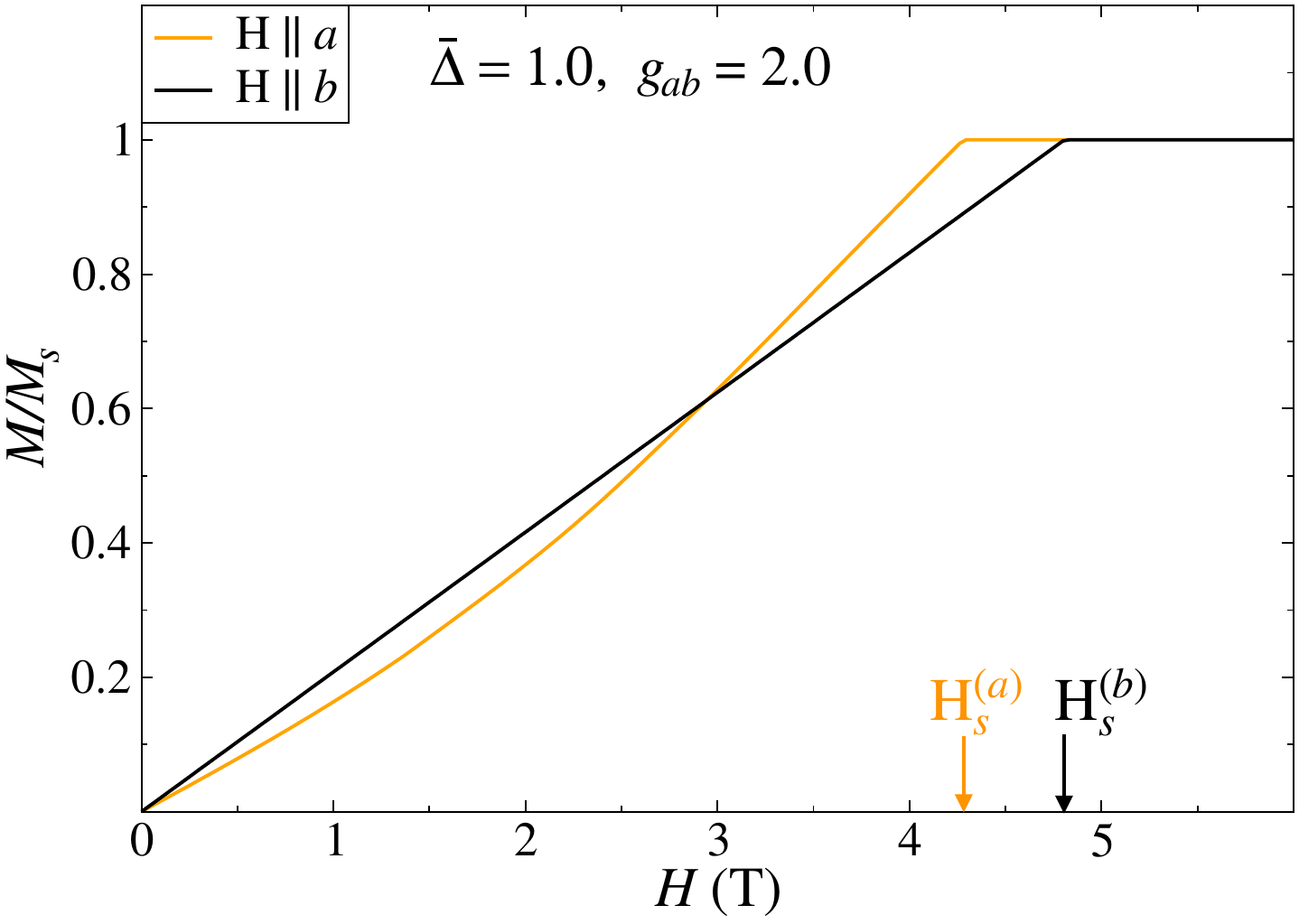}
\caption{The magnetization $M(H)$ in the stripe-{\bf x} phase 
at $T\!=\!0$ for $H\!\parallel\! a$ and $H\!\parallel\! b$. 
Parameters of the Hamiltonian are to match $H_s^{exp}\!\approx\!4.8$~T in the $b$-direction, and
$g$-factors are as discussed in the text.}
\label{fig_MvsHab}
\end{figure}

\begin{figure}[t]
\centering
\includegraphics[width=0.99\linewidth]{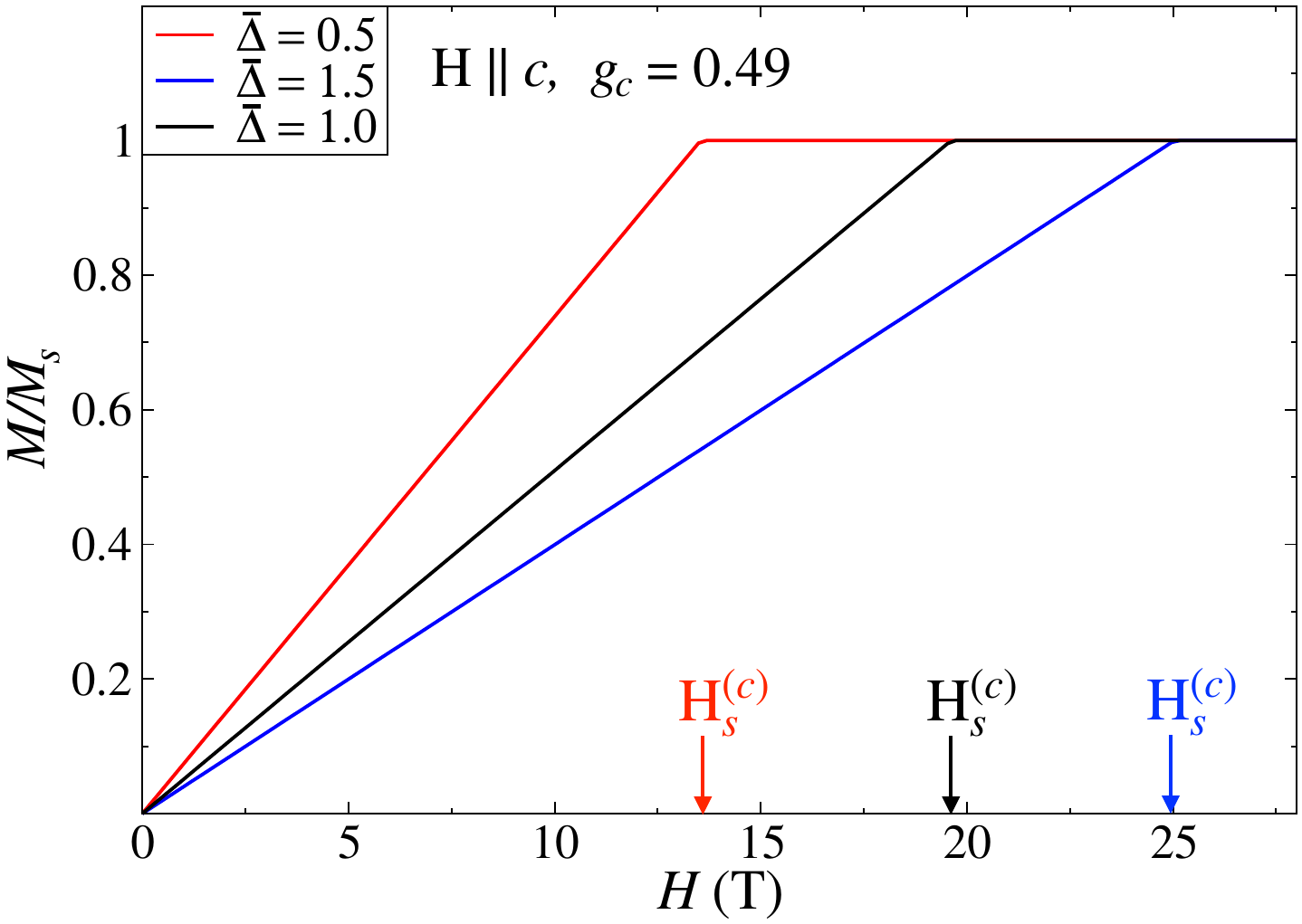}
\caption{Same as in Fig.~\ref{fig_MvsHab} for  $H\!\parallel\! c$ and for several $\bar{\Delta}$.}
\label{fig_MvsHc}
\end{figure}

To demonstrate the implications of these results, we complement them with the magnetization 
field-dependence $M(H)$ in the stripe-{\bf x} phase 
at $T\!=\!0$ for the $a$- and $b$-directions in Fig.~\ref{fig_MvsHab} and 
for the $c$-direction in Fig.~\ref{fig_MvsHc}, respectively. While for the $b$- and $c$-directions
the spin tilting is simple and can be obtained analytically, for the $a$-direction the results are obtained 
from the numerical energy minimization.  In all three directions, the stripe phase continuously 
deforms until reaching  spin saturation at the corresponding critical field $H_s^{(a/b/c)}$ with no 
intermediate transitions.

The parameters that are used to obtain  $M(H)$  in  Fig.~\ref{fig_MvsHab} and 
Fig.~\ref{fig_MvsHc} are discussed in more detail below. At this stage it suffices to say that their choice is
dictated by matching   experimental values of the zero-field gap and of the in-plane saturation 
field  $H_s^{exp}\!\approx\!4.8$~T, with the field  assumed to be in the $b$-direction, and by 
varying the $XXZ$ anisotropy $\bar{\Delta}$ in Fig.~\ref{fig_MvsHc}. 

The difference of the saturation fields $H_s^{(b)}$ and $H_s^{(a)}$ in Fig.~\ref{fig_MvsHab} 
is to be expected from the discussion following Eq.~(\ref{Hs}), with $XXZ$ anisotropy 
having no bearing on the $M(H)$ curve for the field in the $b$-direction.
Although $M(H)$ curves in Fig.~\ref{fig_MvsHc} are featureless, the strong dependence of 
$H_s^{(c)}$ on $\bar{\Delta}$, also in accord with Eq.~(\ref{Hsc}), is obvious. 
Unfortunately, the experimental restrictions on the actual value of 
the saturation field in the $c$-direction  in CeCd$_3$As$_3$ are not tight enough, see \cite{Banda_PRB_2018}. 
While the value as low as $13$~T as for $\bar{\Delta}\!=\!0.5$ is probably a stretch,  
it is difficult to narrow down the shown 
range of $\bar{\Delta}\!=\!0.5$--1.5 more significantly with the available data
given possible uncertainties in the $g$-factors.
The good news is that other quantities of interest that do depend on  
$\bar{\Delta}$ do so rather insignificantly, as we show below. 
Lastly, our choices of the $g$-factors differ somewhat from the results suggested in  
Ref.~\cite{Banda_PRB_2018}, $g_{ab}\!=\!2.0$ instead of 2.38 and $g_{c}\!=\!0.49$ instead of 0.46.
These are to reflect slightly lower in-plane and  higher out-of-plane initial slopes of $M(H)$
in the present study. A smaller $g_{ab}$-factor also seems to be supported by the value of the spin-excitation
gap in the field 9~T, much above the saturation.

\begin{figure*}
\centering
\includegraphics[width=0.75\linewidth]{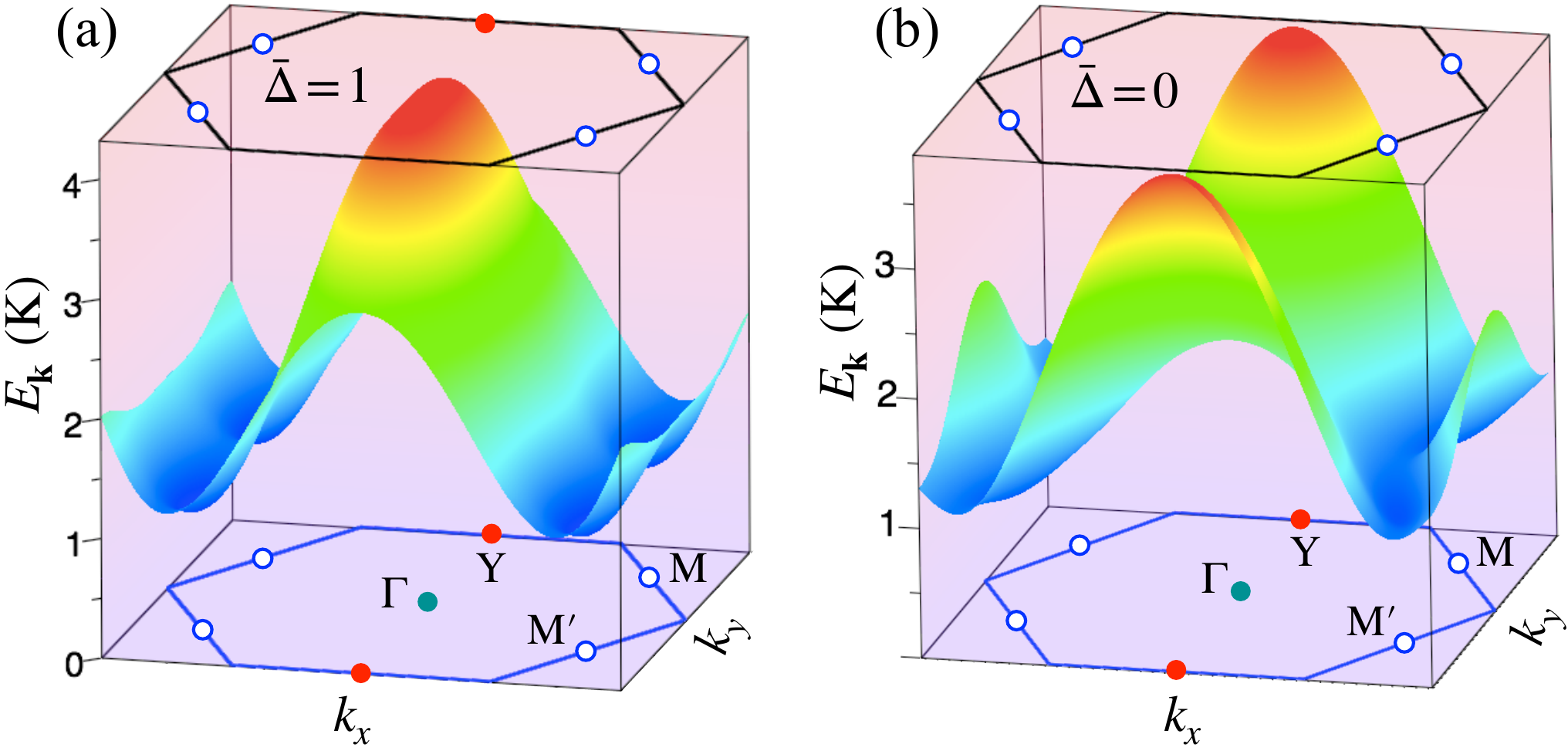}
\caption{The 3D plots of the $H\!=\!0$ magnon energy $\varepsilon_\mathbf{k}$ (in Kelvins) from Eq.~(\ref{Ek})
throughout the Brillouin zone for two representative sets of parameters. 
High-symmetry ${\bf k}$-points are indicated. The $XXZ$ anisotropy is 
(a)  $\bar{\Delta}\!=\!1$ and (b) $\bar{\Delta}\!=\!0$.}
\label{fig_EkH0}
\vskip -0.2cm
\end{figure*}

We add an extra note on the determination of  $\bar{\Delta}$ parameter from the available data.
The estimates on $\bar{\Delta}$ rely on the extrapolation of the magnetization vs field  data for $H||c$, 
that do not reach $H_s^{(c)}$.  
There are two aspects of the extrapolation procedure for $H_s^{(c)}$ that are important. 
If one uses the data for $M(H)$ vs $H||c$ from Ref.~\cite{Liu_Arxiv_2016}, it would imply 
$H_s^{(c)}\!\approx\!30$~T. The slope in our case, Fig.~\ref{fig:Magnetization}, 
is slightly higher, leading to an estimate of $H_s^{(c)}\!\approx\!25$~T. However, both estimates are the 
subject of finite temperature effects, which suppress $M$ vs $H$ slope. For extracting model parameters 
such as $\bar{\Delta}$, we would ideally need $T\!=\!0$ data for $M(H)$. 
Although we cannot access it, we can estimate the effect
of finite $T$ ($\approx\!2$K) from the in-plane $M(H)$ results, for which finite-$T$ $H_s^{(a/b)}$ 
is estimated as $\approx\!7$~T, while the  low-$T$ (``true'') $H_s^{exp}\!\approx\!4.8$~T, a $>\!30$\% 
decrease. For the $H||c$ case, we take a conservative 20\% 
reduction of the finite-$T$ $H_s^{(c)}\!\approx\!25$~T value to give the  low-$T$ 
$H_s^{(c)}\!\approx\!20$~T.
As one can see from our Fig.\ref{fig_MvsHc}, this yields $\bar{\Delta}$ very close to 1.0.
The physically less motivated choices of $H_s^{(c)}\!\approx\!13$~T gives $\bar{\Delta}\!=\!0.5$ 
and the extrapolated $H_s^{(c)}\!\approx\!25$~T gives $\bar{\Delta}\!=\!1.5$.
Thus, our choice of $\bar{\Delta}\!=\!1.0$ is well-justified.

\subsection{Spin-wave theory}

The spin-wave spectrum of the stripe phases in zero field and in fields above the saturation 
have been considered in the past \cite{Zhu_PRL_2017, Maksimov_PRX_2019, Li_Arxiv_2016}. 
Here we develop the linear spin-wave theory 
(LSWT) for the field-induced spin canted state in Fig.~\ref{fig_domains}(b). The per-site 
classical energy is  
\begin{align}
\frac{E_\text{cl}}{S^2}=&\Big(J_1+J_2 \Big) \Big(1-2\cos 2\varphi \Big)+
2 J_{\pm\pm} \Big( 1+\cos 2\varphi \Big)\nonumber\\
\label{Ecl}
&-g_{ab}\mu_B H\sin \varphi/S,
\end{align}
with the canting angle $\varphi$. The energy minimization gives
\begin{align}
 \sin \varphi =H/H^{(b)}_s\equiv h,
\end{align}
see Eq.~(\ref{Hs}) for $H^{(b)}_s$.

Some straightforward algebra with the transformation of spins to a local reference frame 
and a Fourier transform give the LSWT Hamiltonian for $S\!=\!\frac12$
\begin{equation}
\label{H2}
\hat{\cal H}^{(2)}=\frac{3J_1}{2}\sum_{{\bf k}} \bigg(A_{{\bf k}} a^\dagger_{{\bf k}}a^{\phantom{\dag}}_{{\bf k}}
-\frac{B_{\bf k}}{2}
\left(a^{\phantom{\dag}}_{{\bf k}}a^{\phantom{\dag}}_{-{\bf k}}+{\rm H.c.}\right)\bigg),
\end{equation}
where $A_{{\bf k}}$ and $B_{{\bf k}}$ are given by
\begin{align}
A_\mathbf{k}&=\frac{8}{3}(1+\alpha-\eta)h^2+\frac{2}{3}(1+\alpha)(1-4h^2)-\frac{8}{3}(1-h^2)\eta
\nonumber\\
&+\bar{\Delta}\gamma_{{\bf k}}+\bar{\gamma}_{{\bf k}}+ h^2\left(\gamma_{{\bf k}}-\bar{\gamma}_{{\bf k}}\right)
-2\eta\left(\gamma_{{\bf k}}^{\prime}
-h^2\left(\gamma_{{\bf k}}+\bar{\gamma}_{{\bf k}}\right)\right)\nonumber\\
\label{AB}
&+\alpha\left(\bar{\Delta}\gamma_{{\bf k}}^{(2)}+\bar{\gamma}_{{\bf k}}^{(2)}+
h^2\left(\gamma_{{\bf k}}^{(2)}-\bar{\gamma}_{{\bf k}}^{(2)}\right)\right),\\
B_\mathbf{k}&=-\bar{\Delta}\gamma_{{\bf k}}+\bar{\gamma}_{{\bf k}}+ 
h^2\left(\gamma_{{\bf k}}-\bar{\gamma}_{{\bf k}}\right)
-2\eta\left(\gamma_{{\bf k}}^{\prime}
-h^2\left(\gamma_{{\bf k}}+\bar{\gamma}_{{\bf k}}\right)\right)\nonumber\\
&+\alpha\left(-\bar{\Delta}\gamma_{{\bf k}}^{(2)}+\bar{\gamma}_{{\bf k}}^{(2)}+
h^2\left(\gamma_{{\bf k}}^{(2)}-\bar{\gamma}_{{\bf k}}^{(2)}\right)\right),\nonumber
\end{align}
where $\alpha\!=\!J_2/J_1$, $\eta\!=\!J_{\pm\pm}/J_1$, and 
\begin{eqnarray}
\gamma_{{\bf k}} \left[\bar{\gamma}_{{\bf k}} \right]
&=&\frac{1}{3}\bigg(\cos k_x\pm 2\cos\frac{k_x}{2}\cos \frac{\sqrt{3}k_y}{2}\bigg),\nonumber\\
\label{gammas}
\gamma_{{\bf k}}^{(2)}\left[\bar{\gamma}_{{\bf k}}^{(2)}\right] 
&=&\frac{1}{3} \bigg(\cos \sqrt{3}k_y\pm 2\cos\frac{3k_x}{2}\cos \frac{\sqrt{3}k_y}{2}\bigg), \ \ \  \\
\gamma^{\prime}_{{\bf k}} 
&=&\frac{1}{3}\bigg(\cos k_x+\cos\frac{k_x}{2}\cos \frac{\sqrt{3}k_y}{2}\bigg).\nonumber
\end{eqnarray}
The standard Bogolyubov transformation of Eq.~(\ref{H2}) yields the magnon energy for 
$0\!\leq\!H\!\leq\!H_s^{(b)}$
\begin{align}
\label{Ek}
\varepsilon_\mathbf{k}=\frac{3J_1}{2}\sqrt{A_\mathbf{k}^2-B_\mathbf{k}^2}\ .
\end{align}
In Fig.~\ref{fig_EkH0}, we show the 3D plots of the zero-field magnon energy $\varepsilon_\mathbf{k}$
in the stripe-${\bf x}$ phase of Fig.~\ref{fig_domains}(a), throughout the Brillouin zone, and 
for two representative sets of parameters that are chosen to match the experimental zero-field gap 
$\Delta^{exp}_{H=0}\!\approx\!1$~K  and the saturation field of $H_s^{exp}\!\approx\!4.8$~T. 
The main message is that despite a rather drastic difference of the $XXZ$ anisotropy parameter 
between Fig.~\ref{fig_EkH0}(a) and Fig.~\ref{fig_EkH0}(b),
the minima of the spectrum are at the M and M$^{\prime}$-points that are complementary to the 
ordering vector of the ground state, which is at the Y-point. In general, the structure of the low-energy 
part of the spectrum is rather robust to the parameter choices and consists of a quasi-degenerate 
region in ${\bf k}$-space connecting  M and M$^{\prime}$-points, see also 
Refs.~\cite{Zhu_PRL_2017, Maksimov_PRX_2019} 
for unrelated choices of parameters exhibiting the same pattern. 
The major difference between Fig.~\ref{fig_EkH0}(a) and Fig.~\ref{fig_EkH0}(b) is that the 
maximum of the magnon band migrates from the $\Gamma$-point to Y-point upon reducing 
 $\bar{\Delta}$ from 1 to 0.

\subsection{Gaps}
\vskip -0.2cm

From Eqs.~(\ref{AB}) and (\ref{Ek}), one can readily obtain the analytic expressions for the 
magnon energy gaps at the high-symmetry ${\bf k}$-points of interest
\begin{align}
&E_M=E_M^0\sqrt{1+2h^2}\ , \quad\mbox{with}\nonumber\\
\label{EM}
&E_M^0=\sqrt{-2J_{\pm\pm}\left(-4J_{\pm\pm}+\left(J_1+J_2\right)\left(1-\bar{\Delta}\right) \right)} ,\\
&E_Y=E_Y^0\sqrt{1-h^2}\ , \quad\mbox{with}\nonumber\\
\label{EY}
&E_Y^0=\sqrt{h_s^{(b)}\left(-4J_{\pm\pm}+\left(J_1+J_2\right)\left(1-\bar{\Delta}\right) \right)} ,
\end{align}
where $J_{\pm\pm}\!<\!0$, $E_{M(Y)}^0$ are  zero-field gaps,  $h_s^{(b)}$ is from Eq.~(\ref{Hs}), 
and $h\!=\!H/H_s^{(b)}$. 
The ``ordering gap'' at the Y-point vanishes at the critical field for a transition to a 
paramagnetic state, as is expected.
On the other hand, the ``accidental gap'' at the M-point, which is the spectrum minimum in zero field, 
increases with the field. Clearly, one should expect their crossing at some  $H\!<\!H_s^{(b)}$.

\begin{figure}[t]
\centering
\includegraphics[width=0.99\linewidth]{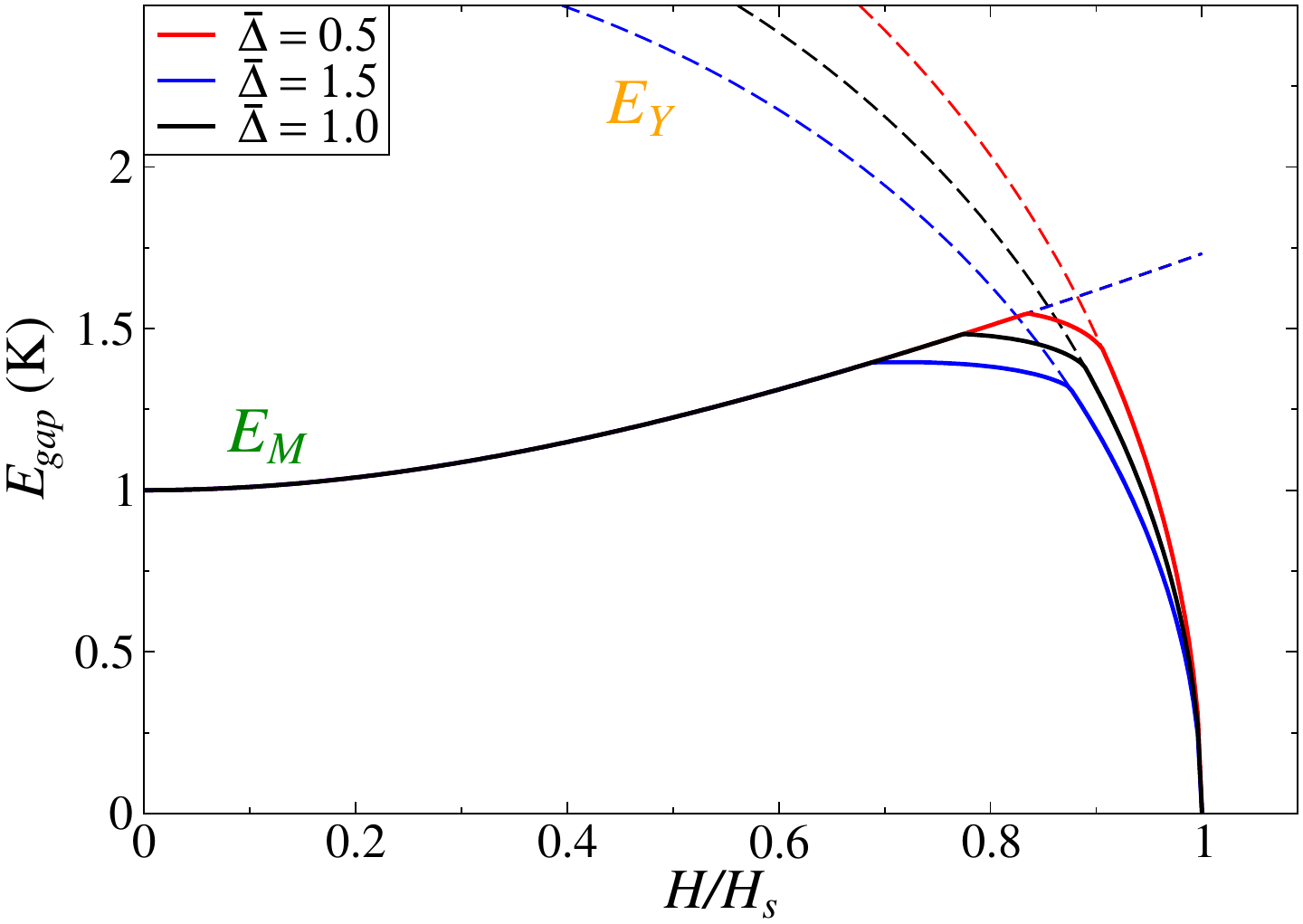}
\caption{Minimum of the spectrum $\varepsilon_{\bf k}$  (solid lines) 
and $E_{M(Y)}$ energies (dashed lines) vs $H/H_s$ for the three parameter sets.}
\label{fig_gap1}
\vskip -0.2cm
\end{figure}
\subsection{Parameter sets}

Before we proceed with the modeling of the CeCd$_3$As$_3$ spectrum, we need to specify parameters 
of the model that meet the phenomenological criteria for it.
As is discussed above, we have already chosen $J_{z\pm}\!=\!0$, 
fixed $J_2\!=\!0.1J_1$, and assumed  $J_{\pm\pm}\!<\!0$.
For the remaining parameters, $J_1$, $\bar{\Delta}$, and $J_{\pm\pm}$, we have 
two strong constraints, the zero-field gap 
$\Delta^{exp}_{H=0}\!\approx\!1$~K  and the in-plane saturation field $H_s^{exp}\!\approx\!4.8$~T,
and one ``soft'' constraint for the out-of-plane saturation field $H_s^{(c)}$ to be within 15--30~T window.

Since the theoretical value for $H_s^{(c)}$ is the only quantity that strongly depends on $\bar{\Delta}$, 
we use the latter constraint to provide us with the broad bounds it. Then, we choose 
several reasonable values of $\bar{\Delta}$ and use the two strong criteria to fix $J_1$ and $J_{\pm\pm}$.

Roughly speaking, the zero-field gap at the M-point from Eq.~(\ref{EM}) fixes $J_{\pm\pm}$ and 
the in-plane saturation field $H_s^{(b)}$ from Eq.~(\ref{Hs}) fixes $J_1$. 
Neglecting $J_{\pm\pm}$ from Eq.~(\ref{Hs}) and setting $\bar{\Delta}\!=\!1$ in Eq.~(\ref{EM}) yield the
estimates $J_1\!\approx\!1.5$~K and $J_{\pm\pm}\!\approx\!0.35$~K.
More precise calculations, using $E_M^0\!=\!\Delta^{exp}_{H=0}$ with Eq.~(\ref{EM}),  
$H_s^{(b)}\!=\!H_s^{exp}$ with Eq.~(\ref{Hs}), and  the in-plane 
$g$-factors $g_{ab}\!=\!2.0$ and $g_{c}\!=\!0.49$ as discussed above, produce three representative sets
\begin{center}
\vskip -0.1cm
\begin{tabular}{c|c|c|| c }
  \hline
     $\bar{\Delta}$ & \ \ \ \ \ \  $J_1$\ \ \  \ \ \   & \ \ \ \ \ \  $J_{\pm\pm}$ \ \ \ \ \ \  &  
     \ \ \ \ \ \ $H_s^{(c)}$ \ \ \ \ \ \ \\   \hline  \hline
\ \ \ \ \ \ 0.5 \ \ \  \ \ \          &   1.21~K               &             -0.280~K                   &              13.0~T \\   \hline
1.0                           &        1.14~K          &              -0.354~K                    &            19.0~T    \\   \hline
1.5                           &        1.07~K          &               -0.435~K                    &  24.6~T  \\   \hline
\end{tabular}
\end{center}
\vskip -0.cm
which we will refer to by their respective choices of the $XXZ$ anisotropy parameter $\bar{\Delta}$
in the first column. The last column shows the out-of-plane saturation field $H_s^{(c)}$ that corresponds 
to each set, giving a sense that the choices of $\bar{\Delta}\!=\!0.5$ and $\bar{\Delta}\!=\!1.5$ are likely 
to be outliers and  $\bar{\Delta}\!=\!1$ is a reasonable choice.


\subsection{Gap vs field results}

We are now set to study the spectral properties of  CeCd$_3$As$_3$. Our Fig.~\ref{fig_gap1}
shows the field dependence of the spectrum minimum for the three parameter sets introduced above;
the energies are in K and the field is in units of $H_s^{(b)}$.
The solid line traces the true ``minimal gap'' of the spectrum $\varepsilon_{\bf k}$ in Eq.~(\ref{Ek}), while
the dashed lines track the energies of the M and Y-points, Eqs.~(\ref{EM}) and (\ref{EY}), respectively.
The dashed curves intersect  below $H_s$, as anticipated. 
The overall behavior of the minimal gap is notable: a gradual increase followed by a rather abrupt
transition to a steep decrease and closing at the critical point, with the results that only moderately 
depend on the $XXZ$ anisotropy in the allowed range.

\begin{figure}[t]
\centering
\includegraphics[width=0.99\linewidth]{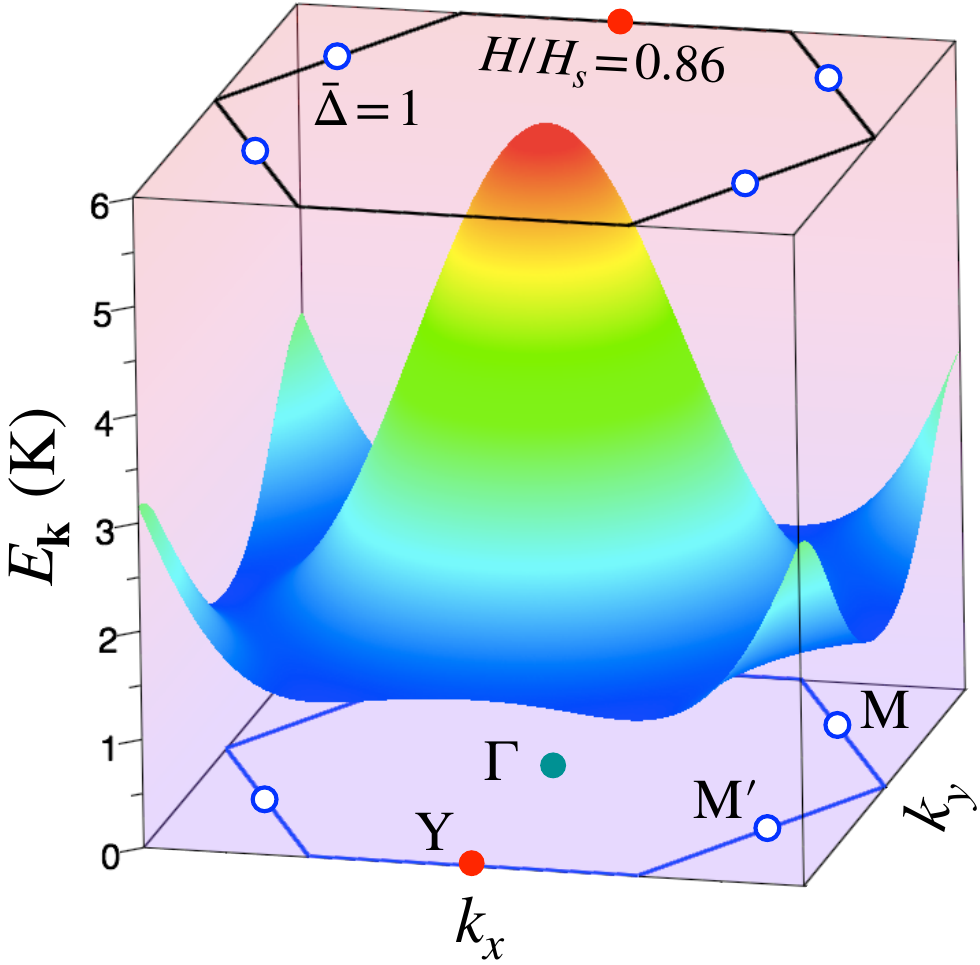}
\caption{$\varepsilon_{\bf k}$ at $H\!=\!0.86H_s$ for the $\bar{\Delta}\!=\!1$ parameter set.}
\label{fig_Ekh}
\vskip -0.2cm
\end{figure}

The true minimal gap in Fig.~\ref{fig_gap1} has some field region where it is neither at M nor at Y-point, but  
at the ${\bf k}$-points that are intermediate between them, the situation that is 
illustrated in Fig.~\ref{fig_Ekh} for the field $H\!=\!0.86H_s$ and for the $\bar{\Delta}\!=\!1$ parameter set,
for which the spectrum minimum is nearly degenerate along a contour that includes M, M$^{\prime}$, and Y-points.  

\begin{figure}[t]
\centering
\includegraphics[width=0.99\linewidth]{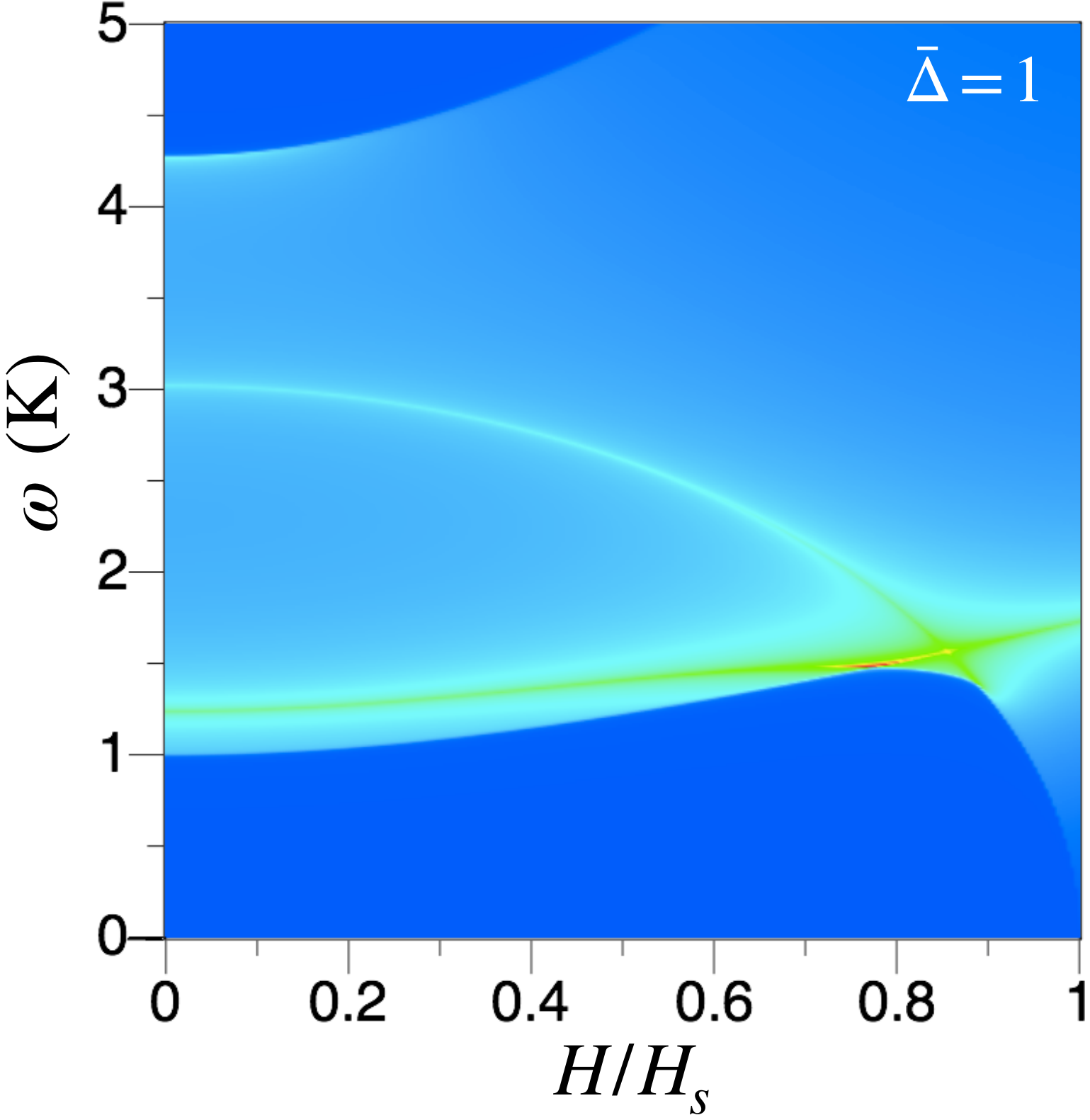}
\caption{Magnon DoS vs $H/H_s$ for the $\bar{\Delta}\!=\!1$ parameter set.}
\label{fig_DoS}
\vskip -0.2cm
\end{figure}

A complementary prospective is also offered by the magnon density of states (DoS)
in Fig.~\ref{fig_DoS}, the quantity 
that is directly related to the specific heat. This figure explicitly shows the 
field-induced spectral weight redistribution due to the gap crossing and Van Hove singularities 
associated with the spectrum degeneracies. It  
suggests that the higher density of states may lead to additional features in the specific heat. 

\subsection{Polarized phase}

\vskip -0.2cm

Above the saturation field, spins' quantization axis aligns with the field direction 
and the spin-wave algebra simplifies considerably \cite{Zhu_PRL_2017, Li_Arxiv_2016}.
For the field  $H\!\geq\!H_s^{(b)}$ in the $b$-direction, 
the  $A_{{\bf k}}$ and $B_{{\bf k}}$ terms in the LSWT Hamiltonian 
(\ref{H2}) are given by
\begin{align}
\widetilde{A}_\mathbf{k}&=\frac{8}{3}(1+\alpha-\eta)h-2(1+\alpha)\nonumber\\
\label{ABh}
&+\left(1+\bar{\Delta}\right)\left(\gamma_{{\bf k}}+\alpha\gamma_{{\bf k}}^{(2)}\right)
+2\eta\gamma_{{\bf k}}^{\prime\prime},\\
\widetilde{B}_\mathbf{k}&=\left(1-\bar{\Delta}\right)\left(\gamma_{{\bf k}}+\alpha\gamma_{{\bf k}}^{(2)}\right)
+2\eta\gamma_{{\bf k}}^{\prime\prime},\nonumber 
\end{align}
where $\alpha\!=\!J_2/J_1$, $\eta\!=\!J_{\pm\pm}/J_1$, $h\!=\!H/H_s^{(b)}$ as before,
$\gamma_{{\bf k}}$ and $\gamma_{{\bf k}}^{(2)}$ given in Eq.~(\ref{gammas}) and 
\begin{eqnarray}
\label{gammapp}
\gamma^{\prime\prime}_{{\bf k}} 
=\frac{1}{3}\bigg(\cos k_x-\cos\frac{k_x}{2}\cos \frac{\sqrt{3}k_y}{2}\bigg).
\end{eqnarray}
We note that in the case of $H\!\parallel\! a$ the expression is the same up to the change 
$J_{\pm\pm} \!\rightarrow \! -J_{\pm\pm}$.

The energy spectrum has the same form as in (\ref{Ek})
\begin{align}
\label{Ekh}
\varepsilon_\mathbf{k}=\frac{3J_1}{2}\sqrt{\widetilde{A}_\mathbf{k}^2-\widetilde{B}_\mathbf{k}^2}\ .
\end{align}
At the saturation field, Eqs.~(\ref{Ek}) and (\ref{Ekh}) yield the same result.
The spectrum has a gapless mode at the Y-point that has an acoustic character, i.e.,
$\varepsilon_\mathbf{k}\!\propto\! |\delta{{\bf k}}|$, where $\delta{{\bf k}}\!=\!{\bf k}-{\bf k}_Y$. 
This is the behavior that is generally expected for the transitions in anisotropic systems, see 
more discussion in Sec.~\ref{Sec_asympt}.
 
Above the saturation field, the gap at the Y-point reopens, but the spectrum does not experience 
any significant transformations aside from a roughly uniform, Zeeman-like shift of the spectrum 
as a whole. That is, the Y-point remains a minimum for all $H\!\geq\!H_s^{(b)}$ with the gap
\begin{align}
\label{EYh}
&\widetilde{E}_Y=\sqrt{\left(\widetilde{h}-h_s^{(b)}\right)\left(\widetilde{h}-
\left(J_1+J_2\right)\left(3+\bar{\Delta}\right) \right)} ,
\end{align}
where $h_s^{(b)}$ is from Eq.~(\ref{Hs}) and $\widetilde{h}\!=\!g_{ab}\mu_B H$. 

\begin{figure}[t]
\centering
\includegraphics[width=0.99\linewidth]{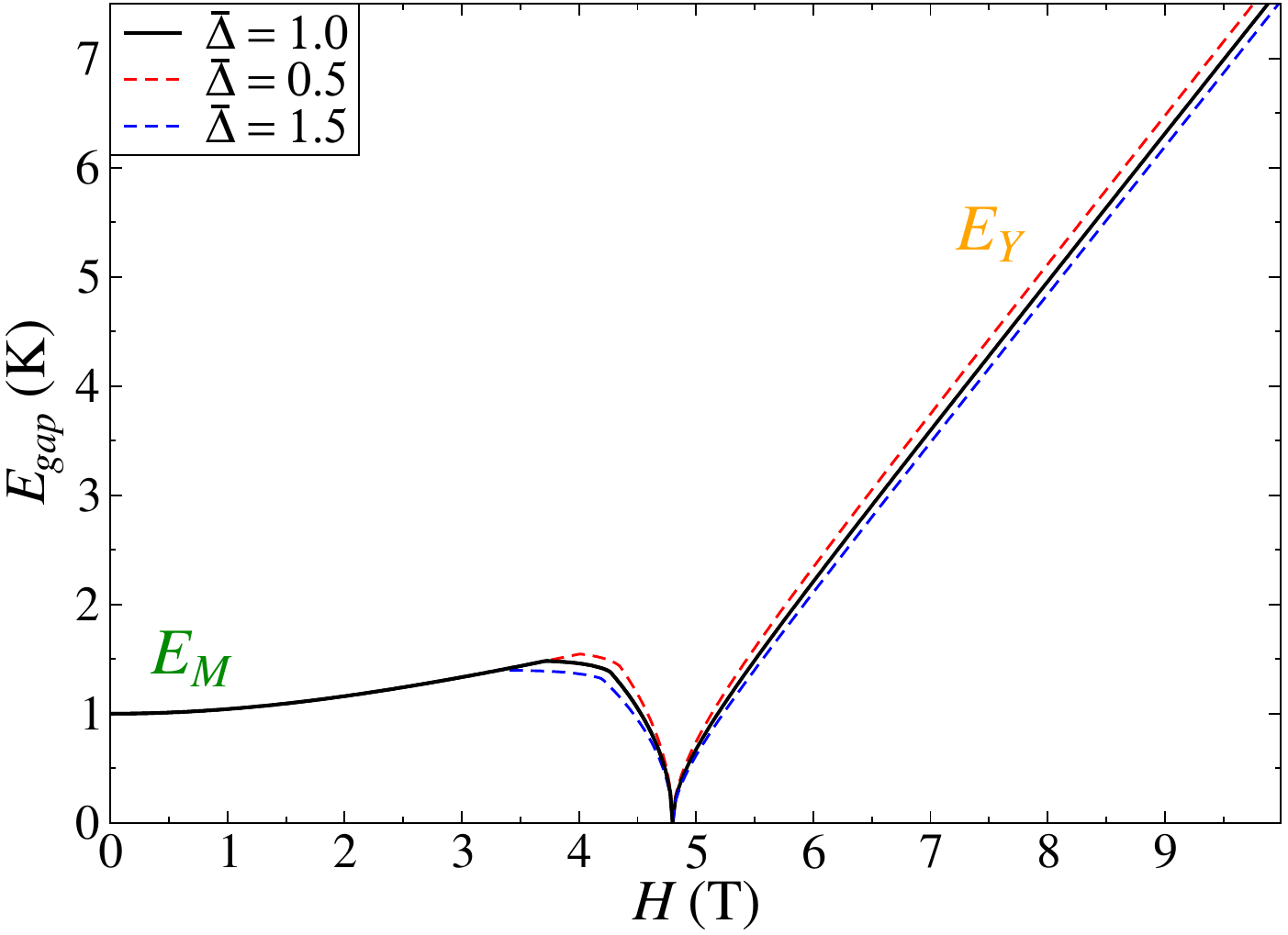}
\vskip -0.2cm
\caption{Minimum of the spectrum $\varepsilon_{\bf k}$   vs $H$ for the three parameter sets,
and $g_{ab}\!=\!2.0$.}
\label{fig_gap2}
\vskip -0.2cm
\end{figure}

In Fig.~\ref{fig_gap2}, we present the 
field-dependence of the spectrum minimum  in both regions, $H\!\leq\!H_s^{(b)}$ and 
$H\!\geq\!H_s^{(b)}$, for the three parameter sets and  $g_{ab}$ discussed above.

\begin{figure}[t]
\centering
\includegraphics[width=0.99\linewidth]{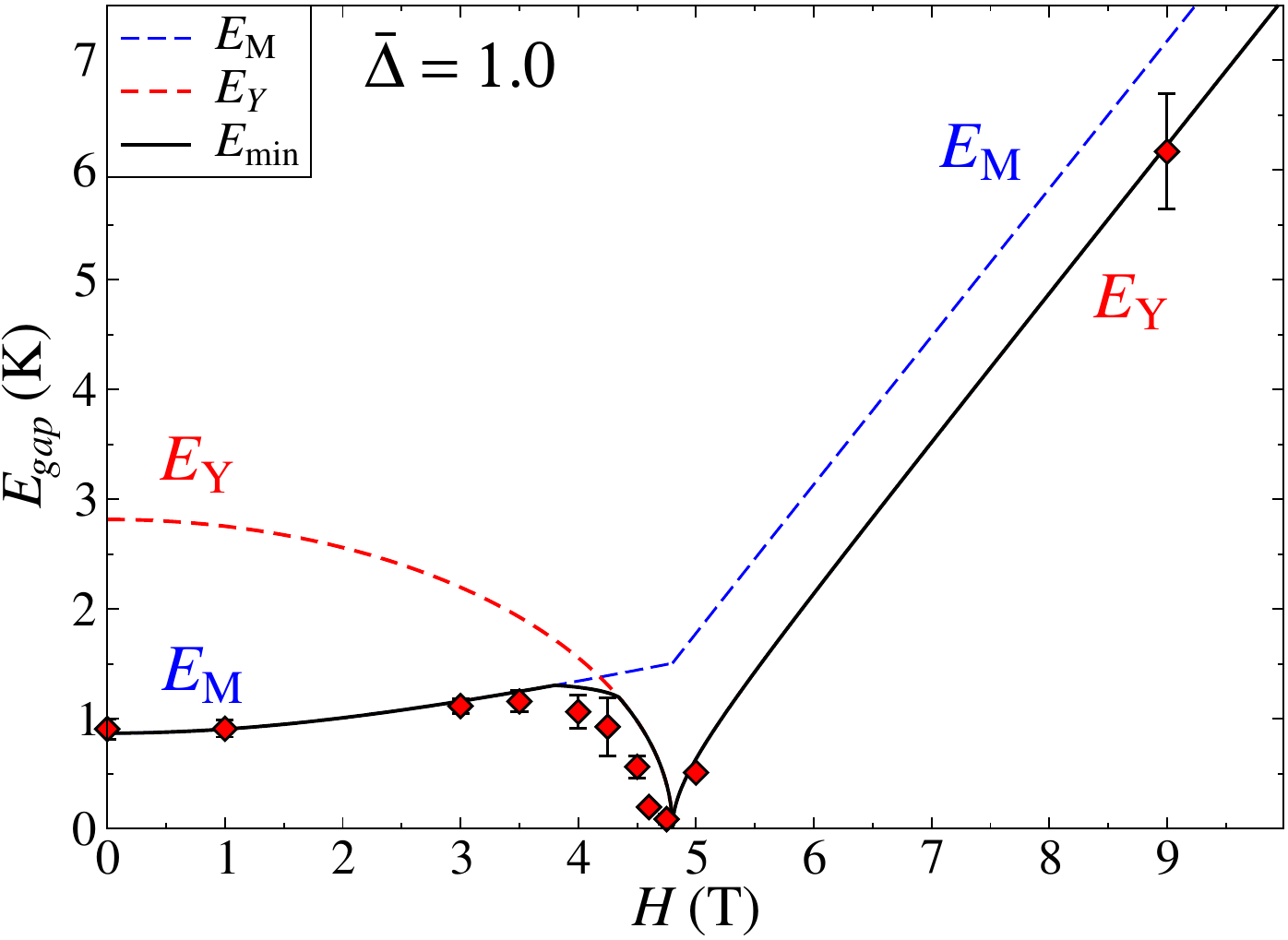}
\vskip -0.2cm
\caption{Fig.~3 of the main text on the linear scale.}
\label{fig_gap_exp}
\end{figure}

Our Fig.~\ref{fig_gap_exp} reproduces Fig.~3 of the main text on the linear scale. It shows the 
``minimal gap'' together with the gaps at the M- and Y-points throughout the entire field regime  
for the $\bar{\Delta}\!=\!1$ parameter set and together with experimental data. 
We point out again that after $\bar{\Delta}$ and $J_2/J_1$ are fixed, the remaining parameters
of the model, $J_1$ and $J_{\pm\pm}$, are fully constrained by the saturation field
value $H_s^{exp}$ and the zero-field gap $E^{exp}_{gap}(H\!=\!0)$. There are no free parameters left. 
Yet the theoretical curve goes right through the value of the gap at 9~T with no fitting.

\subsection{Ising phase gaps}
\label{Sec_gaps}
\vskip -0.2cm

As was discussed earlier, the only feasible alternative to the gapped stripe state is an Ising-like state 
that does not need bond-dependent terms at all. However, on a triangular lattice, the simplest 
candidate, the nearest-neighbor-only phase, is a gapless and ferrimagnetic.  It is a 
relative of the $120{\degree}$ phase with the plane containing the triad of spins turned perpendicular 
to the basal plane and spins forming a deformed ``Y'' structure with a net magnetic moment \cite{Miyashita_JPSJ_1986}.  

The gapped state is reached with the help of a finite $J_2$ and is a stripe-like phase with spins pointing along the
$z$-axis, referred to as a stripe-${\bf z}$ phase. It is known to exhibit a staircase of the spin-flop transitions
starting at low fields in the $c$-direction \cite{Seabra_PRB_2011}, 
the features that are not observed in the $M(H)$ data CeCd$_3$As$_3$.

\begin{figure}[t]
\centering
\includegraphics[width=0.99\linewidth]{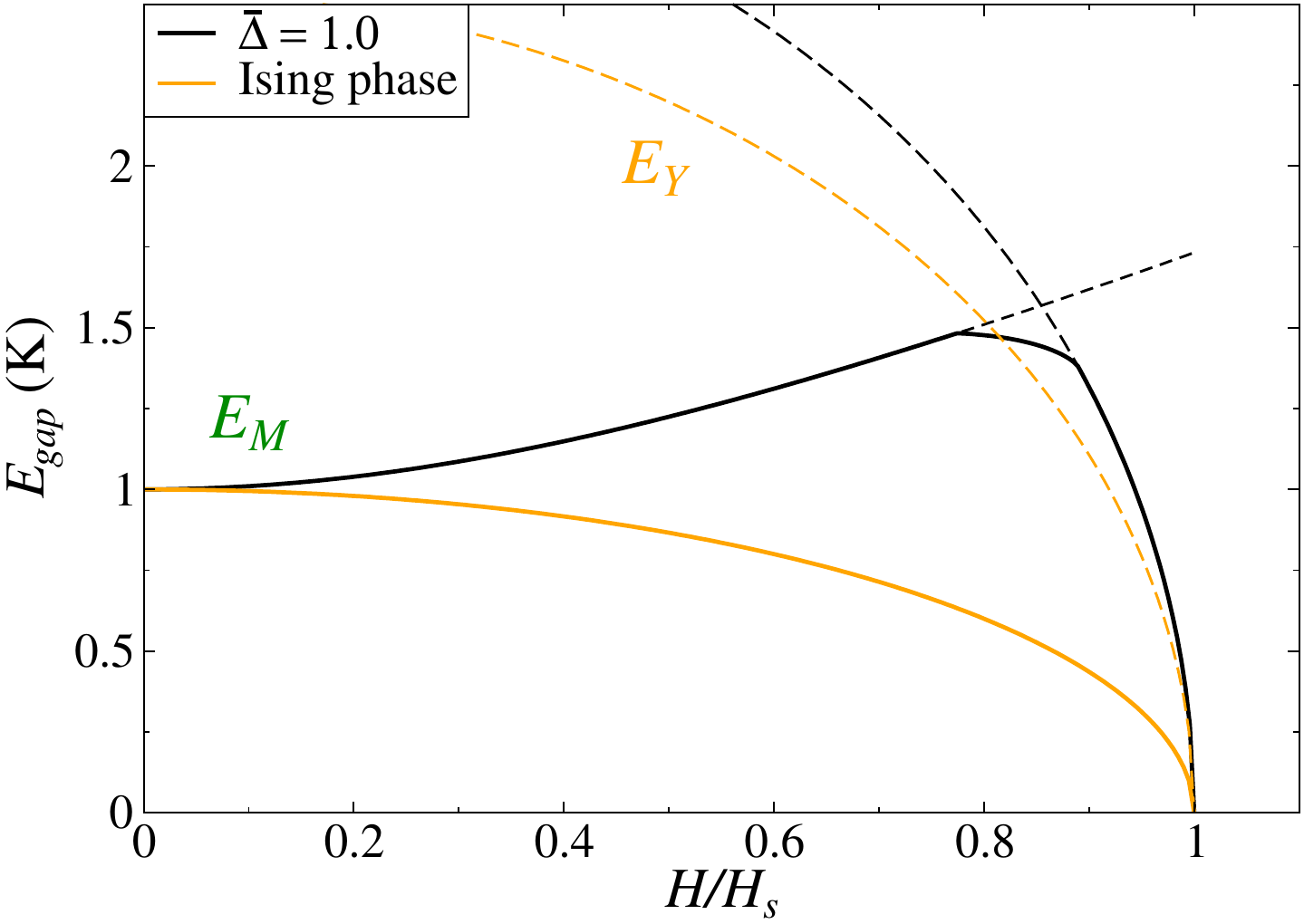}
\vskip -0.2cm
\caption{Same as Fig.~\ref{fig_gap1} for the Ising-like stripe-${\bf z}$ phase.}
\label{fig_Ising}
\end{figure}

Nevertheless, we have volunteered to provide a study of the field-dependence of the spectrum gap 
in this phase for the in-plane field. Our Fig.~\ref{fig_Ising} shows the 
the excitation energies of the ``accidental'' and ``ordering'' M and Y-points for a 
representative parameter set from this phase: $J_1\!=\!1.13$~K, $J_2\!=\!0.2J_1$,  $J_{\pm\pm}\!=\!0$,
and $\bar{\Delta}\!=\!1.735$.  This set of parameters matches the zero-field gap value of 1~K and the
saturation field value of 4.8~T with the same $g$-factor as before. Having $J_{\pm\pm}\!\neq\!0$
does not change the field-dependence of the gaps in the stripe-${\bf z}$ phase in any quantitative way.
These results are also compared to the ones for the $\bar{\Delta}\!=\!1.0$ 
stripe-${\bf x}$ parameter set from Fig.~\ref{fig_gap1}.

One can see a markedly different behavior of the gaps in the stripe-${\bf z}$ phase. 
The gap at the M-point remains an absolute minimum of the spectrum for the entire field range,  is 
monotonically decreasing vs field, and  vanishes at the critical point together with the gap at the ordering vector
(Y-point). This is because for the stripe-${\bf z}$ state at the full polarization point, all three stripe domains of 
different orientation are degenerate. There is no level-crossing in the spectrum and no abrupt change in the 
minimal gap behavior vs field. Needless to say, this is inconsistent with the CeCd$_3$As$_3$ phenomenology.


\section{N\'eel temperature}

We have modified the so-called self-consistent random-phase approximation (RPA) 
to calculate N\'{e}el ordering temperature and its field-dependence for the selected parameter sets for the 
stripe state. The self-consistent RPA approach is based on the mean-field decoupling of the 
equations of motion for the spin Green's functions and has been recently employed in the context of the 
anisotropic-exchange systems, see Refs.~\cite{Maksimov_PRR_2020, Maksimov_PRX_2019} for details. 

In our case, the RPA approach for the ordering temperature  $T_N$ 
needed to be modified to account for the order parameter corresponding to the 
component of the ordered moment that is transverse to the external field. The result 
is particularly simple
\begin{align}
\frac{1}{T_N}=\frac{3J_1\cos\varphi}{N}\sum_{\mu,\mathbf{k}} \frac{A_{\mathbf{k}}}{\varepsilon^2_{\mathbf{k}}},
\label{eq_tn}
\end{align}
where $\varphi$ is the spin canting angle in Fig.~\ref{fig_domains}(b).

\begin{figure}[t]
\centering
\includegraphics[width=0.99\linewidth]{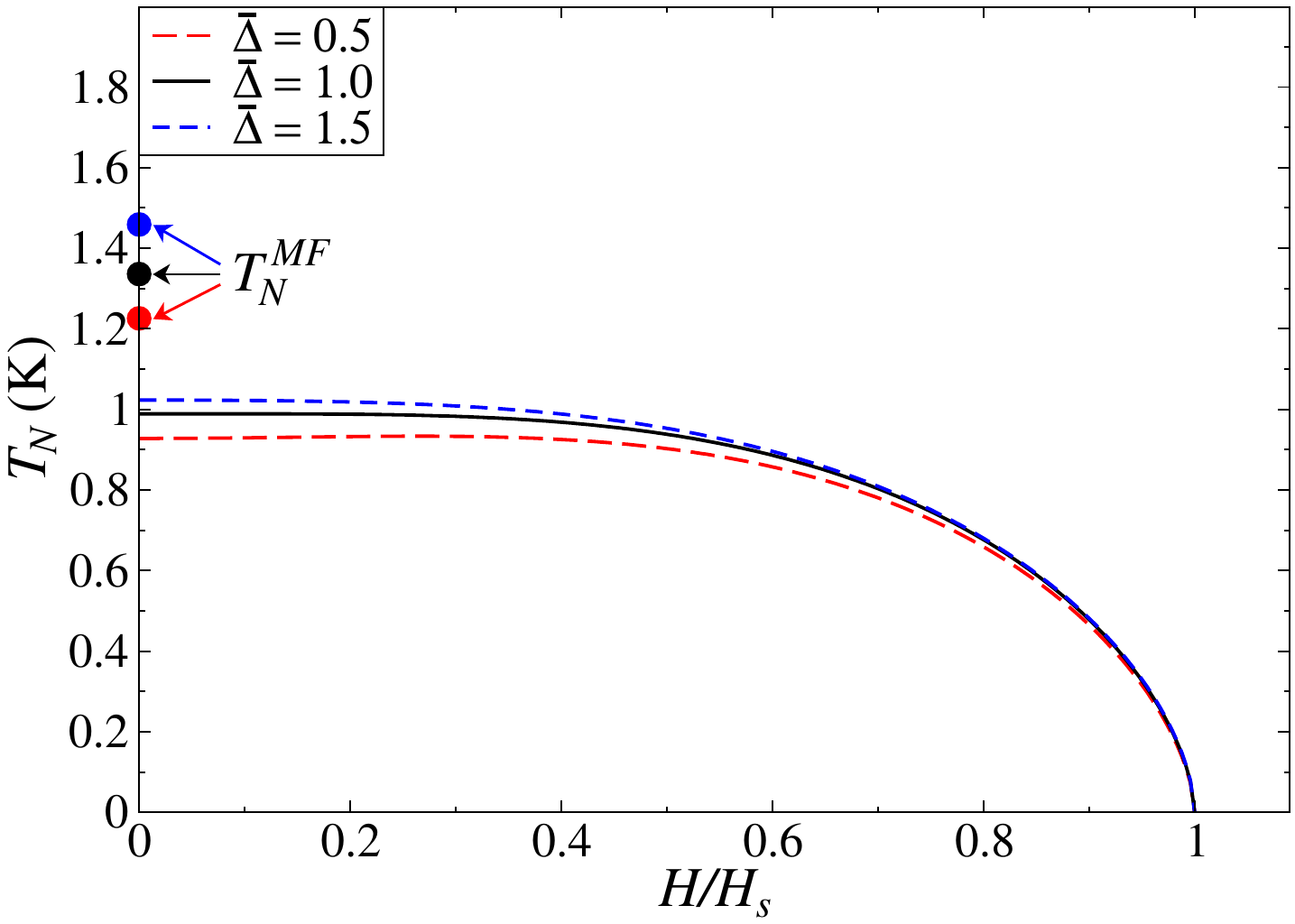}
\vskip -0.2cm
\caption{$T_N$ vs field by the self-consistent RPA approach for the three parameter sets.
Mean-field (MF) results for zero field are shown for comparison.}
\label{fig_TN}
\end{figure}

Our Fig.~\ref{fig_TN} presents the results of such calculations.
The overall shape of the $H$--$T$ phase diagram is in a general accord with the data for 
CeCd$_3$As$_3$. However, there are two significant differences. While the RPA method offers  
a significant improvement over the ``bare'' mean-field values \cite{Maksimov_PRX_2019}, the absolute values of $T_N$ 
are still at least a factor of 
two larger than in the experiment. The second crucial difference is the lack of the notable initial increase in the
$T_N$ vs field in the RPA results compared with the experimental data, with the latter suggesting a linear slope, 
$\delta T_N\propto\!|H|$. The observations in a different Ce-based compound also indicate a 
similar increase \cite{Bastien_Arxiv_2020}. 

Both discrepancies are of the same origin.
The quantitative successes and failures of the RPA method are known \cite{Komleya_Arxiv_2020}, but the reason for them 
is not properly discussed. The key issue is that this method is 
based on the picture of thermal reduction of the local order parameter and, thus, 
is just a glorified Lindemann criterion for melting of a given spin order. 
It is quite successful quantitatively in the 2D systems with continuous symmetries, in which the ordering happens 
when the 2D correlation length is very large and the fluctuating component into the 3D ordering is small. 
For all other cases, since it does not include critical fluctuations into consideration, it fails. 
We can also state that since the RPA results for 
$T_N$ (\ref{eq_tn}) are based on the spectrum  $\varepsilon_{\bf k}$ from Eq.~(\ref{Ek}), 
they can only contain terms $\propto\!H^2$, but not the linear term.
Therefore, the origin of the latter is most likely in the field-induced suppression of the critical fluctuations.
At larger fields, the order parameter reduction due to spin canting dominates and the agreement improves.


\section{Specific heat and various asymptotes}
\label{Sec_asympt}

At low temperatures, population of spin excitations is low and can be approximated by a 
bosonic statistics. This is also a description of such excitations within the spin-wave approach.
Then, the specific heat is given by 
\begin{align}
\label{eq_Cv}
C(T)=\frac{\partial E}{\partial T}=\sum_{\bf k} \left(\frac{\varepsilon_{\bf k}}{T}\right)^2
\frac{e^{\varepsilon_{\bf k}/T}}{(e^{\varepsilon_{\bf k}/T}-1)^2}\ .
\end{align}
The results obviously depend on the spectrum $\varepsilon_{\bf k}$ and on the dimensionality of the system.
We use this expression to obtain the leading, 
asymptotically correct terms in the specific heat  $T$-dependence   in 2D
for both gapped and gapless spectra at low temperatures. 
Since the results are expected to be generic, the Debye approximation, that uses the low-energy form of the 
spectrum and a cut-off Debye momentum,  suffices.


\subsection{Gapped spectrum}

For the gapped spectrum near the minimum, spin-excitation energy can be approximated as 
\begin{align}
\label{eq_Ek2}
\varepsilon_{\bf k}\approx\Delta +J  {\bf k}^2 \ ,
\end{align}
with the ``kinetic'' energy   that can be related to the bandwidth $W\!=\!J k_D^2$, where $k_D$
is the Debye momentum. In 2D, this yields   $C(T)$ in the Debye approximation
\begin{align}
\label{eq_Cvk2}
C(T)=\frac{T}{W}\int_{x_0}^{x_{m}} dx \ x^2\, \frac{e^{x}}{(e^{x}-1)^2}\ ,
\end{align}
where $x_0\!=\!\Delta/T$ and  $x_{m}\!=\!x_0+W/T$. In the limit $T\!\ll\!\Delta$, $T\!\ll\!W$, 
a simple algebra gives
\begin{align}
\label{eq_Cvk2_final}
C(T)\approx \frac{T}{W}\left(\frac{\Delta^2}{T^2}+\frac{2\Delta}{T}+2\right)e^{-\Delta/T}\,
\end{align}
with the omitted terms of order $O\left(e^{-2\Delta/T};e^{-\Delta/T-W/T}\right)$.

A different, ``relativistic'' form of the gapped spectrum can be of interest, especially close to the 
critical field
\begin{align}
\label{eq_Ekr}
\varepsilon_{\bf k}\approx\sqrt{\Delta^2 +J^2  {\bf k}^2} \ .
\end{align}
Following the same steps as above results in 
\begin{align}
\label{eq_Cvkr_final}
C(T)\approx AT\left(\frac{\Delta^2}{T^2}+\frac{3\Delta}{T}+6+\frac{6T}{\Delta}\right)e^{-\Delta/T}\,
\end{align}
with the leading term coinciding with the one in Eq.~(\ref{eq_Cvk2_final}),
both corresponding to an activated behavior of the specific heat with the leading  
$T^{-d+1}$ prefactor.
These expressions allow to extract the value of the lowest gap in the spectrum
from the activated behavior of $C(T)$. 


\subsection{Gapless spectrum at the critical point}

The field-induced transitions of an antiferromagnet to a saturated, paramagnetic state are 
common. For the systems with the continuous spin symmetries, the magnetic field 
coupling is to a conserved total magnetization and the transition is of the Bose-Einstein 
condensation type \cite{Batyev_SPJETP_1984, Zapf_RMP_2014}. In this case, the dispersion relation of the bosonic
excitations at the QCP  is $\varepsilon_{\bf k}\!\propto\!{\bf k}^2$ (dynamical exponent $z\!=\!2$).
Above the saturation field the spectrum is Zeeman-shifted, but otherwise unmodified. 

In our case, because of the symmetry-breaking anisotropic terms, magnetization is not a conserved 
quantity and the transition is of a different universality class, necessarily resulting in an acoustic-like 
$\varepsilon_{\bf k}\!\propto\!|{\bf k}|$ (dynamical exponent   $z\!=\!1$), see Ref.~\cite{Zapf_RMP_2014}.
Following the derivation that is analogous to the textbook one for phonons, 
the  leading term in the specific heat is  a power law
\begin{align}
\label{eq_Cvl_final}
C(T)\approx AT^2\,
\end{align}
with the power $d\!=\!2$ and $A$ is a constant.

Tracking the field-dependent specific heat should allow to identify the field value that yields 
such a behavior at low temperatures. In practice, it may be difficult to locate such a point exactly, so 
a different ``double asymptotic'' expansion may be useful. One needs to consider specific heat (\ref{eq_Cv}) 
for the relativistic dispersion of Eq.~(\ref{eq_Ekr}) in the limit of $T\!\gg\!\Delta$, but still  $T\!\ll\!W$. 
A straightforward algebra yields a gap-dependent correction to (\ref{eq_Cvl_final})
\begin{align}
\label{eq_Cvl_as}
C(T)\approx AT^2\left(1-\frac{\Delta^2}{\alpha T^2}\right)\,
\end{align}
where $\alpha\!=\!12\zeta(3)$ with $\zeta(3)\!\approx\!1.2$. This 
approximation should be valid down to $T\!\sim\!\Delta/2$.

In the main text we presented specific heat data for CeCd$_3$As$_3$ 
for two fields near the QCP and their fits using
the asymptotic expressions of Eqs.~(\ref{eq_Cvl_as}) and (\ref{eq_Cvl_final}).
The fits suggest
that the 4.75~T field is at or very close to the QCP as it is well-fit by the $T^2$ power-law, 
while the 4.6~T low-temperature data are well-fit by the asymptote in Eq.~(\ref{eq_Cvl_as}) 
with a small gap of $\Delta\!=\!0.16$~K.
We would like to point out that the success of these fits may be fortuitous 
as the ``real'' LSWT dispersions from Eq.~(\ref{Ek}) also contain the non-linear terms that can
affect the asymptotic behavior  in this temperature regime.


\subsection{Other anomalies in the specific heat}

In the CeCd$_3$As$_3$, in the field region between $\sim\!4$~T and $4.6$~T approaching the QCP,  
specific heat data demonstrates additional feature besides the one that is associated with the proper 
phase transition, potentially suggesting a sliver of another phase. 

As was discussed in Sec.~\ref{Sec_stripe_theory}~G, one possible scenario of the origin of this feature is in the 
transformation of the spin-excitation spectrum, which occurs in a similar field range and 
is responsible for additional Van Hove singularities that may contribute to the specific heat. 
Unfortunately, the characteristic temperatures of these features are such that the 
bosonic approximation for the specific heat may be unreliable. We, therefore, cannot 
substantiate this scenario as unbiased numerical methods are needed.  
We point out that some frustrated models on the triangular lattice consistently demonstrate a two-peak structure 
in their specific heat for some parts of the phase diagram as obtained by exact 
diagonalization \cite{Schmidt_PR_2017}. 
Their origin may be related to the one suggested above.

The second scenario involves real-life complications.
The difference of the critical fields in the $a$ and $b$ directions in Eq.~(\ref{Hs}) and in Fig.~\ref{fig_MvsHab} 
translates into $\Delta H_s\!\alt\!0.5$~T for the realistic parameters that we study. 
This is about the same as the width 
of the suggested additional phase in the $H$--$T$ phase diagram of CeCd$_3$As$_3$. Thus, one cannot exclude that 
the additional features are associated with  different transition temperatures  in  
different stripe domains of the type discussed in Sec.~\ref{Sec_stripe_theory}~B and shown in 
Fig.~\ref{fig_domains}, which survive because of  pinning by disorder.
Within this scenario, it needs to be explained, though, why they do not show up at the lower fields.

Lastly, another rare-earth triangular-lattice material, YbMgGaO$_4$, 
has also been suggested to belong to a stripe phase, which, however,  is ``molten'' by an inter-site 
mixing of the non-magnetic Mg and Ga ions that affects magnetism in Yb$^{+3}$ layers, 
leading to a spin-liquid ``mimicry'' \cite{Zhu_PRL_2017}.
The key differences of CeCd$_3$As$_3$ and its stripe phase are the following. 
First, there is no Cd/As site-mixing in the case of CeCd$_3$As$_3$, only a partial 
occupation of sites, with the latter  not changing the local charge environment in the 
immediate neighborhood of Ce sites, see Fig.~\ref{fig:Crystal}.
Thus, the partial occupation of  the  Cd and As sites can be 
expected to be much less disruptive than a random charge distribution due to Mg/Ga mixing 
in YbMgGaO$_4$, which is the likely source of disorder in its exchange parameters as the 
low-energy doublet structure is affected by random component of   crystal field 
\cite{Paddison_NatPhys_2019, Zhu_PRL_2018}.
The second important difference is that the proposed stripe-phase in YbMgGaO$_4$ is nearly 
gapless and belongs to a different region of the phase diagram in a close proximity of the 120$\degree$
region. This is also corroborated by an indication of plateau-like transitions in external field \cite{Steinhardt_2020}.
In case of CeCd$_3$As$_3$, a strong gap ``protects'' from randomization of exchange parameters.
This may also make a disorder relevant to the appearance of the ``intermediate'' phase upon 
suppression of the gap with the field.

\bibliography{Bibby}